\documentclass[aps, rmp,reprint, amsmath,amssymb, superscriptaddress]{revtex4-1}

\usepackage{graphicx}
\usepackage{dcolumn}
\usepackage{bm}
\usepackage{lineno}

\usepackage{color,graphicx,subfigure}
\usepackage{enumitem}
\usepackage{wrapfig}
\usepackage{soul}
\usepackage{amsmath}
\usepackage{titlesec}
\usepackage{multirow}
\usepackage{textcomp}
\usepackage{gensymb}
\usepackage[]{hyperref}
\hypersetup{colorlinks=true,linkcolor=blue,citecolor=blue,urlcolor=blue,pdfpagemode=UseNone}
\usepackage{xcolor}
\usepackage{etoolbox}
\usepackage{floatrow}
\usepackage{braket}

\usepackage[normalem]{ulem}

\DeclareUnicodeCharacter{2009}{}
\DeclareUnicodeCharacter{2212}{}
\DeclareUnicodeCharacter{0081}{}

\begin{document}

\title{Time- and Angle-Resolved Photoemission Studies of Quantum Materials}

\author{Fabio\,Boschini}
\email[]{fabio.boschini@inrs.ca}
\affiliation{Institut National de la Recherche Scientifique -- \'{E}nergie Mat\'{e}riaux T\'{e}l\'{e}communications Varennes QC J3X 1S2 Canada}
\affiliation{Quantum Matter Institute University of British Columbia Vancouver BC V6T 1Z4 Canada}
\author{Marta\,Zonno}
\email[]{marta.zonno@lightsource.ca}
\affiliation{Quantum Matter Institute University of British Columbia Vancouver BC V6T 1Z4 Canada}
\affiliation{Canadian Light Source Saskatoon SK S7N 2V3 Canada}
\author{Andrea\,Damascelli}
\email[]{damascelli@physics.ubc.ca}
\affiliation{Quantum Matter Institute University of British Columbia Vancouver BC V6T 1Z4 Canada}
\affiliation{Department of Physics $\&$ Astronomy University of British Columbia Vancouver BC V6T 1Z1 Canada}

\begin{abstract}
    Angle-resolved photoemission spectroscopy (ARPES) -- with its exceptional sensitivity to both the binding energy and momentum of valence electrons in solids -- provides unparalleled insights into the electronic structure of quantum materials. Over the last two decades, the advent of femtosecond lasers, which can deliver ultrashort and coherent light pulses, has ushered the ARPES technique into the time domain. Now, time-resolved ARPES (TR-ARPES) can probe ultrafast electron dynamics and the out-of-equilibrium electronic structure, providing a wealth of information otherwise unattainable in conventional ARPES experiments. This paper begins with an introduction to the theoretical underpinnings of TR-ARPES followed by a description of recent advances in state-of-the-art ultrafast sources and optical excitation schemes. It then reviews paradigmatic phenomena investigated by TR-ARPES thus far, such as out-of-equilibrium electronic states and their spin dynamics, Floquet-Volkov states, photoinduced phase transitions, electron-phonon coupling, and surface photovoltage effects. Each section highlights TR-ARPES data from diverse classes of quantum materials, including semiconductors, charge-ordered systems, topological materials, excitonic insulators, van der Waals materials, and unconventional superconductors. These examples demonstrate how TR-ARPES has played a critical role in unraveling the complex dynamical properties of quantum materials. The conclusion outlines possible future directions and opportunities for this powerful technique. 
\end{abstract}

\date{\today}

\maketitle

\tableofcontents

\begin{figure*}
\centering
\includegraphics[scale=1]{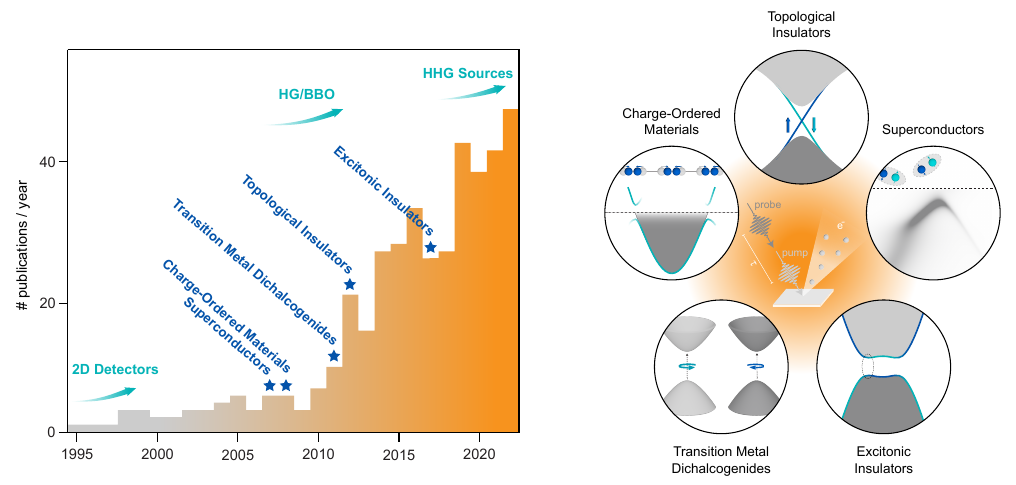}
\caption[Intro]{The rapid adoption of time- and angle-resolved photoemission spectroscopy. Left: increase of the approximate number of annual TR-ARPES publications over the past three decades. Exponential technological progress (milestone advancements in cyan arrows) and interest in new materials (blue stars marking the initial TR-ARPES publications for various classes of materials) have fueled the growth of the TR-ARPES technique over the past two decades. Right: Illustration of key classes of quantum materials extensively studied via TR-ARPES.}
\label{Intro}
\end{figure*}

\section{Introduction} \label{INTRO_section}
The physics of quantum materials -- solids whose macroscopic properties directly manifest quantum mechanical effects \cite{keimer2017physics} -- lies at the heart of current condensed matter research and holds the potential to revolutionize many aspects of everyday life, ranging across energy, transportation, medical and advanced technology applications. In many quantum materials, interactions among electrons themselves or with other degrees of freedom (\emph{e.g.}, orbital, spin, charge and lattice) give rise to various collective excitations. These interactions lead to a number of exotic phenomena, such as high-temperature superconductivity or symmetry-protected topological states. Understanding how electron interactions result in the emergence of novel quantum phases of matter is a central question in the field. Among the experimental techniques sensitive to fermionic quasiparticles, angle-resolved photoemission spectroscopy (ARPES) stands out due to its powerful capability to resolve the one-particle spectral function, which to a first approximation maps the electronic band structure with momentum resolution over the entire 3D Brillouin zone (BZ). Relying on the photoelectric effect, ARPES is a mature photon-in/electron-out technique – employing both table-top light sources and synchrotron facilities – that has been successfully applied to the study of a wide variety of quantum materials over the last few decades and has been extensively reviewed [see for instance \cite{damascelli2003angle,damascelli2004probing,gedik2017photoemission,lu2012angle,lv2019angle,sobota2021angle,zhang2022angle}]. In recent years, ARPES has been adapted in different directions to tackle the specific needs of various fast-growing domains within quantum materials research, including achieving increasingly better spatial resolution (micro- or nano-ARPES), imaging of spin degree of freedom (spin-ARPES) and characterizing the out-of-equilibrium regime (time-resolved ARPES). 

In particular, by extending ARPES into the time domain, time- and angle-resolved photoemission spectroscopy (TR-ARPES) not only provides access to materials' dynamical properties, but also facilitates identification of energetically entangled interactions directly in momentum space. A series of technological advancements – optimization of laser systems based on harmonic generation in nonlinear crystals, high-harmonic-generation, and improvements in 2D/3D electron detectors – have enabled TR-ARPES to mature rapidly into a powerful technique. The concurrent discovery of new classes of materials that have particularly benefitted from study by TR-ARPES – such as topological insulators and two-dimensional transition metal dichalcogenides – have also played an important role in affirming TR-ARPES as a key experimental technique for unraveling the complex dynamical properties of quantum materials, ranging from high-temperature superconductors to excitonic insulators (see Fig.\,\ref{Intro}).

TR-ARPES is now maturing in a manner similar to the way that ARPES has floursished over the past three decades. By building upon and expanding on previous reviews on the topic \cite{sobota2021angle,smallwood2016nonequilibrium,huang2022high}, this review aims to explain the foundations for the forthcoming development of the TR-ARPES technique by offering a panoramic view of how TR-ARPES has played a crucial role in advancing our knowledge of quantum materials via a detailed discussion of several paradigmatic examples. To this end, the different sections are organized by scientific questions and needs rather than by materials, so discussion of TR-ARPES research into specific materials is often spread across different sections. Section\,\ref{TR_ARPES_section} introduces the TR-ARPES technique, summarizes the physics principles on which it is based, describes the information encoded in the transient photoemission signal, covers the main technical aspects of state-of-the-art TR-ARPES systems around the world, and discusses the physical properties, phenomena and scientific questions that TR-ARPES measures and addresses. Section\,\ref{Mapping_section} discusses the mapping of unoccupied and/or transient states in semiconductors and topological insulators. Section\,\ref{Phase_section} reviews photoinduced phase transitions in charge ordered systems, Mott insulators, excitonic insulators and high-temperature superconductors. Section\,\ref{ElPhon_section} discusses how TR-ARPES can qualitatively or quantitatively evaluate the electron-phonon interaction, with particular emphasis on topological insulators and graphene/graphite. Section\,\ref{SPV_section} highlights how TR-ARPES can directly assess the emergence of the surface photovoltage phenomenon in semiconductors and topological insulators. Section\,\ref{Spin_section} illustrates a few paradigmatic examples of how spin-resolved and/or dichroic TR-ARPES can explore light-induced evolution of the spin degree of freedom in ferromagnets and topologically-protected materials. Finally, Section\,\ref{Conclusions_section} offers some concluding remarks and outlines future opportunities to apply TR-ARPES to further expand our knowledge of quantum materials.

\begin{figure*}
\centering
\includegraphics[scale=1]{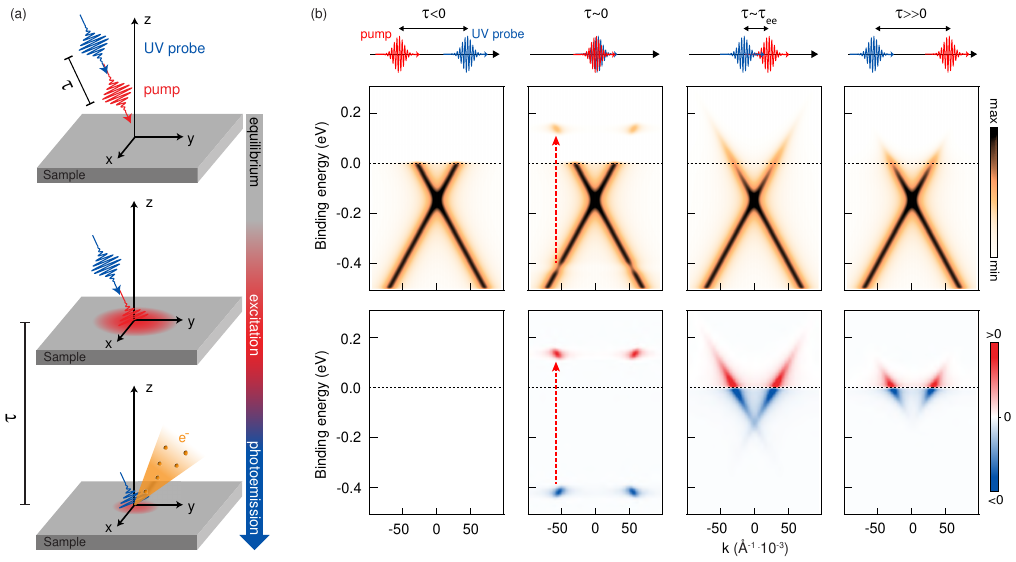}
\caption[IntroTI]{Working principles of TR-ARPES. (a) Illustration of the pump-probe approach in TR-ARPES using two ultrafast laser pulses separated by time interval $\tau$: the pump pulse drives the system out of equilibrium, and the probe pulse triggers the photoemission process. (b) Simplified sketch of the pump-induced evolution of the TR-ARPES signal for a Dirac-cone dispersion. Top row: TR-ARPES maps for different pump-probe delays $\tau$ ($\tau_{\text{ee}}$ defines the electron-electron thermalization time). The red dashed arrow indicates the pump photon energy and related vertical optical transition from occupied into unoccupied states. Bottom row: differential TR-ARPES maps highlight the pump-induced redistributions of spectral weight.}
\label{IntroTI}
\end{figure*}

\section{Time- and Angle-Resolved Photoemission} \label{TR_ARPES_section}
TR-ARPES offers unique access to the electronic dynamics and interactions on sub-picosecond timescales with both time and momentum resolution \cite{huang2022high,smallwood2016ultrafast,sobota2021angle}. Similarly to other time-resolved techniques, TR-ARPES takes advantage of the so-called \textit{pump-probe} approach \cite{giannetti2016ultrafast}. As illustrated in Fig.\ref{IntroTI}(a), two ultrafast laser pulses are directed onto the sample, separated in time by a tunable delay: the intense pump pulse (commonly in the visible-to-infrared spectral range) excites the system into an out-of-equilibrium condition, whereas the ultraviolet (UV, $\sim$\,6\,eV) or extreme ultraviolet (XUV, $>$\,10\,eV) probe pulse elicits the material to emit photoelectrons. The detector captures these photoelectrons with angular resolution, allowing us to visualize the transient evolution or modification of the electronic band structure and its occupation. By varying the pump-probe delay, TR-ARPES can effectively record a \textit{movie} of the light-induced ultrafast electron dynamics of the system under study.

A simplified and idealized introduction to the principles of the TR-ARPES technique is given in Fig.\,\ref{IntroTI}(b), where an n-doped (\textit{i.e.}, electron-doped) Dirac-like dispersion centered at $\textbf{k}$=0 momentum provides a simple illustrative example that resembles the dispersion of the surface state of a 3D topological insulator. The top row displays the TR-ARPES energy-momentum maps for four different characteristic pump-probe delays ($\tau$). Negative time delays ($\tau < 0$, far left) correspond to the scenario in which the pump pulse reaches the sample after the probe pulse, so the excitation effect does not appear in the photoemission signal (unless events lasting longer than the separation of two consecutive pump pulses, \textit{e.g.} local heating of the sample, or related to the generation of electric fields at the sample's surface, such as surface photovoltage discussed in Sec.\,\ref{SPV_section}, are present). 
When the pump and probe pulses overlap ($\tau$\,$\sim$\,0, central left), the TR-ARPES map shows a new electronic distribution determined by the availability of vertical optical transitions driven by the pump excitation, \emph{i.e.}, the optical joint density of states (assuming that the pump pulse does not modify the underlying electronic band structure). Next, on timescales comparable to the electron-electron scattering time ($\tau$\,$\sim$\,$\tau_\text{ee}\sim$10--100\,fs central right), pump-induced hot electrons tend to thermalize into a quasi-equilibrium Fermi-Dirac distribution characterized by an effective electronic temperature. Finally, for longer positive delays ($\tau$\,$\gg$\,0, $\sim$100\,fs--10\,ps, far right), the system returns back to its equilibrium conditions by transferring energy from the electronic bath into lattice vibrations and/or other collective modes.

The bottom row of Fig.\,\ref{IntroTI}(b) presents differential TR-ARPES maps, computed as $I(\text{k},\omega)_{\tau}-I(\text{k},\omega)_{\tau<0}$ (where $I(\text{k},\omega)_{\tau}$ represents the photoemission intensity at momentum $\text{k}$, energy $\epsilon$, and pump-probe delay $\tau$), for the same positive time delays as in the top row. The red (blue) color highlights an increase (decrease) of photoemission intensity with respect to the equilibrium counterpart, and thus is commonly referred to as population (depletion).
In the early days of TR-ARPES, differential maps were extensively employed to highlight the pump-induced transfer of spectral weight within the band structure. However, this straightforward interpretation may be challenged by several concurrent transient contributions to the photoemission intensity, as discussed in the following sections.

Before moving into a more detailed description of the technique and the science attainable with it, we emphasize that this review focuses mainly on TR-ARPES, defined as a technique in which the photoemission process is elicited by the probe pulse, and it is not assisted by the pump pulse. In most cases, the pump pulse drives collective excitations and/or promotes electrons into unoccupied states. In the latter case, once the photoexcited electron thermalizes and undergoes decoherence scattering processes, its subsequent photoemission by the probe pulse can be treated as a \textit{single photon} process. In this regard, note that a related approach such as two-photon photoemission spectroscopy (2PPE), in which two photons are needed to overcome the material’s work function, represents a powerful alternative for exploring electron dynamics of metals and semiconductors \cite{damascelli1996multiphoton,bartoli1997nonlinear,hofer1997time,ferrini1999linear,cui2014transient,petek1997femtosecond,petek2000real}, as well as for mapping unoccupied states \cite{sobota2013direct,hao2010nonequilibrium,yang2017revealing}. In some cases, TR-ARPES and 2PPE may be indistinguishable. However, this review does not discuss two-photon processes where the first photon excites an electron into a virtual state (\emph{i.e.}, with zero lifetime), or the phase of the intermediate state plays an important role in the subsequent photoemission process elicited by the second photon.

\subsection{General description}
Since TR-ARPES extends conventional ARPES into the time domain, a description of the transient photoemission signal requires an introductory overview of its \emph{equilibrium} counterpart.

\begin{figure}
\centering
\includegraphics[scale=1]{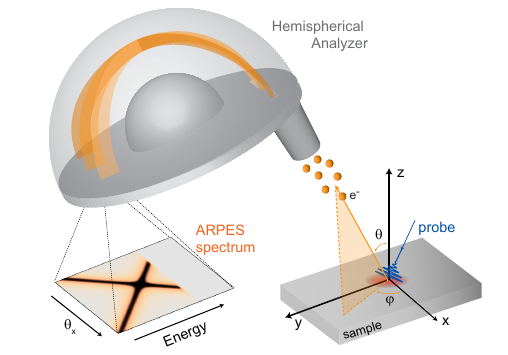}
\caption[IntroARPES]{Schematic configuration of an ARPES experiment. Probe pulses in the ultraviolet (UV) to extreme ultraviolet (XUV) spectral range prompt the photoemission of electrons from the sample. Photoelectrons are then collected by an electron analyzer (a Hemispherical Analyzer in the sketch), which discriminates them both by kinetic energy and angle of emission, thus generating the 2D-ARPES spectrum.}
\label{IntroARPES}
\end{figure}

ARPES is a photon-in/electron-out spectroscopic technique that grants direct insight into the electronic properties of matter and has already been extensively discussed in detail in several reviews \cite{damascelli2003angle,damascelli2004probing,gedik2017photoemission,lu2012angle,lv2019angle,sobota2021angle,zhang2022angle}. As depicted in Fig.\,\ref{IntroARPES}, incoming light with photon energy $h\nu$ greater than the material's work function ($\phi$, typically around 4-5\,eV) may eject electrons from the sample's surface via the photoelectric effect. A fraction of such photoelectrons are collected and discriminated by a photoemission spectrometer to map the material's electronic structure. By taking advantage of the momentum and energy conservation laws, one can relate the kinetic energy $E_k$ and angles of emission ($\theta$, $\varphi$) of photoelectrons to the binding energy $E_B$ and crystal momentum $\hbar\textbf{k}$ of electrons inside the solid.
In particular, the retrieved component of the crystal momentum parallel to the material's surface is proportional to the square root of the photoemitted electron's kinetic energy, $\hbar\text{k}_{||}=\sqrt{2m E_k}\cdot \text{sin}\theta$. This relation naturally implies that a larger photon energy (and consequently larger kinetic energy of the photoelectrons: $E_k=h\nu -\phi -|E_B|$) grants access to a larger portion of momentum space within or beyond the first BZ. The perpendicular component of the photoelectron wavevector $\textbf{k}_{\perp}$ is not conserved during the photoemission process due to the breaking of translational symmetry at the sample's surface, but it can be experimentally estimated via photon-energy dependent ARPES measurements.

While ARPES is commonly associated with the investigation of \emph{static} (or \emph{equilibrium}) electronic properties of matter and TR-ARPES is associated with the exploration of ultrafast (within the femtosecond to picosecond timescales) electron dynamics, this simplified dichotomy is not entirely correct: \emph{equilibrium} ARPES itself offers valuable information on intrinsic electrodynamics in materials since removing an electron (\emph{i.e.} generation of a photohole) leaves the system in an excited state. The propagation of an electron (or photohole) with momentum $\textbf{k}$ and energy $\omega$ through an interacting system is captured by the Green's-function formalism, which characterizes the probability that a system will evolve into a different state upon addition/removal of an electron.
Following the already existing ARPES reviews, it can be shown that the ARPES signal is intimately related to the system's Green function G($\textbf{k},\omega$), and thus to fundamental dynamical interactions. Under the sudden approximation and for a single band system (and neglecting energy and momentum resolutions, and extrinsic background contributions), the equilibrium ARPES intensity can be calculated via Fermi's Golden rule as
\begin{equation} \label{IntARPES_EQ}
    \text{I}(\textbf{k},\omega)=A(\textbf{k},\omega) \cdot |M(\textbf{k},\omega)|^2 \cdot f(\omega),
\end{equation}
where $A(\textbf{k},\omega)$\,=\,-(1/$\pi$)Im G($\textbf{k},\omega$) is the one-particle removal spectral function, which encompasses both the single-particle band structure $\epsilon_{\textbf{k}}$, as well as all many-body correlation effects embedded in the electron self-energy $\Sigma(\textbf{k},\omega)$ \cite{mahan2000many}. Note that $\text{I}(\textbf{k},\omega)$ also depends on two other terms, namely the photoemission matrix element $M(\textbf{k},\omega)$, which connects the initial and final state of the photoemitted electron, and the Fermi-Dirac distribution function $f(\omega)$.

In the hypothetical scenario of a non-interacting system, upon the removal of an electron with energy $\epsilon_{\textbf{k}}$, the system left behind would not undergo any relaxation process, and the spectral function would consist of a single line centered at the band energy $\epsilon_{\textbf{k}}$; \textit{i.e.}, $A(\textbf{k},\omega)=\delta(\omega-\epsilon_{\textbf{k}})$; in other words, the particle at $\epsilon_{\textbf{k}}$ would have an infinite lifetime. Considering instead an interacting system, all the corrections to the non-interacting Green's function are taken into account by introducing the electron self-energy $\Sigma(\textbf{k},\omega)=\Sigma'(\textbf{k},\omega)+i\Sigma''(\textbf{k},\omega)$. The spectral function is then written as \cite{mahan2000many}:
\begin{equation} \label{spectral}
    A(\textbf{k},\omega)=-\frac{1}{\pi}\frac{\Sigma''(\textbf{k},\omega)}{[\omega-\epsilon_{\textbf{k}}-\Sigma'(\textbf{k},\omega)]^2+[\Sigma''(\textbf{k},\omega)]^2}.
\end{equation}
Here, $\Sigma'(\textbf{k},\omega)$ and $\Sigma''(\textbf{k},\omega)$ encompass all energy renormalization and lifetime changes with respect to the non-interacting case, respectively. Intuitively, $\Sigma'$ leads to a shift of the bare dispersion $\epsilon_{\textbf{k}}$ to a new position $\tilde{\epsilon}_{\textbf{k}}$=$\epsilon_{\textbf{k}}+\Sigma'(\textbf{k},\omega)$, while $\Sigma''$ broadens the delta function from the non-interacting case. The single-particle lifetime $\tau_s(\textbf{k})$, which describes all relaxation properties of an excited single particle with energy $\epsilon_{\textbf{k}}$, is inversely proportional to the imaginary part of the electron self-energy, $\tau_s(\textbf{k})=\hbar / [2 \Sigma''(\textbf{k},\omega)]$, and can be estimated by a detailed evaluation of the ARPES lineshape. To this end, two main strategies are often combined, focusing on the energy- and the momentum-dependence of the photoemission signal. The first approach is to analyze the energy-dependence of the photoemission intensity at a fixed momentum \textbf{k}$^*$ in the so-called Energy Distribution Curve (EDC). The second focuses instead on the momentum-dependence at a fixed binding energy $\omega^*$ via the so-called Momentum Distribution Curve (MDC). Generally, by combining the analysis of EDCs and MDCs in global fitting and/or self-consistent procedures one can more rigorously model and interpret ARPES data, \textit{e.g.} see \cite{kurleto2021two,li2018coherent}.

Given that \emph{equilibrium} ARPES already provides insights into fundamental single-particle electrodynamics, one may wonder what additional dynamical information TR-ARPES provides. In contrast with its static counterpart, TR-ARPES observes changes of the sample's electronic structure upon pump-pulse excitation. Such optical excitations typically (i) modify the electronic distribution (beyond the equilibrium Fermi-Dirac statistics); (ii) generate specific collective excitations (such as phonons); and, consequently, (iii) lead to the emergence of new nonequilibrium and/or metastable phases of matter with no equilibrium counterpart. Thus, TR-ARPES does not simply access fundamental single-particle dynamics, but maps how out-of-equilibrium electrons interact among themselves and with underlying collective excitations; TR-ARPES also maps how the electronic properties of matter are affected by selective excitation of specific collective modes on their intrinsic timescales. 

To further reinforce the critical role of TR-ARPES for advanced investigations of quantum materials, it is worth emphasizing that the design and engineering of new materials for applied purposes strongly relies on their dynamical (\emph{i.e.} nonequilibrium) response rather than on their equilibrium properties.
Indeed, every time we apply a voltage (\emph{e.g.}, in a microchip) or shine light (\emph{e.g.}, in a photovoltaic cell) to generate current, a nonequilibrium electronic distribution emerges and its dynamical behaviour determines the material's macroscopic working parameters, such as resistivity and energy-harvesting efficiency. Therefore, TR-ARPES is an effective technique not only for fundamental studies, but also for characterizing out-of-equilibrium properties of quantum materials for future applications.

The following section provides a general description of the TR-ARPES intensity, both in a quasi-equilibrium scenario and a more strictly out-of-equilibrium framework, providing an overview of the wealth of information that can be extracted by analyzing the transient photoemission intensity, from a qualitative to a quantitative level (Section\,\ref{TR_ARPES_Int}). Afterwards, a brief review of state-of-the-art TR-ARPES systems covers how UV and XUV probe pulses are generated and coupled to ARPES systems, as well as what optical excitation schemes are commonly employed (Section\,\ref{TR_ARPES_systems}).

\begin{figure*}
\centering
\includegraphics[scale=1]{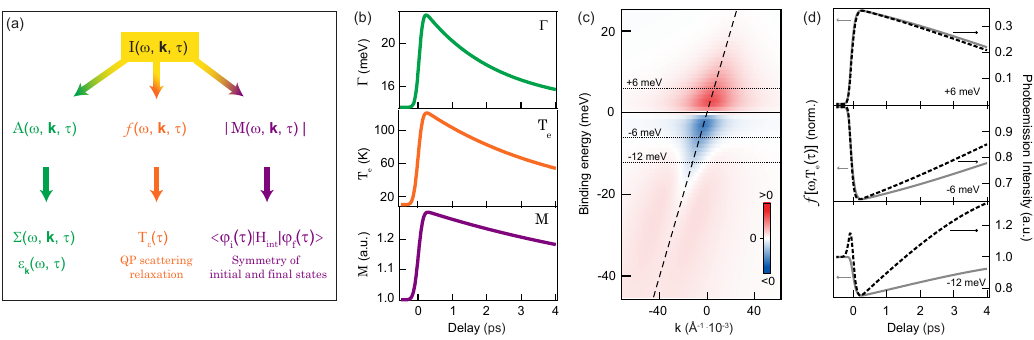}
\caption[TRARPESInt]{The different components that comprise the transient photoemission intensity, with an illustration of how their intertwined contributions may be reflected in TR-ARPES data. (a) Diagram of the different contributions, and corresponding encoded information, that characterize the transient photoemission intensity $\text{I}(\textbf{k},\omega,\tau)$. (b) Hypothetical temporal evolution of the spectral broadening $\Gamma$, electronic temperature T$_{\text{e}}$, and matrix element $|M|$. $\Gamma$ is assumed to have a Fermi-Liquid form, $\Gamma(\tau)=\Gamma_{\text{imp}}+\beta [\omega^2+\pi^2 k_{\text{B}}^2 \text{T}_{\text{e}}^2(\tau)]$, where $\Gamma_{\text{imp}}$=14\,meV and $\beta$=8. The temporal evolution of $\Gamma$ is displayed for $\omega$=0. (c) Simulated differential ARPES spectrum at $\tau$\,=\,3\,ps of a linear dispersion (dashed black line) assuming the transient parameters displayed in (b) and accounting for the $\omega$-dependence of $\Gamma$. (d) Comparison between the transient evolution of the photoemission intensity at three different binding energies from (c) [black dashed lines, right axis] and the evolution of the electronic distributions given solely by T$_{\text{e}}(\tau)$ in (b) [gray solid lines, left axis], highlighting the discrepancy in their dynamics.}
\label{TRARPESInt}
\end{figure*}

\subsection{TR-ARPES intensity} \label{TR_ARPES_Int}

\subsubsection{Quasi-equilibrium approximation} \label{quasiEq_ArpesInt}
The description of a system's transient evolution upon perturbation by light is far from straightforward, especially when considering the effects of the pump pulse's electric field, the potential appearance of new transient states of matter (see for instance Section\,\ref{Floquet}), as well as other highly non-thermal processes \cite{randi2017bypassing,sentef2013examining,kemper2017review,kemper2018general}. However, in most cases, after the initial light-induced excitation of the electronic bath, one can approximately assume that electrons thermalize into a quasi-equilibrium state within a time span of a few tens-to-hundreds of femtoseconds (see Section\,\ref{TTM}), and it has been shown that the adiabatic band picture (\textit{i.e.} the unperturbed band structure) survives well even at zero pump-probe delay and high pump field intensities \cite{neufeld2022time}. In such a scenario, the TR-ARPES signal for a single band system can be approximated by extending the equilibrium formalism of Eq.\,\ref{IntARPES_EQ} into the time domain $\tau$ [also known as generalized Kadanoff–Baym approximation \cite{freericks2021two}]:
\begin{equation}
    \label{TRARPES_int_EQ}
    \text{I}(\textbf{k},\omega,\tau) \simeq A(\textbf{k},\omega,\tau) \cdot |M(\textbf{k},\omega,\tau)|^2 \cdot f(\textbf{k},\omega,\tau).
\end{equation}
Simple visual inspection of Eq.\,\ref{TRARPES_int_EQ} reveals that the transient evolution of the photoemission intensity may arise from the complex interplay of several contributions, as depicted schematically in Fig.\,\ref{TRARPESInt}. 
The parallel evolution of all the terms, namely the electronic distribution $f(\textbf{k},\omega,\tau)$, the photoemission matrix elements $|M(\textbf{k},\omega,\tau)|^2$, and the spectral function $A(\textbf{k},\omega,\tau)$, can make the interpretation of the $\text{I}(\textbf{k},\omega,\tau)$ challenging. Figure\,\ref{TRARPESInt} highlights how changes in $A$ (encoded into the change of spectral broadening $\Sigma''$), T$_{\text{e}}$ and $|M|$ can lead to a transient evolution of $\text{I}(\textbf{k},\omega,\tau)$ that cannot easily be explained in terms of a single contribution. 
Figure\,\ref{TRARPESInt}(d) compares the transient photoemission intensity at three different binding energies (black dashed curves) with the evolution of the electronic distribution based solely on the changes in T$_{\text{e}}$ (solid gray).
The evolution of the TR-ARPES signal does not necessarily mimic the transient electronic distribution (see panel d, bottom), nor the single-particle lifetime $\tau_s$ (here described by $\Gamma$ in panel b). Instead, it results from the complex interplay of the different contributions depicted in Fig.\,\ref{TRARPESInt}(a). 
The following discussion explains the transient evolution of all the terms of Eq.\,\ref{TRARPES_int_EQ} and how to interpret their intertwined contributions.

In the conventional ARPES scenario, the Fermi-Dirac (FD) distribution $f(\omega)=(1+e^{\omega/k_B T_e})^{-1}$ does not have any momentum dependence and accounts for the fact that photoemission probes solely the occupied states. In contrast, in a TR-ARPES experiment, the pump excitation promotes valence electrons into unoccupied states in specific regions of the 3D BZ, potentially leading to an electronic distribution that is highly anisotropic in momentum space, at least for very short time delays $\tau$\,$\sim$\,0. At later times, following thermalization processes (such as electron-electron, electron-impurity, and electron-boson scattering), hot electrons redistribute over the whole 3D momentum space, and $f(\textbf{k},\omega,\tau$\,$\gg$\,0) may be approximated by a $\textbf{k}$-independent FD distribution with a time-dependent effective electronic temperature $T_e (\tau)$. The light-induced emergence of a non-thermal electronic distribution at short pump-probe delays, together with the momentum-resolution of TR-ARPES, provides a valuable approach to map the band dispersion of unoccupied states of quantum materials (\emph{e.g.}, the topological surface state of 3D topological insulators) and a method to extract electronic band gaps and test \emph{ab-initio} evaluations of the band structure (\emph{e.g.}, the electronic bandgap of transition metal dichalcogenides), as discussed in Section\,\ref{Mapping_section}. Moreover, Section\,\ref{ElPhon_section} showcases how, by tracking how carriers photo-injected into the unoccupied states thermalize and return back to equilibrium conditions, TR-ARPES offers qualitative and quantitative information on multi-particle scattering processes (\emph{e.g.}, electron dynamics in graphene), ultrafast changes of the screening and electronic dispersion (\emph{e.g.}, the detection of coherent phonons), as well as fundamental electron-boson coupling constants (\emph{e.g.}, by detecting quantized electron decay processes).

Next, under the sudden approximation the one-electron dipole matrix element takes a simple form, $M$\,$\propto$\,$ <\phi_i^\textbf{k}|H_{int}|\phi_f^\textbf{k}>$, and connects the initial (\emph{i}) and final (\emph{f}) state of the photoemitted electron via the light-matter interaction Hamiltonian $H_{int}$\,$\propto$\,$\textbf{A} \cdot \textbf{p}$ (where $\textbf{p}$ is the electron momentum operator and $\textbf{A}$ is the electromagnetic vector potential). $|M|^2$ encodes the dependence of the photoemission intensity on the polarization and photon energy of the probe beam, as well as the experimental geometry (\textit{e.g.}, angle of incidence, sample orientation, and orientation of the entrance slit for hemispherical electron analyzers). In this regard, detailed polarization and photon-energy dependent ARPES measurements have been proven able to provide direct information on the orbital character of the initial photoemission state and its spatial extent \cite{schuler2022polarization,day2019Chinnok,gierz2011illuminating,cao2013mapping,zhu2013layer,moser2017experimentalist}. 
In TR-ARPES investigations, the transient evolution of the photoemission matrix element has commonly been neglected. However, possible transient modifications of  $M(\textbf{k},\omega,\tau)$ have recently been acknowledged theoretically \cite{freericks2016constant}, and demonstrated experimentally \cite{boschini2020role}. In particular, in multi-orbital systems where the initial state wavefunction may result from a complex interplay of different or layer-dependent orbital characters, $M(\textbf{k},\omega,\tau)$ can have a non-negligible impact on the transient evolution of $\text{I}(\textbf{k},\omega,\tau)$. As a paradigmatic example, a light-induced modulation of the atomic distances via emission of coherent and/or incoherent phonons can dynamically change the orbital mixing of the probed states, and hence $|M|^2$. Therefore, unless single-band systems with a well-defined orbital character are studied, the possibility that the transient evolution of $\text{I}(\textbf{k},\omega,\tau)$ may be linked to the evolution of $M(\textbf{k},\omega,\tau)$ should always be considered. Indeed, \onlinecite{boschini2020role} have recently shown that a probe-polarization-dependent study of the TR-ARPES signal can offer direct evidence of the transient changes of matrix elements. We foresee several future TR-ARPES investigations able to precisely track light-induced modifications of the symmetry of the initial photoemission state wavefunction by performing detailed polarization/dichroism measurements \cite{schuler2022polarization,schuler2022probing}.

Lastly, the third term of Eq.\,\ref{TRARPES_int_EQ} is the spectral function $A(\textbf{k},\omega,\tau)$, which accounts for all transient changes of the band structure and density of states (DOS), as well as many-body interactions. Note that an optical pump may modify the bare electron dispersion in the adiabatic limit by driving coherent or incoherent phonons (see Section\,\ref{CoherentPhon}, \emph{e.g.} periodic modulations of the electronic dispersion in topological insulators and Fe-based superconductors), or transiently changing many-body interactions, \emph{i.e.} $\Sigma(\tau)$, with the filling of the superconducting gap of cuprates as an emblematic example. Transient evolution of the self-energy can be the byproduct of various effects ranging from transient modification of the screening and scattering phase-space to light-induced phase-transitions (Sections\,\ref{Mapping_section},\ref{Phase_section}, and\,\ref{ElPhon_section}). All these different contributions to $A(\textbf{k},\omega,\tau)$ are often deeply intertwined and difficult to disentangle. 

\begin{figure*}
\centering
\includegraphics[scale=1]{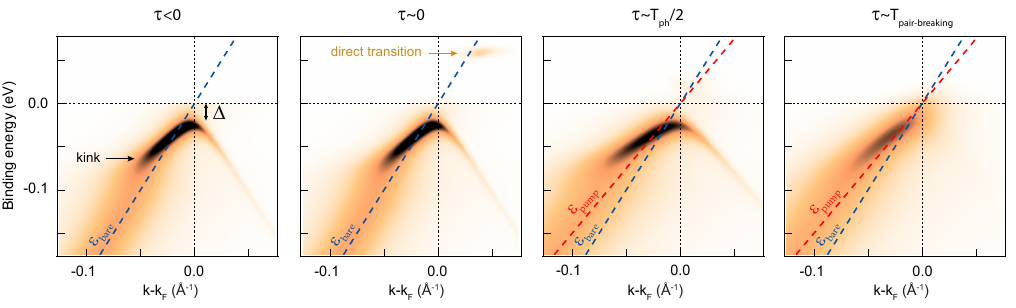}
\caption[IntroSC]{Toy model depicting some of the challenges associated with the interpretation of the transient photoemission intensity for the case of a superconductor with gap $\Delta$ and a strong electron-phonon coupling that renormalizes the electronic dispersion (\emph{a.k.a.}, kink). Multiple effects may occur, such as direct optical transitions, transient modifications of the bare dispersion from the unperturbed $\varepsilon_{\mathrm{bare}}$ to $\varepsilon_{\mathrm{pump}}$ after excitation (dashed blue and red lines, respectively), enhancement of the electronic temperature, filling of the superconducting gap, as well as softening of the electron-phonon band structure renormalization. These effects are often intertwined, complicating the interpretation of the TR-ARPES signal.}
\label{IntroSC}
\end{figure*}

As a representative thought experiment that emphasizes the plethora of possible distinct contributions to $A(\textbf{k},\omega,\tau)$ consider a superconductor with s-wave gap $\Delta$, critical temperature T$_c$, and linear dispersion in the normal state, as depicted in Fig.\,\ref{IntroSC}. Assuming that the electronic bath is strongly coupled to an Einstein phonon with energy $\Omega_0 \sim 2\Delta$ (similar to the phenomenology in the cuprates), a renormalization of the band structure occurs, also known as \emph{kink} \cite{lanzara2001evidence}, as shown by the unperturbed electronic band dispersion at low temperature (T$<$T$_c$) for $\tau < 0$ (far left panel). The optical excitation populates unoccupied states at $\tau \sim 0$, thus creating a nonequilibrium electronic distribution. As a consequence of the transient redistribution of the electronic density, excitation of phonon modes may occur: coherent phonons triggered via displacive mechanisms or incoherent phonons emitted through the thermalization process of the photoinduced non-thermal electronic distribution or via anharmonic decay of coherent phonons. Such an increase of the phonon temperature (on timescales comparable to $\frac{1}{2}$ of the phonons' period) can transiently change the electronic dispersion (discussed in Sec.\,\ref{ElPhon_section}), as pictured by the red line in the right panels of Fig.\,\ref{IntroSC}. Afterwards, the superconducting gap might fill up on a timescale determined by pair-breaking events (\textit{i.e.}, inelastic electron scattering breaking Cooper pairs). In addition, the transient change of the density of states (DOS) at the Fermi level due to this filling of the superconducting gap enhances the available electronic scattering phase space, consequently changing the kink renormalization strength (see Sec.\,\ref{SelfEn} and Eq.\,\ref{EqSigmaEPh}). 

Overall, the examples of Fig.\,\ref{TRARPESInt} and Fig.\,\ref{IntroSC} highlight in a simplified fashion some of the challenges associated with the interpretation of TR-ARPES data for a complex/intertwined system, emphasizing once more that TR-ARPES does not simply probe population dynamics but also transient many-body interactions encoded in the one-particle spectral function.
Within the quasi-equilibrium approximation, the spectral function and related self-energy can still be explored via \emph{equilibrium} analysis methods (\textit{e.g.}, extracting $\Sigma$ via analysis of the ARPES lineshape). Moreover, after the initial thermalization time (which depends on the specific sample and excitation parameters), a transient electronic temperature $T_e (\tau)$ can be defined. As discussed later in detail in Section\,\ref{TTM}, $T_e (\tau)$ can differ from the temperature of the lattice, and in general of other bosonic modes (such as phonons, magnons, etc...), over a delay range determined by intrinsic electron-boson scattering processes. Within this time interval, TR-ARPES can be used in concert with equilibrium ARPES to determine whether lifetime broadening and band structure renormalization effects arise from pure electron-electron effects or interactions involving bosons. For example, \onlinecite{zonno2021ubiquitous} recently compared the equilibrium and transient temperature dependencies of the quasiparticle peak of Bi-based cuprates, demonstrating that Fermi-liquid theory (\emph{i.e.} solely electron-electron interactions) properly accounts for the low-energy part of the self-energy in the gapless nodal direction.

In conclusion, the TR-ARPES technique innately correlates transient redistributions of the carrier density with changes to the many-body interactions, electronic and bosonic DOS, and the symmetry of the initial and final photoemission states. To the extent that transient contributions to Eq.\,\ref{TRARPES_int_EQ} can be disentangled, TR-ARPES can directly and unambiguously identify and demystify light-induced phenomena in condensed matter.

\begin{figure*}
\centering
\includegraphics[scale=1]{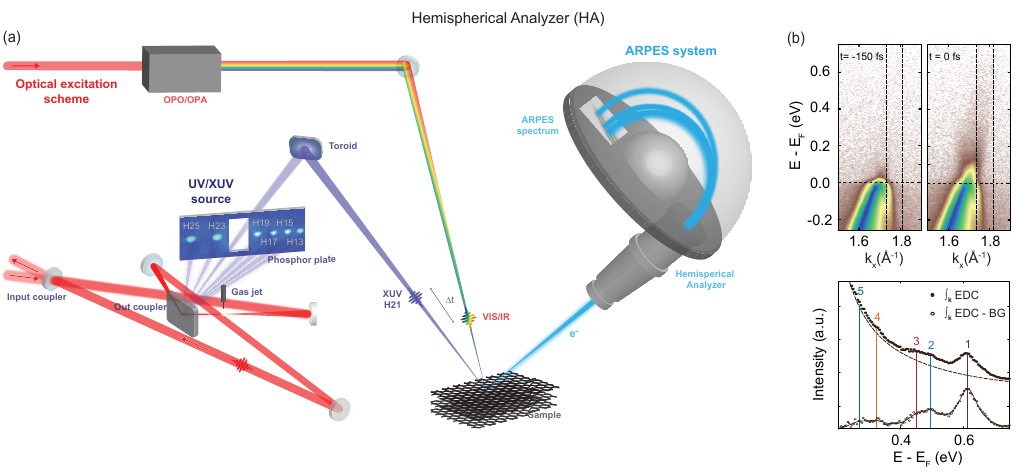}
\caption[Intro_systems]{Illustration of a TR-ARPES system based on a Hemispherical Analyzer along with an example dataset that showcases its capabilities. (a) Illustration of the three main components of a state-of-the-art TR-ARPES system: the optical schemes for the pump excitations, the generation of the ultraviolet/extreme ultraviolet probe pulses and the coupling to the ARPES system. (b) TR-ARPES data of graphite acquired with 25\,eV probe and 1.2\,eV pump. The high energy and momentum resolution, as well as dynamic range, achievable with a Hemispherical Analyzer allowed the detection of multiple peaks, corresponding to both direct transitions and photon-induced replicas. In fact, although (top) these peaks are not obvious in the ARPES intensity maps, (bottom) they stand out from the quasi-thermalized electron background when the EDC is integrated in momentum, at energies up to 0.6\,eV above E$_{\text{F}}$. Adapted from \onlinecite{thesisNa_2022}.}
\label{Intro_systems}
\end{figure*}

\subsubsection{Out-of-equilibrium description}
The quasi-equilibrium approximation of the TR-ARPES signal discussed in the previous section is a valid approach to describe the transient evolution of systems when pump and probe are not overlapped. The evaluation of the TR-ARPES signal becomes increasingly challenging when one wants to account fully for light-induced dynamical effects and finite time width of the probe pulse \cite{freericks2009theoreticalPRL,schuler2021theory}. In particular, assuming a single-band system and constant matrix elements, the TR-ARPES intensity is computed from the probe-pulse weighted relative-time Fourier transform of the lesser Green's function $\text{G}^{<}_{\textbf{k}(t,t')}$ as:
\begin{equation}
    \label{TRARPES_int_onestep}
    \text{I}(\textbf{k},\omega,\tau) = Im \int dt dt' p(t,t',\tau) e^{i \omega(t-t')} \text{G}^{<}_{\textbf{k}}(t,t'),
\end{equation}
where $p$ corresponds to a two-dimensional Gaussian probe centered at $(t,t')=(\tau,\tau)$.

This review does not encompass theoretical challenges related to the interpretation of the TR-ARPES signal and we instead refer the reader to the relevant literature \cite{freericks2009theoreticalPRL,freericks2009theoretical,tao2010theory,sentef2013examining,kemper2014effect,braun2015one,kemper2017review,xu2019efficient,perfetto2020time,deGiovannini2022first,marini2022coherence,eckstein2021time,schwarz2020momentum}. However, it is important to recognize that the similarities between \emph{equilibrium} ARPES and TR-ARPES discussed in the previous section may be deceptive for time delays comparable to the system's characteristic dephasing times, or when new electonic states with no equilibrium counterpart (\textit{e.g.}, Floquet-Bloch states) emerge. Although the TR-ARPES signal is commonly evaluated by assuming that the various many-body contributions are linearly proportional to the overall population decay rate, that decay rate does not originate simply from the self-energy and it is often not possible to separate scattering rates in the TR-ARPES signal \cite{kemper2018general}. Following experimental evidence of \onlinecite{yang2015inequivalence}, \onlinecite{kemper2018general} showed that population dynamics are controlled by the difference between the distributions for the Green's function and the self-energy\footnote{The distribution functions for the Green's function G and self-energy $\Sigma$ are respectively $f^{\text{G}}(\tau,\omega)$\,=\,$i \text{G}^{<}_{\textbf{k}}(\tau,\omega)/ 2 Im \text{G}^{R}_{\textbf{k}}(\tau,\omega)$ and $f^{\Sigma}(\tau,\omega)$\,=\,$i \Sigma^{<}_{\textbf{k}}(\tau,\omega)/ 2 Im \Sigma^{R}_{\textbf{k}}(\tau,\omega)$. In equilibrium, $f^{\text{G}}$ and $f^{\Sigma}$ converge to the Fermi-Dirac distribution. 
However, it can be shown that the population decay $\frac{df_{\textbf{k}}(\tau)}{d \tau} \propto$ [$f^{\text{G}}(\tau,\varepsilon_{\textbf{k}}) -- f^{\Sigma}(\tau,\varepsilon_{\textbf{k}})$].}.
Most of the studies reviewed here rely on the assumptions that it is possible to separate different relaxation processes in the time-domain or the self-energy alone governs the population decay rate. Although these are not universally true, as discussed in Section\,\ref{ElPhon_section}, we emphasize that (i) when many-body interactions are dominated by a single-scattering process, the relaxation rate is proportional to the self-energy, and (ii) when scattering processes take place on sufficiently different timescales, it is generally possible to separate relaxation rates  \cite{kemper2018general}.

\subsection{State-of-the-art TR-ARPES} \label{TR_ARPES_systems}
Generally speaking, in order to perform a TR-ARPES experiment, one has to generate pump and ultraviolet pulses and couple them into the ARPES chamber, as schematically depicted in Fig.\,\ref{Intro_systems}(a). For this reason, the design and development of the TR-ARPES technique has relied on close cooperation between the photoemission and ultrafast laser communities. This section briefly reviews methods and approaches to generate pump and ultraviolet pulses for TR-ARPES, as well as the commonly used ARPES systems, detectors and coupling methods. 

\subsubsection{UV and XUV sources}
Generation of ultraviolet (UV) and extreme ultraviolet (XUV) probe pulses to photoemit electrons has historically relied on three approaches: (i) frequency conversion with nonlinear crystals, (ii) highly cascaded harmonic generation in the gas phase, and (iii) high-order harmonic generation (HHG). Although HHG has been applied to photoemission spectroscopy since the 1990s \cite{haight1994tunable}, the advent of reliable and commercially available ultrafast laser sources used to generate UV light via nonlinear crystals \cite{gauthier2020tuning} has significantly accelerated development of TR-ARPES. In the latter case, 6-eV UV probe pulses are obtained by quadrupling the fundamental output of Ti:Sa lasers ($\sim$1.5\,eV photon energy) \cite{faure2012full,smallwood2012ultrafast,carpene2009versatile} or via sum-frequency generation in Yb-based laser ($\sim$1.2\,eV fundamental photon energy) \cite{boschini2014innovative} via a cascade of nonlinear crystals, specifically $\beta$-BaB$_2$O$_4$ (BBO). The technical simplicity of low photon energy optical setups has greatly favoured early investigations of ultrafast electron dynamics in the proximity of the Brillouin zone center of quantum materials, such as topological insulators. In addition, 6-eV TR-ARPES systems have historically granted easy access to high-repetition-rate operation (100\,kHz-100\,MHz) considerably improving the signal-to-noise of TR-ARPES spectra, and have provided broad tunability of the UV bandwidth enabling high temporal ($<$60\,fs) or energy ($<$10\,meV) resolutions depending on the particular experimental needs \cite{gauthier2020tuning}. The ability to continuously tune the UV photon energy in the $\sim$5.3-7\,eV range by combining different nonlinear crystals has also been recently demonstrated by \onlinecite{yang2019time} and \onlinecite{bao2022ultrafast}.

XUV-based TR-ARPES systems grant access to ultrafast electron dynamics over the entire momentum space of quantum materials \cite{dakovski2010tunable,heyl2012high,frietsch2013high,mathias2007angle,buss2019setup,heber2022multispectral,eich2014time,rohde2016time,plotzing2016spin}. However, the challenges related to generation of stable and bright XUV sources operating at high-repetition rate, as well as to coupling XUV light into ARPES systems, have considerably slowed down the operation of XUV-based TR-ARPES system. The advent of high repetition rate ($>$100\,kHz) and narrow-bandwidth XUV-based TR-ARPES systems \cite{chiang2012high,mills2019cavity,sie2019time,puppin2019time,lee2020high,peli2020time,guo2022narrow} has significantly improved the signal-to-noise ratio and enabled access to the transient evolution of low-energy spectral features previously hidden to low repetition rate large-bandwidth XUV-based TR-ARPES systems.
Moreover, the use of high-repetition rate systems helps to reduce space-charge effects, improving the energy resolution and determination of the position of the Fermi level \cite{hellmann2009vacuum,passlack2006space,corder2018ultrafast}. For a more comprehensive review of the light sources employed in TR-ARPES systems,  we refer the reader to a forthcoming review \cite{Na2023Review}.

\begin{figure}
\centering
\includegraphics[scale=1]{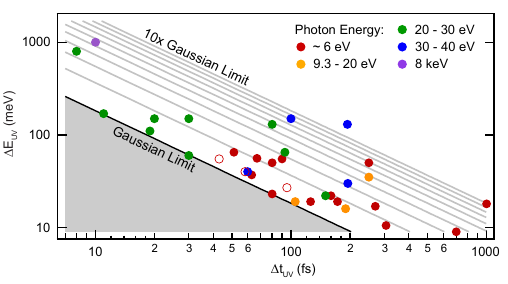}
\caption[DeltaEtime]{Pulse duration and bandwidth of UV/XUV probes from different TR-ARPES studies. The black line indicates the Gaussian Fourier limit, while gray lines represent integer multiples of this fundamental limit up to n\,=\,10. Adapted from \onlinecite{gauthier2020tuning}.}
\label{DeltaEtime}
\end{figure}

Finally, the development of high repetition rate free-electron laser facilities in principle paves the way to fully-tunable TR-ARPES experiments in a broad spectral range from XUV to soft x-ray \cite{rossbach201910,oloff2016time,kutnyakhov2020time}. These new facilities will enable exploration of ultrafast electron dynamics in highly-dispersive 3D systems, as well as enable element-specific studies via resonant TR-ARPES. 

Early TR-ARPES works have historically focused on temporal resolution at the expense of energy resolution, partially impairing the ability to resolve energy gaps and renormalization effects. Independently of how UV/XUV light is generated, over the past two decades we have witnessed different compromises between the time and energy resolutions (whose Fourier limit for Gaussian pulses can be expressed as $\Delta$E$\Delta\tau \geq 4 \hbar$ ln2 $\sim$1.825\,eV\,fs, see Fig.\,\ref{DeltaEtime}), which have enabled diverse TR-ARPES investigations ranging from tracking ultrafast electron dynamics with sub-15\,fs temporal resolution \cite{rohde2016time,rohde2018ultrafast} to the detection of sub-20\,meV superconducting gaps \cite{smallwood2012tracking,parham2017ultrafast,boschini2018collapse}.

\subsubsection{Optical Excitation Scheme}
TR-ARPES has primarily relied on the fundamental output of Ti:Sapphire or Yb-based laser sources (1.55\,eV and 1.2\,eV, respectively) or their second harmonics (3.1\,eV and 2.4\,eV, respectively) for the pump excitation. This historical focus has led to the initial groundbreaking demonstrations of the capabilities of the TR-ARPES technique in a variety of different systems, as discussed throughout this review. 
Although there are several specific cases where near-infrared/visible light is needed (such as the excitation of bright excitons or the photodoping of charge-transfer insulators), the continued almost exclusive reliance on near-infrared pump excitations does not take advantage of the full spectrum of opportunities to employ optical excitation to manipulate electron properties and prompt the emergence of metastable phases of matter. Indeed, near-IR/visible light usually redistributes electrons and holes over the entire momentum space -- potentially leading to quasi-thermal states with effective electronic temperatures as high as thousands of K -- and does not allow selective coupling to specific low-energy excitation without affecting the electronic bath.

Recent all-optical investigations have demonstrated that intense long-wavelength light excitation can selectively drive the emergence of new quantum phenomena in quantum materials with minimal coupling to the electronic bath \cite{liu2020pump,fausti2011light,mciver2020light,mitrano2016possible,disa2021engineering}. To date, the development of TR-ARPES systems with long-wavelength excitation capabilities remains at an early stage, but pioneering studies have demonstrated the feasibility of these experiments by showing groundbreaking results with mid-infrared pump pulses on topological insulators, graphene, black phosphorus and charge-ordered systems \cite{wang2013observation,gierz2015phonon,kuroda2016generation,monney2016revealing,chavez2019charge,ito2023build,zhou2023pseudospin}, as well as THz excitation of topological insulators \cite{reimann2018subcycle}.

Finally, one should always note that a ubiquitous effect of any pump excitation is to increase the sample's average temperature. Such thermal effects can be mitigated by decreasing the repetition rate for fixed absorbed fluence, or by pumping in a spectral range for which the material is (quasi)transparent. Nevertheless, maintaining the sample temperature in the sub-10\,K range for negative pump-probe delays is nearly always challenging. 

\begin{figure}
\centering
\includegraphics[scale=1]{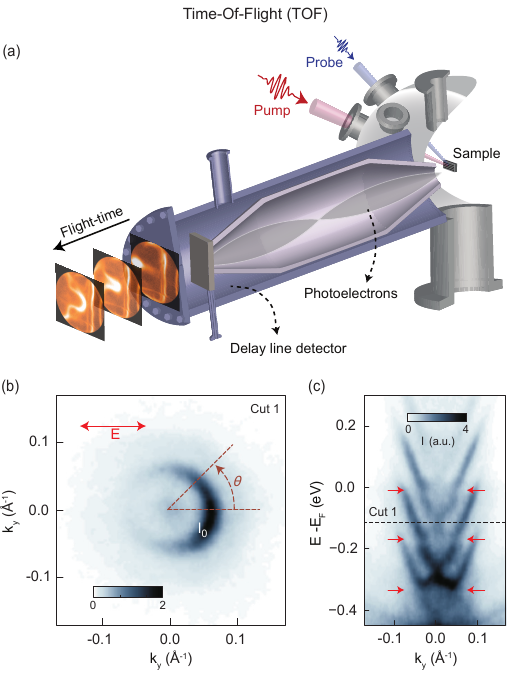}
\caption[TOF]{Operating principles of a time-of-flight (TOF) spectrometer. (a) Sketch of a TOF detector; adapted from \onlinecite{Zong2021}. (b) One-shot constant energy cut at -0.12\,eV and (c) ARPES spectrum along the k$_y$ direction acquired via TOF on Bi$_2$Se$_3$ upon mid-IR light excitation; adapted from \onlinecite{mahmood2016selective}. These maps show the appearance of Floquet-Volkov states when the impinging electric field is oriented along the k$_x$ direction (discussed in Sec.\,\ref{Floquet}) and exemplify the ability of TOF detectors to measure the photoemission intensity simultaneously over 2D momentum space, \emph{e.g.} I(E,k${_x}$,k${_y}$).}
\label{TOF}
\end{figure}

\subsubsection{ARPES systems and coupling to light sources}
\begin{figure}
\centering
\includegraphics[scale=1]{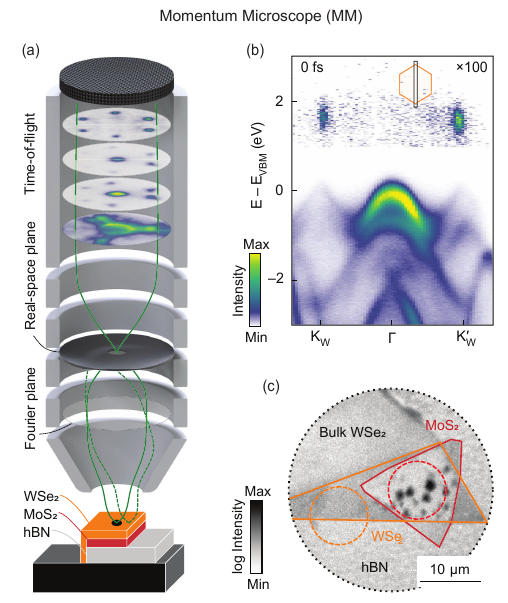}
\caption[MM]{Experimental configuration of a momentum microscope (MM) setup, which can locate the heterostructure in the real-space acquisition mode with micrometer resolution. (a) Illustration of a MM detector studying a WSe$_2$/MoS$_2$/hBN sample. (b) The acquired ARPES map at 0\,fs shows the valence bands of WSe$_2$, MoS$_2$ and hBN, as well as the bright A-excitons of WSe$_2$. (c) Regions of interest in the study indicated by the red and orange 10-$\mu$m diameter circles. Adapted from \onlinecite{schmitt2022formation}.}
\label{MM}
\end{figure}

TR-ARPES systems rely on state-of-the-art ARPES technology and we refer the reader to equilibrium ARPES reviews for comprehensive descriptions \cite{sobota2021angle,gedik2017photoemission,zhang2022angle,lv2019angle}. The sample is commonly prepared by cleaving in ultra-high-vacuum (UHV, commonly $p$\,$<$\,5\,$\cdot$\,10$^{-10}$\,mbar) conditions, and positioned in front of the electron detector via five- or six-axis manipulators. To investigate ultrafast electron dynamics in low-temperature phases, sub-10\,K cryogenic cooling is commonly available, although the omnipresent pump-induced thermal effects (even in perturbative pump regimes) often limit the achievable base temperature. Nowadays, one of three main electron detectors is used: time-of-flight (TOF) spectrometers, momentum microscopes (MM), or hemispherical analyzers (HA) \cite{sobota2021angle,zhang2022angle}. Figures\,\ref{Intro_systems},\,\ref{TOF}, and\,\ref{MM} illustrate the advantages of these three different electron detectors, respectively. In particular, TOF spectrometers allow the detection of iso-energy contours simultaneously over 2D momentum space; MMs can operate both in spatial and momentum modes, enabling detection of photoelectrons emitted from specific $\mu$m--regions (such as exfoliated samples); while HAs, though they require mechanically moving the sample or deflecting the captured electrons via voltages to map the whole 2D momentum space, generally provide higher energy and momentum resolutions, as well as a larger dynamic range in photoemission intensity.

TR-ARPES systems operating solely with UV probe light (\emph{e.g.}, 6 eV) do not require differential pumping stages to deliver UV into the chamber: UV light can be generated in air and delivered to the sample through windows transparent in the UV spectral range (\emph{e.g.}, MgF$_2$ or CaF$_2$). However, the trajectories of photoelectrons generated by 6-eV probe light, which have low kinetic energy, may be affected by stray fields inside the ARPES chamber, such as the electric fields caused by the difference in work function between the sample and its sample holder, and correction techniques may be required \cite{fero2014impact,gauthier2021expanding}.
On the other hand, XUV pulse generation and propagation need to take place in vacuum, and different strategies have been implemented to connect HHG generation chambers (operating commonly in the 10$^{-2}$/10$^{-4}$\,mbar range) to ARPES endstations, such as metallic windows or differential pumping lines with multiple pumping stages and/or capillaries. Moreover, XUV beamlines comprise not only HHG chambers but also monochromator systems to select specific harmonics. Various methods with different levels of complexity are employed to sort harmonics, ranging from simple metallic filters to multi-layer mirrors, Brewster plates, and gratings \cite{mills2019cavity,sie2019time,puppin2019time,lee2020high,dakovski2010tunable,frietsch2013high,mathias2007angle,buss2019setup,heber2022multispectral,poletto2010time}. 

Finally, the recent development of XUV light sources with full polarization control (linear and circular) \cite{comby2022ultrafast} in combination with the rise of high efficiency spin detectors \cite{gotlieb2013rapid} have now positioned the community to add spin/dichroic degree of freedom to the TR-ARPES technique, promoting exciting studies of spin and orbital dynamics of quantum materials over the entire momentum space (see Sec.\,\ref{Spin_section}).

\section{Mapping unoccupied and transient states} \label{Mapping_section}
TR-ARPES is frequently employed to map non-thermal electronic distributions in unoccupied states and to explore light-driven modifications of the electronic band dispersion. 
The mapping of unoccupied states is probably the most straightforward use of TR-ARPES: by taking advantage of pump pulses (commonly in the near-infrared/visible range) to redistribute the electronic population above the Fermi level and across momentum space, TR-ARPES grants access to the dispersion of unoccupied electronic states unattainable in conventional ARPES experiments. Note that although inverse photoemission spectroscopy would be an alternative technique to map the dispersion of unoccupied states, its poor energy-resolution and low cross-section make it unsuitable to explore low-energy ($<$1\,eV) unoccupied states.
Moreover, TR-ARPES enables momentum-resolved investigations of how intense light excitation may induce both sizable renormalizations of the electronic dispersion or appearance of electronic states with no equilibrium counterpart. 

Sections\,\ref{CB_semi} and\,\ref{TIs} present a few emblematic examples of how TR-ARPES maps the unoccupied conduction band of III-V and III-VI semiconductors and 2H transition metal dichalcogenides, as well as the topological surface state in 3D topological insulators.
Next, Section\,\ref{exciton_map} reviews the recent observation of bright and dark excitons in momentum space. Finally, Section\,\ref{Floquet} discusses the emergence of Floquet-Bloch states.

\subsection{Conduction band of semiconductors} \label{CB_semi}
\subsubsection{III-V group and III-VI semiconductors}
\begin{figure}
\centering
\includegraphics[scale=1]{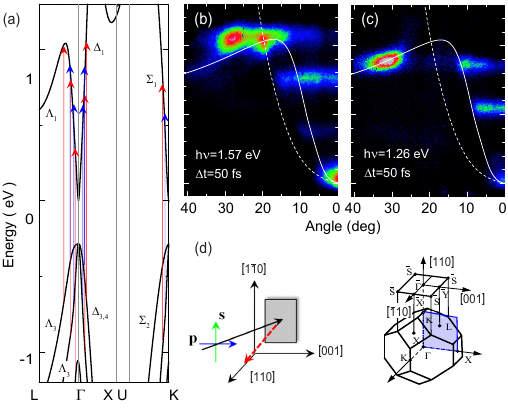}
\caption[MapInSb]{TR-ARPES mapping of the unoccupied states of InSb semiconductor shows how different pump energies access distinct optical transitions. (a) Band structure of InSb with available inter-band optical transitions with 1.57\,eV (solid lines) and 1.26\,eV (dashed lines) photons. TR-ARPES spectra of InSb(110) acquired 50\,fs after $p$-polarized (b) 1.57\,eV and (c) 1.26\,eV pump excitation. Solid and dashed white lines represent the CB dispersion along the high-symmetry directions $\Gamma$-L and $\Gamma$-X, respectively. (d) Illustration of the surface and bulk BZ for InSb(110). Adapted from \onlinecite{tanimura2015ultrafast}.}
\label{Map_InSb}
\end{figure}

GaAs and InSb are prototypical III-V group semiconductors extensively used in optoelectronic devices and sensors. Therefore, precise mappings of their unoccupied conduction band (CB), available optical transitions, as well as ultrafast scattering processes over the whole Brillouin zone are of crucial importance. Taking advantage of different pump photon energies in the visible range of the spectrum, multiple groups have precisely mapped optical transitions into the unoccupied CB and subsequent intra- and inter-valley scattering processes in both GaAs and InSb via TR-ARPES \cite{kanasaki2014imaging,tanimura2015ultrafast,tanimura2016formation,sjakste2018energy,tanimura2021ultrafast}. TR-ARPES data acquired on InSb by \onlinecite{tanimura2015ultrafast} provide a key illustrative example. Figure\,\ref{Map_InSb} shows the mapping of the unoccupied states at 50\,fs pump-probe delay for two different excitation schemes: p-polarized 1.57\,eV and 1.26\,eV pumps, respectively. The solid and dashed lines in panels (b)-(c) highlight the dispersion of the CB along two crystallographic directions, namely $\Gamma$--L and $\Gamma$--X. 
The differing distributions of the intensity hotspots observed in Fig.\,\ref{Map_InSb} showcase the capability to access distinct optical transitions from the heavy-hole, light-hole and split-off valence bands, depending on the pump photon energy utilized. By tracking the temporal evolution of these photo-injected electrons over the entire Brillouin zone, it is possible to study intra- and inter-valley relaxation processes. In the specific case of InSb, TR-ARPES data show that electrons injected in the $\Gamma$-valley with an energy exceeding the minimum of the L-valley undergo inter-valley scattering on a sub-50\,fs scale and relax from L back to $\Gamma$ within $\sim$0.8\,ps. Instead, if the energy of photoexcited electrons does not exceed the minimum of the L-valley, their decay within the $\Gamma$-valley is governed by impact ionization processes \cite{tanimura2015ultrafast}.

As another example, InSe is a III-VI group semiconductor layered chalcogenide, with a direct bulk band gap of $\sim$1.25\,eV. Similarly to what has been discussed above for InSb, TR-ARPES at different excitation energies has allowed \onlinecite{chen2018ultrafast} to accurately map its unoccupied states and track how non-thermal carriers accumulate at the bottom of CB at $\Gamma$ or scatter into a valley at the M point. More recently, they reported the ultrafast response of a quasi-2D electron gas induced on the surface of InSe via Cs doping \cite{chen2020ultrafast}. This study represents the first ultrafast mapping of the light-induced population of a quasi-2D electron gas, and by accessing the dynamics of the photoemission intensity and photo-injected excess energy on a ps timescale, it also offers a qualitative estimate of the electron-phonon scattering processes as a function of the Cs doping (see also Section\,\ref{TTM}). In particular, they discuss the observed $\sim$\,0.5\,ps relaxation time in terms of the coupling of optically excited quasi-2D electrons with optical phonons via the Frohlich mechanism \cite{chen2020ultrafast}.

\begin{figure*}
\centering
\includegraphics[scale=1]{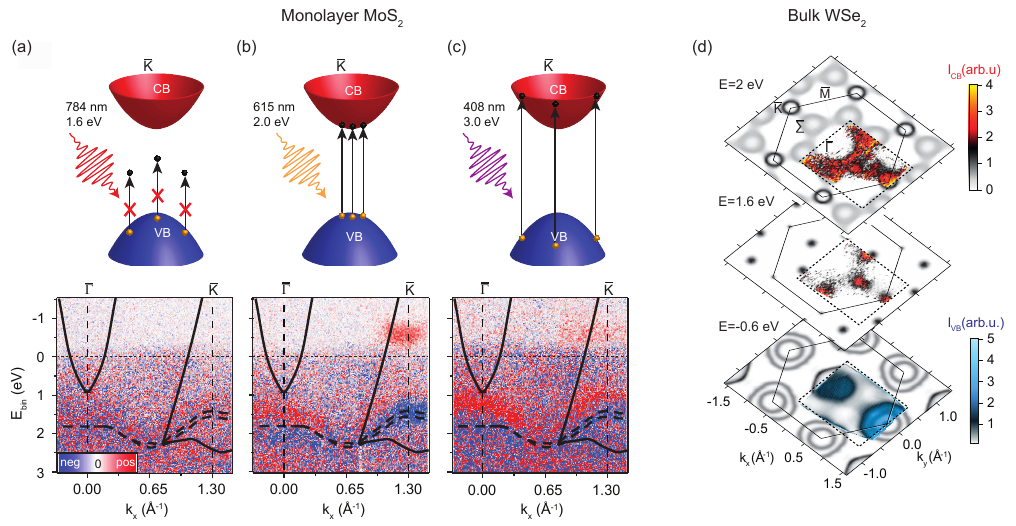}
\caption[MapMoS2]{Mapping of the conduction band of transition metal dichalcogenides. (a)-(c) Sketch of the available transitions and TR-ARPES differential mapping acquired with 25\,eV photons for monolayer MoS$_2$ around $\overline{\mathrm{K}}$ at the peak of the optical excitation by using three different pump energies: (a) below the CB onset; (b) resonant to the band gap; (c) above the CB onset. By tuning the pump photon energy, the authors estimated the value of the direct bandgap to 1.95\,eV. Solid lines in (a)-(c) mark the boundaries between the projected bulk band continuum and gaps of Au(111), while dashed curves are the equilibrium fitted dispersion. Adapted \onlinecite{cabo2015observation}. (d) Photoelectron intensity distribution as a function of parallel momentum for three binding energies at a pump-probe delay of 100\,fs acquired for bulk 2H WSe$_2$ with 21.7\,eV probe and 3.1\,eV pump. The higher pump energy allows the mapping of the CB within a wider energy range, which favors a more comprehensive comparison with \textit{ab-initio} calculated band structure. Here, the experimental data is collected in the region delimited by the dashed line; outside this region, the results of calculations are displayed, and the theoretical band dispersion along the k$_{\text{z}}$ direction was integrated. Adapted from \onlinecite{puppin2022excited}.}
\label{Map_TMD}
\end{figure*}

\subsubsection{Transition Metal Dichalcogenides} \label{TMD_map}
Transition Metal Dichalcogenides (TMDs), especially in the 2D limit, have attracted considerable interest for the development of novel valleytronic devices.
While centrosymmetric bulk TMDs (2H phase) present an indirect bandgap between the top of the VB at $\Gamma$ and the bottom of the CB at the corner of the hexagonal Brillouin zone (\emph{i.e.} K-point), monolayer TMDs exhibit a direct bandgap at K. When considering the monolayer limit, the lack of inversion symmetry in concert with strong spin-orbit coupling lifts the degeneracy of two adjacent K points, commonly identified as K and K'. The different valley pseudospins associated with K and K' open the possibility for valley-selective electronic excitations with circularly polarized light \cite{mak2012control,zeng2012valley}. While this effect should be forbidden in bulk materials, a hidden pseudospin texture \cite{zhang2014hidden} may appear as a consequence of the local breaking of inversion symmetry in some sections of the unit cell, leading to dichroic effects even in 2H-TMDs (see Section\,\ref{dichroic_TRARPES}).

TR-ARPES investigations of various monolayer and bulk TMDs have precisely identified their electronic bandgap and ultrafast relaxation processes and have also explored the light-induced valley-polarization \cite{cabo2015observation, bertoni2016generation, hein2016momentum, ulstrup2016ultrafast, ulstrup2017spin, wallauer2016intervalley, puppin2022excited,lee2021time,wallauer2020momentum,beyer201980}.
The work by \onlinecite{cabo2015observation} focused on light-induced occupation of the CB and subsequent recombination dynamics in monolayer MoS$_2$ epitaxially grown on Au(111), as a function of the pump photon energy. Figure\,\ref{Map_TMD}(a)-(c) displays TR-ARPES differential maps computed at the peak of the optical excitation for three distinct excitation energies with respect to the bandgap. By fitting the dispersion of the occupied VB and of the light-populated CB, the authors extracted a 1.95\,eV direct bandgap, which differs by several hundreds of meV from estimates via equilibrium ARPES upon alkali doping, thus emphasizing how external perturbations may severely impact the electronic properties of monolayer TMDs \cite{cabo2015observation}.
A large transient redistribution of the carriers, as a pump excitation may induce, can lead to changes in the screening potential that may significantly renormalize the electronic gap \cite{puppin2022excited,andreatta2019transient,liu2019direct}. For instance, \onlinecite{ulstrup2016ultrafast} reported a renormalization of the electronic gap as large as $\sim$400\,meV in monolayer MoS$_2$ grown on graphene due to the combination of pump-induced carrier redistribution and the weak screening in the monolayer MoS$_2$/graphene heterostructure. 
Figure\,\ref{Map_TMD}(d) shows a recent mapping of the occupied VB and unoccupied CB of 2H-WSe$_2$, illustrating how TR-ARPES data can be directly compared to the electronic dispersion predicted by \textit{ab-initio} calculations \cite{puppin2022excited}.

\begin{figure}
\centering
\includegraphics[scale=1]{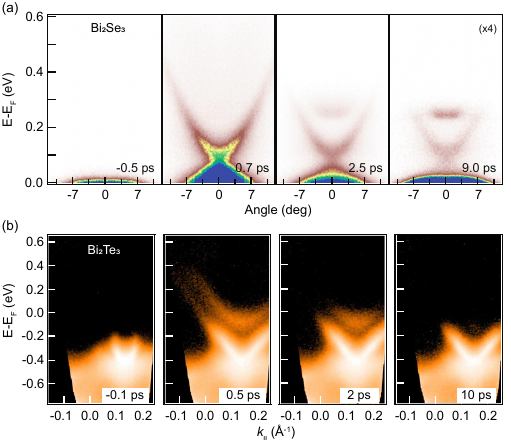}
\caption[Map1TI]{TR-ARPES mapping of the unoccupied states acquired near the $\Gamma$ point with 6\,eV upon a 1.5\,eV pump of two topological insulators: (a) Bi$_2$Se$_3$ and (b) Bi$_2$Te$_3$. The near-IR excitation promotes carriers into the topological Dirac-like surface state and the CB. Subsequent relaxation processes lead to an accumulation of hot carriers at the bottom of the bulk CB. Panel (a) is adapted from \onlinecite{sobota2014ultrafast}; panel (b) is adapted from \onlinecite{hajlaoui2012ultrafast}.}
\label{Map1_TI}
\end{figure}

TR-ARPES papers on monolayer TMD/graphene heterostructures have reported an ultrafast and opposite-in-energy band shift of the VB of WS$_2$ and $\pi$-band of graphene \cite{aeschlimann2020direct}, as well as charge and energy transfer between WSe$_2$ and graphene \cite{dong2021observation}, evidence of a charge-separated transient state with photoexcited electrons and holes located in the TMD and graphene layer, respectively.
More recently, \onlinecite{majchrzak2021spectroscopic} have comprehensively characterized the quasiparticle band gap in monolayer and bilayer MoS$_2$ and WS$_2$ on different substrates: graphene, Au(111), and Ag(111). They reported a clear influence of the substrate on the electronic gap, as well as on the relaxation dynamics of the photoexcited carriers. In the specific case of bulk MoS$_2$, \onlinecite{wallauer2020momentum} showed that the surface exhibits an electronic gap larger than the bottom layers by employing tunable visible excitations across the different gap amplitudes. In particular, by pumping carriers solely into the bottom-layer's K points, they observed an interlayer charge transfer into the $\Sigma$-valley of the topmost layer on a $\sim$\,20\,fs timescale.
These papers demonstrate the capability of TR-ARPES both to access the unoccupied band structure in perturbative regimes, and to map out-of-equilibrium light-induced renormalizations of the electronic properties of quantum materials.

TR-ARPES studies have also demonstrated the selective population of K/K' valleys in 2H-WSe$_2$ and single-layer WS$_2$ via dichroic excitation \cite{bertoni2016generation,ulstrup2017spin,beyer201980,volckaert2019momentum} (see more details in Section\,\ref{Spin_section} and Fig.\,\ref{Spin_Dic_TMD}), as well as the population of the $\Sigma$-valley (which is the CB minimum in 2H-TMDs; also denoted as $\Delta$ and Q in the literature) on sub-100\,fs timescales \cite{bertoni2016generation,hein2016momentum,wallauer2016intervalley,wallauer2020momentum}. This comprehensive momentum-resolved mapping of the out-of-equilibrium redistribution of the electronic population via intra- and inter-band scattering processes in TMDs further attests to the capability of TR-ARPES to reveal fundamental electrodynamics that remain hidden to other equilibrium and time-resolved techniques.

\begin{figure*}
\centering
\includegraphics[scale=1]{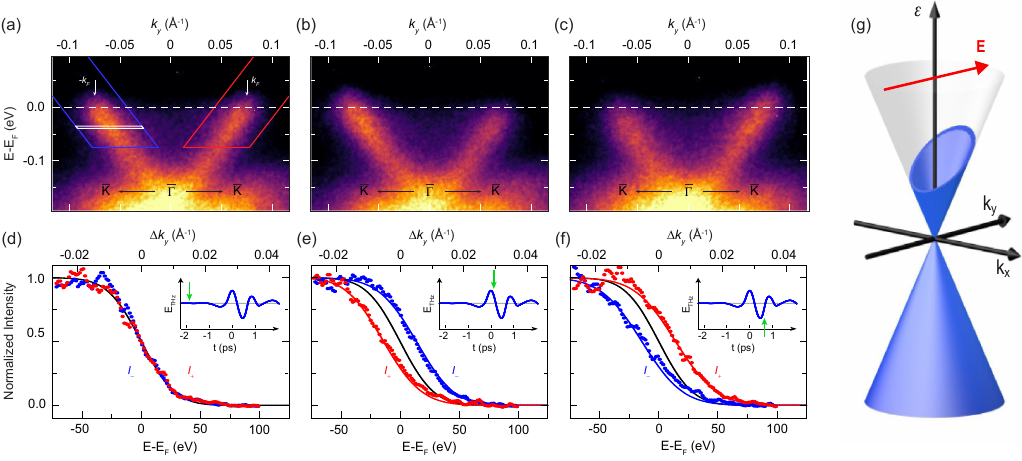}
\caption[Map2THz]{Probing the effect of THz pump excitation on the topological surface state of Bi$_2$Te$_3$ in momentum space. TR-ARPES spectra along $\Gamma$K acquired (a) before the THz pump, (b) at the electric field node with the largest negative gradient and (c) at the field node with the largest positive gradient, respectively. (d)-(f) Photoemission intensity of the left (blue circles) and right (red) branches of the Dirac cone integrated inside the blue and red boxes shown in (a)-(c). Red and blue solid lines represent simulations based on the Boltzmann equation, while the black line is the Fermi-Dirac distribution convolved with the experimental resolution. Insets in (d)-(f) illustrate the evolution of the electric field where the green arrows mark the time delay of each snapshot. The observed asymmetric population (\emph{i.e.} a transient momentum shift of the electrons) arising when the electric field is applied indicates the emergence of a transient surface current driven by the ultrashort THz pump pulse. (g) Sketch of how the acceleration of Dirac fermions in the topological surface state can shift the Fermi surface via an electric field. Adapted from \onlinecite{reimann2018subcycle}.}
\label{Map2_THz}
\end{figure*}

\subsection{Topological Insulators (TIs)} \label{TIs}
Three-dimensional topological insulators (TIs) are materials in which the strong spin-orbit interaction leads to the emergence of a metallic topological surface state (TSS), which lies within a bulk gap and has a Dirac-like dispersion with a helical spin structure \cite{fu2007topological,hasan2010colloquium}. 
An extensive experimental campaign to study TIs via TR-ARPES began in 2012 for both technical and scientific reasons. First, since the TSS is commonly centered at $\Gamma$ in these materials, it is easily accessible to a low photon-energy source; also, the scientific community was interested in evaluating the dispersion of the Dirac cone and bulk bands in the unoccupied states, as well as the intrinsic scattering processes of Dirac electrons with topological protection.
As a result, these compounds served as an excellent stimulus for commissioning and characterizing the performance of new TR-ARPES systems, significantly advancing this ultrafast spectroscopy technique.
\onlinecite{sobota2012ultrafast}, followed closely by \onlinecite{wang2012measurement}, \onlinecite{hajlaoui2012ultrafast}, and \onlinecite{crepaldi2012ultrafast}, first mapped the unoccupied dispersion of the TSS and bulk bands for the prototypical TIs Bi$_2$Se$_3$ and Bi$_2$Te$_3$.
The vanishing of the density of states at the Dirac point, together with the inefficient coupling between bulk and surface states, leads to relaxation times ranging from a few hundreds of fs for unoccupied bulk bands 0.5\,eV above E$_F$ to several tens-hundreds of ps for the bottom of the conduction band and TSS (see also Section\,\ref{TTM_TI}). These relatively long decay rates favor the mapping of unoccupied states in TIs. Figure\,\ref{Map1_TI} shows the unoccupied band structure of Bi$_2$Se$_3$ and Bi$_2$Te$_3$ \cite{hajlaoui2012ultrafast,sobota2012ultrafast,sobota2014ultrafast}. After the pump photoexcitation from the occupied states (bulk and surface) into the unoccupied bulk bands, the carriers' thermalization is characterized by the accumulation at the bottom of the bulk CB and the slow bulk-to-surface diffusion, leading to delayed ($\sim$500\,fs) transient population of the TSS \cite{sobota2012ultrafast,hajlaoui2012ultrafast,hedayat2018surface,sanchez2016ultrafast}.
For Bi$_2$Se$_3$, a parabolic dispersion is populated before time zero and decays toward negative delays \cite{sobota2012ultrafast}. This feature does not belong to the band structure of Bi$_2$Se$_3$, but results from the transient population of the first image potential state -- a bound state between an electron in vacuum in front of a metallic surface and the photohole left behind; it is populated by the 6-eV beam and photoemitted by the 1.55-eV pulse \cite{luth2013surfaces,petek1997femtosecond}.

Following these initial studies, the TR-ARPES community has widely applied the technique to probe the unoccupied electronic structure of a variety of TIs \cite{reimann2014spectroscopy,kanasaki2014imaging,zhu2015ultrafast,sterzi2017bulk,neupane2015gigantic,sumida2017prolonged,hedayat2018surface,xu2019magnetic,maklar2022ultrafast}. \onlinecite{sobota2013direct} employed two-photon photoemission with two 6-eV beams and \onlinecite{niesner2012unoccupied} employed circularly-polarized optical excitations to reveal the presence of a second topological surface state in the unoccupied bulk electronic gap of Bi$_2$Se$_3$ and Bi$_2$Te$_2$Se, respectively. Furthermore, the advent of spin-resolved TR-ARPES allowed both to map the intrinsic spin-polarization of the unoccupied TSS and also to track the evolution of optically-induced spin-polarized electronic population, as well as to reveal the presence and spin-polarization of surface resonance states \cite{cacho2015momentum, jozwiak2016spin, sanchez2016ultrafast, sanchez2017subpicosecond} (see Section\,\ref{Spin_section}).
\onlinecite{kuroda2016generation,kuroda2017ultrafast, reimann2018subcycle} have shown that the use of mid-IR light (sub 0.5\,eV) and THz radiation can generate ultrafast photocurrents in the TSS. In particular, the seminal work of \onlinecite{reimann2018subcycle} was the first to successfully implement THz excitations as pumping scheme in a TR-ARPES experiment. Figure\,\ref{Map2_THz} shows the ARPES images (top row) and corresponding integrated photoemission intensity for the two branches of the Dirac cone (bottom) acquired on Bi$_2$Te$_3$ at different pump-probe delays, after correcting for the THz streaking of photoemitted electrons. These experimental results indicate that the in-plane electric field of the THz pulse induces a transient displacement of the Fermi surface dependent on the field's direction, \emph{i.e.} a transient surface current \cite{ashcroft1976solid}. Moreover, by comparing the photo-induced current density at the surface to the THz field, \onlinecite{reimann2018subcycle} were able to infer that TSS electrons are weakly scattered ($\sim$1 ps as lower bound). This prominent work demonstrates that the mapping of ARPES spectra with subcycle resolution may provide a new way to track transport processes in complex materials.

Finally, in recent years magnetic TIs have attracted substantial attention because of the coexistence of magnetism and topology \cite{otrokov2019prediction,li2019intrinsic}, which may prompt novel quantum phenomena such as the quantum anomalous Hall effect \cite{deng2020quantum}. Only a few TR-ARPES papers have explored ultrafast dynamics in magnetic TIs to date, in MnBi$_{2n}$Te$_{3n+1}$ and EuSn$_2$As$_{2}$ \cite{nevola2020coexistence,zhong2021light,li2019dirac,lee2023layer}. \onlinecite{zhong2021light} employed TR-ARPES with $\mu$m resolution to specifically resolve one of the four surface terminations of MnBi$_8$Te$_{13}$ and reported a long-lasting ($>$\,10\,ns) light-induced filling of the hybridization gap originating from the hybridization between the MnBi$_2$Te$_4$ septuple layer and Bi$_2$Te$_3$ quintuple layer. This light-induced filling of the hybridization gap was discussed in terms of a transient change of the interlayer coupling due to charge redistribution.

\subsection{Excitons in semiconductors} \label{exciton_map}

Excitons are correlated bound states composed of electrons photoexcited in the conduction band and holes left behind in the valence band, and they are commonly observed in semiconductors. Exciton physics is of particular importance in the context of semiconducting 2D TMDs as they dominate their optical properties, thus making the fundamental research of how excitons emerge and interact essential for future applications in optoelectronics and valleytronics \cite{wang2018colloquium}. Exciton physics is commonly explored via all-optical techniques with no momentum resolution, which access mainly bright excitons. These excitons consist of pairs of electrons and holes with the same crystal momentum, and can thus be created via direct optical excitation (for instance, by photoexciting an electron from the VB to the CB at the K point of monolayer TMDs). However, all-optical techniques are commonly (to the first order) blind to excitons with a non-zero total momentum, \emph{a.k.a.} dark excitons, which are comprised of bound electrons and holes located in different valleys in momentum space. For this reason, a time- and momentum-resolved technique able to directly probe dark excitons, as well as to track how and on what timescale photoexcited electrons and holes are bound into excitonics states, has been of great interest. 

Excitons have been studied via time-resolved photoemission for the past two decades \cite{weinelt2004dynamics,weinelt2005electronic}, but only in recent years have theoretical studies discussed how TR-ARPES, which accesses the transient evolution of the single-particle band structure (see Section\,\ref{TR_ARPES_Int}), could serve as a probe to study excitons, which are two-fermion quasiparticles \cite{perfetto2016first,rustagi2018photoemission,christiansen2019theory}. Experimental evidence of excitons in momentum space has been reported in Si(100) \cite{weinelt2004dynamics,weinelt2005electronic}, ZnO \cite{deinert2014ultrafast}, and Cu$_2$O \cite{tanimura2019dynamics} semiconductors. Following the recent development of momentum microscopes that allow the investigation of $\mu$m-sized samples, \onlinecite{madeo2020directly} reported the direct observation in momentum space of bright and dark excitons in monolayer WSe$_2$. By employing TR-ARPES with micrometer spatial resolution and an optical excitation resonant to the first bright exciton (namely the A-exciton at 1.72\,eV), the authors observed an accumulation of intensity $\sim$\,400-500\,meV below the bottom of the CB developing instantly at the K-valley, and with few hundreds fs delay at the Q-valley. The appearance of such transient spectral weight and its position in k-space are direct signatures of the excitonic state in WSe$_2$, and allowed them also to extract the $\sim$\,400\,fs intrinsic timescale associated with the formation of the dark Q-valley excitons through scattering from the K-valley. By using sub-50\,fs visible pump pulses, \onlinecite{wallauer2021momentum} subsequently showed that the formation of dark excitons in WS$_2$ takes place within a few tens of femtoseconds via exciton-phonon scattering of the electron-hole polarization cloud at the K points.

\begin{figure}
\centering
\includegraphics[scale=1]{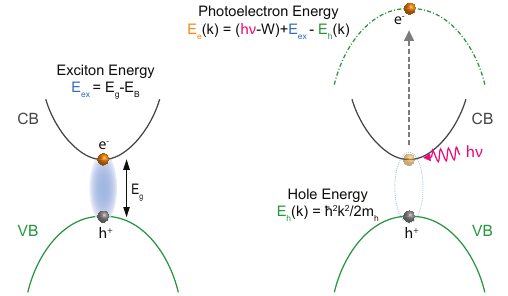}
\caption[MapIXL]{Schematic illustration of the dispersion of an electron photoemitted from an exciton. Following conservation laws, an electron photoemitted from an excitonic bound state exhibits an energy-momentum dispersion that mimics that of the valence band, with a spectral intensity that diminishes when moving away from the initial momentum location of the bound state.}
\label{Map_ExcitonScheme}
\end{figure}

\begin{figure*}
\centering
\includegraphics[scale=1]{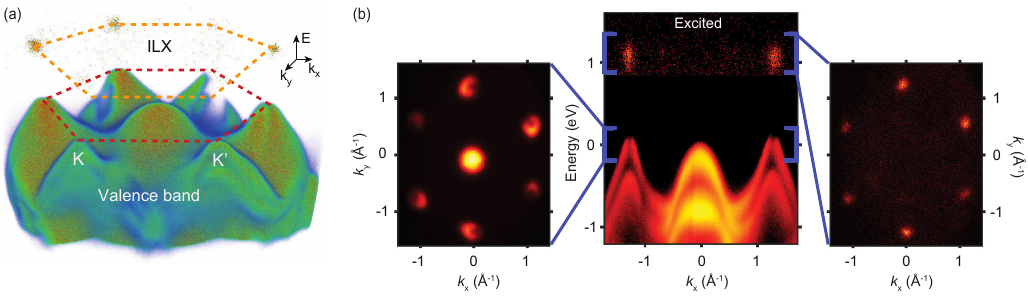}
\caption[MapIXL]{Employing TR-ARPES to map excitons in momentum-space. (a) 3D rendering of the TR-ARPES mapping of the occupied and unoccupied band structure of a WSe$_2$/MoS$_2$ heterostructure, $\sim$\,2$^o$ twist angle, with $\mu$m-resolution via momentum microscopy, featuring the presence of an interlayer exciton (ILX) above E$_F$. (b) TR-ARPES spectrum along the K--$\Gamma$ direction 25\,ps after optical excitation, along with iso-energy maps displaying the momentum location of (left) the top of the VB and
(right) of the ILX. Note the decrease of spectral weight in the VB accompanying the formation of the ILX. These maps are
generated by integrating over the energy ranges specified by the blue brackets. Adapted from \onlinecite{karni2022structure}.}
\label{Map_IXL}
\end{figure*}

The momentum dispersion of the electron photoemitted from the A-exciton (at K) in monolayer WSe$_2$ was later obtained in follow-on TR-ARPES studies by \onlinecite{man2021experimental} and \onlinecite{dong2021direct}. When an exciton is annihilated through a photoemission process, the dispersion of the previously bound electron mimics the dispersion of the VB \cite{man2021experimental,rustagi2018photoemission,christiansen2019theory,dong2021direct}, in agreement with the exciton dispersion reported in Cu$_2$O \cite{tanimura2019dynamics}. Figure\,\ref{Map_ExcitonScheme} shows how energy and momentum conservation laws determine the dispersion of an electron photoemitted from a bright exciton. In addition, assuming a direct relation between the k-dependence of the TR-ARPES spectral weight and the square of the excitonic wave function, \cite{man2021experimental,dong2021direct} computed via Fourier transform an estimate of the spatial extent of the excitonic wavefunction itself, obtaining a root mean square value of $\approx$\,2\,nm for the A-exciton of WSe$_2$.
These studies have demonstrated and established how the TR-ARPES technique can map excitons, thus paving the way towards the first observation and estimate of the spatial extent of interlayer excitons (IXL) in Moir\'{e} WSe$_2$/MoS$_2$ heterostructures \cite{karni2022structure,schmitt2022formation}. Figures\,\ref{MM} and\,\ref{Map_IXL} show the ARPES mapping of the entire BZ (via momentum microscope) of WSe$_2$/MoS$_2$ heterostructures with a twist angle of $\sim$10$^o$ and $\sim$2$^o$, respectively, highlighting the appearance of transient spectral weight $\sim$1\,eV above the maximum of the VB. For the $\sim$2$^o$-twist heterostructure, TR-ARPES maps along the $\Gamma$-K direction shown in Fig.\,\ref{Map_IXL}(b) demonstrate the presence of IXL-bound holes and electrons at the K valleys (holes are located at the maximum of the valence band while electrons at 1\,eV binding energy) \cite{karni2022structure}.  
A TR-ARPES investigation on monolayer MoS$_2$ has shown how excitonic correlations may renormalize the band dispersion of both the CB and VB. In particular, the simultaneous enhancement of both the band gap and the effective mass may act as a direct probe of excitonic many-body correlations \cite{lin2022exciton}.

\begin{figure*}
\centering
\includegraphics[scale=1]{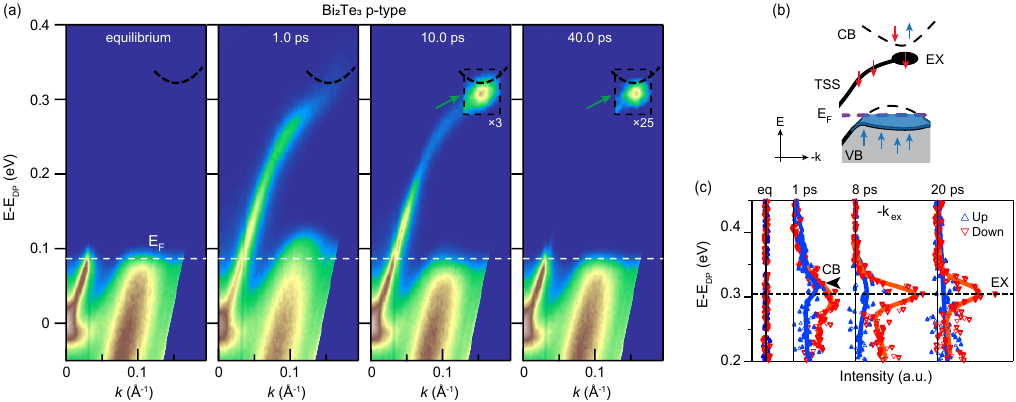}
\caption[MapTopoEX]{Revealing spatially indirect excitons in 3D topological insulators. (a) TR-ARPES spectra along $\Gamma$-M of p-doped Bi$_2$Te$_3$ at equilibrium and three different pump-probe delays. The energy scale refers to the position of the Dirac point in energy. (b) Sketch of the experimental observation, highlighting the spin-polarization and formation of the excitonic state in the intensity buildup and the topological surface state. (c) Spin-resolved spectra of the exciton intensity buildup at the exciton momentum k$_{\text{EX}}$, measured at a delay time of 20\,ps. Adapted from \onlinecite{mori2023spin}.}
\label{Map_TopologialEx}
\end{figure*}

Recently, \onlinecite{mori2023spin} reported evidence of a build-up of intensity below the conduction band of Bi$_2$Te$_3$, as shown in Fig.\,\ref{Map_TopologialEx}(a).
By tracking the transient evolution of the spectral weight below the conduction band and verifying its spin polarization via spin-resolved TR-ARPES, as well as by reporting changes in the topological surface state dispersion, the authors attributed this intensity build-up to a long-lived spatially indirect spin-polarized exciton, in which the photo-hole resides at the top of the bulk valence band and the electron is in a mixed state between the topological surface state and the conduction band, thus somewhat localized at the surface [see Fig.\,\ref{Map_TopologialEx}(b)].

Overall, these recent studies demonstrate the ability of TR-ARPES to access and map the momentum-dispersion of excitonic wavefunctions, as well as track the excitons' ultrafast formation and interactions. These results open the door to the use of TR-ARPES for investigating many-particle excitations, such as excitons, biexcitons and trions in TMDs, topological systems, and quantum materials in general.

\subsection{Floquet-Bloch states} \label{Floquet}
Breaking or modifying the underlying symmetries of materials can have remarkable effects on their electronic properties. In recent years such an approach has attracted the interest of the broad scientific community and has led to prominent discoveries. Twisted bilayer graphene is an exemplary case \cite{cao2018correlated,cao2018unconventional} in which novel electronic properties emerge from the fine tuning of its spatial translational symmetry via a Moir\'{e} pattern. Floquet engineering promises a similar level of control over electronic properties by taking advantage of the temporal periodicity of the electric field of light excitation and can potentially be applied to nearly any material \cite{rudner2020band,oka2019floquet}. 

\begin{figure}[b]
\centering
\includegraphics[scale=1]{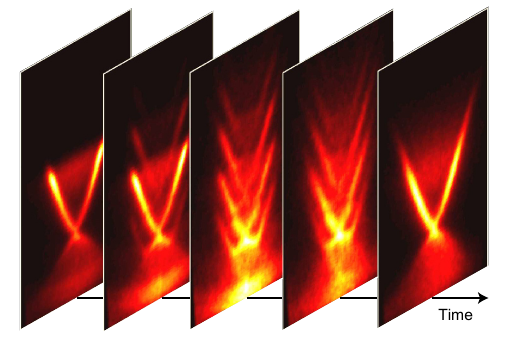}
\caption[Map3Floq]{Tracking the emergence of Floquet-Bloch states on the surface of Bi$_2$Se$_3$ induced by 160\,meV, 300\,fs light pulse. The coherent interaction between the time-periodic potential of the light pulse and the electrons in the Dirac cone results in Floquet–Bloch states that appear as replicas in energy of the original Dirac cone. Dynamical gaps in the electronic structure open up at positions where the replica cones intersect the original cone. The replica bands disappear once the driving field dies off.
Adapted from \onlinecite{gedik2017photoemission} and \onlinecite{mahmood2016selective}.}
\label{Map3_Floq}
\end{figure}

The Floquet theorem is a time analog of the Bloch theorem \cite{ashcroft1976solid} and states that when a Hamiltonian $H(\tau)=H(\tau+\text{T})$ is periodic in time, new \emph{quasistatic} energy levels (\emph{a.k.a.} Floquet levels) emerge with a spacing of $2 \pi / \text{T}$ \cite{faisal1997floquet}. Since Bloch states are intrinsically present in crystalline materials, Floquet states in solids are commonly referred to as Floquet-Bloch states. 
In principle, Floquet-Bloch states can be driven by any generic time-periodic perturbation. However, at the moment, successful experimental investigations of Floquet physics have relied solely on taking advantage of the intrinsic time-periodicity of the electric field of light pulses, with emblematic examples in molecules and photonic crystals \cite{rechtsman2013photonic,bandrauk1981photodissociation}.
In solids, the crossing of Floquet-Bloch replicas with original bands may lead to hybridization and opening of electronic gaps, thus effectively engineering the systems' electronic properties. Figure.\,\ref{Map3_Floq} illustrates the emergence and temporal evolution of such non-equilibrium states induced by the pump excitation in a topological insulator.

\onlinecite{wang2013observation} first reported the appearance of Floquet-Bloch states in Bi$_2$Se$_3$ with energy- and momentum-resolution. As a function of the polarization of the impinging mid-IR field, energy gaps may (or may not) open along different momentum directions \cite{oka2009photovoltaic,zhou2011optical}. In the particular case of Bi$_2$Se$_3$, a linearly polarized mid-IR excitation prompts a gap opening only along the $k_y$ direction. In contrast, the use of circularly polarized light with photon energy comparable with the bulk energy gap breaks the intrinsic time-reversal symmetry of the system and opens a full gap at the Dirac point, as well as Floquet-Bloch hybridization gaps along all momentum directions\cite{wang2013observation}.

\begin{figure*}
    \centering
    \includegraphics[scale=1]{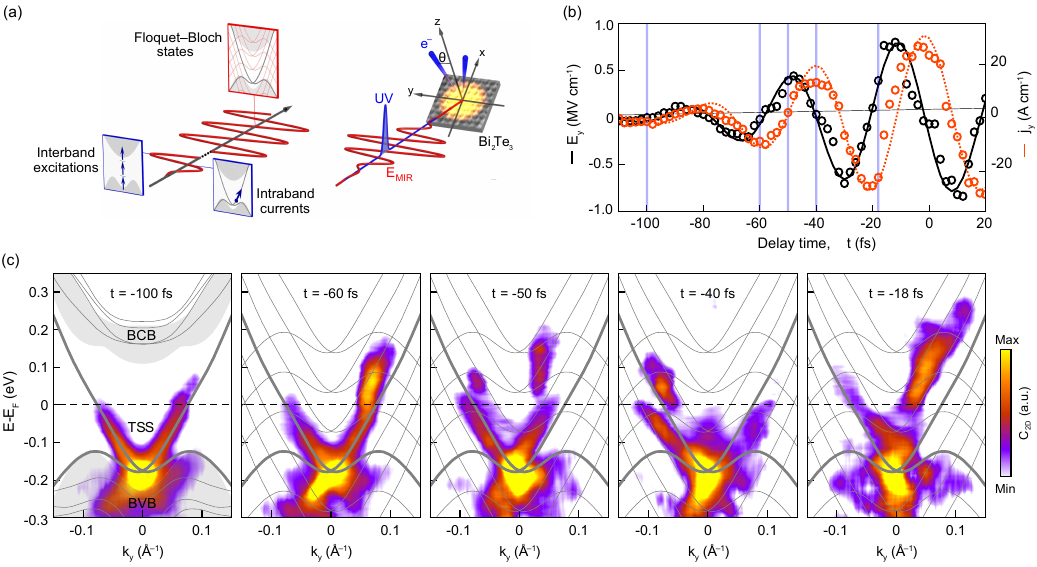}
    \caption[]{Floquet-Bloch states on subcycle timescales in Bi$_2$Te$_3$. (a) Left: illustration of electronic processes driven via mid-IR excitation, such as multiphoton interband excitation and intraband currents, which may also result in the emergence of Floquet-Bloch states. Right: schematic experimental strategy to investigate the emergence of Floquet-Bloch states on subcycle timescales. (b) Electric-field waveform from mid-IR-induced momentum streaking (black markers) and current density (red markers). The black solid line shows the analytic function of the electric field with a peak amplitude of 0.8 MV/cm, while the dashed red line is the electric current induced by this electric field according to the Boltzmann model \cite{reimann2018subcycle}. (c) Curvature-filtered TR-ARPES maps recorded at different pump-probe delays during the first half of the mid-IR driving pulse. Gray lines display the band structure calculated by DFT. The electronic distribution splits into multiple branches the follow the Floquet-Bloch replicas of the topological surface state (TSS). Adapted from \onlinecite{ito2023build}.}
\label{MapIto_Floquet}
\end{figure*}

\begin{figure*}
\centering
\includegraphics[scale=1]{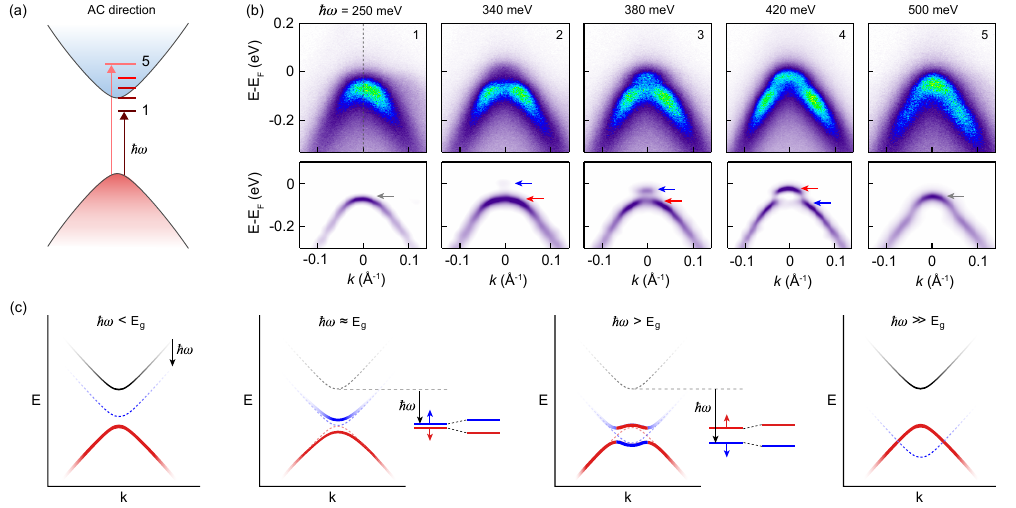}
\caption[Map3Floq]{Floquet engineering of black phosphorus. (a) Schematic of the band structure of semiconducting black phosphorus (with electronic gap E$_g \sim$ 340 meV), with the different mid-IR photon energies used. (b) TR-ARPES maps at zero pump-probe delay (top) and corresponding second-derivative images (bottom) for different mid-IR photon energies. (c) Sketch of the light-induced modifications of the electronic band structure for four different pump photon energies: sub gap, comparable to the gap, slightly above gap, and above gap. Adapted from \onlinecite{zhou2023pseudospin}.}
\label{MapNew_FloquetBlack}
\end{figure*}

Although both Floquet physics and the laser-assisted photoemission (LAPE) phenomenon\footnote{LAPE is commonly used to find the zero pump-probe delay of TR-ARPES systems and characterize their temporal resolution \cite{eich2014time}. Operationally, LAPE appears only when the pump electric field has an out-of-plane component relative to the sample surface.} \cite{saathoff2008laser} lead to the appearance of replica bands along the energy direction, they originate from different physics. Light-induced Floquet-Bloch states emerge from the dressing of a bound electron \emph{inside} the solid by a photon, whereas LAPE describes the photon dressing of a free electron \emph{outside} the solid and results in Volkov states \cite{mahmood2016selective}. Therefore, band hybridization and consequent gapping of the electronic band structure can originate only from Floquet-Bloch states and not from Volkov states. The different contributions to the light-induced band replicas observed in ARPES images arising from Floquet-Bloch and Volkov states can be distinguished via a careful assessment of the intensity and gapping of the light-induced replicas as a function of the momentum direction and polarization of the driving light pulse (see Fig.\,\ref{TOF}) \cite{mahmood2016selective,schuler2022probing}. 

A recent TR-ARPES work has offered further insights into the subcycle formation of Floquet-Bloch replicas \cite{ito2023build}. By using probe pulses with a temporal duration shorter than the optical cycle of the mid-IR driving pulse [see Fig.\,\ref{MapIto_Floquet}(a)], they reported the emergence of Floquet-Bloch replica bands in Bi$_{2}$Te$_{3}$ within a single optical cycle of the driving mid-IR pulse. Figure\,\ref{MapIto_Floquet}(c) shows how intraband currents fill the Floquet sidebands at opposite momenta depending on the orientation of the driving electric field until interband electron scattering destroys Floquet bands \cite{ito2023build}.

In systems other than TIs,\onlinecite{reutzel2020coherent} and \onlinecite{aeschlimann2021survival} succeeded in observing Floquet-Bloch replica via four-photon photoemission and TR-ARPES in Cu(111) and WSe$_2$, respectively. 
While evidence of Floquet-engineering in exfoliated graphene has been reported in a time-resolved transport experiment \cite{mciver2020light}, subsequent TR-ARPES exploration of the effects of strong mid-IR excitation on graphene did not observe any gap opening. Instead, \onlinecite{aeschlimann2021survival} observed a pronounced broadening of the spectral features, which they attributed to efficient decoherence scattering, pointing out the possible practical limitations of Floquet engineering for systems with intrinsically short scattering times (see section\,\ref{TTM}).

Recently, \onlinecite{zhou2023pseudospin} has provided the first demonstration of Floquet engineering of the electronic band structure of black phosphorus. Figure\,\ref{MapNew_FloquetBlack} displays the hybridization of the valence band with the first Floquet replica of the conduction band as the function of the mid-IR photon energy. By varying the photon energy of the driving field, the Floquet-driven hybridization gap moves across the valence band, effectively tuning the position in energy of the light-induced electronic band gap [see the scheme of panel (c) -- note that these data also suggest a possible band inversion between the valence band and conduction band edges by increasing the pump photon energy].

Overall, the recent works of \onlinecite{ito2023build} and \onlinecite{zhou2023pseudospin} establish Floquet engineering as a promising approach to control the electronic properties of quantum materials, even when non-negligible dissipation processes are present.

\subsection{Summary and Outlook}
TR-ARPES has undoubtedly become the preferred technique for mapping the momentum dispersion of unoccupied states, such as the conduction band of semiconductors or the topological surface state of 3D topological insulators, as well as for observing the emergence of electronic states with no equilibrium counterpart, such as excitons in TMDs and TIs or Floquet-Bloch states. Although we have not comprehensively reviewed all the many different systems in which TR-ARPES offers direct access to unoccupied/transient states [\textit{e.g.,} Dirac semimetals \cite{gatti2020light,bao2022population} and transition metal pentatellurides \cite{zhu2022comprehensive,manzoni2015ultrafast}, as well as graphene \cite{gierz2013snapshots} and charge-ordered systems \cite{nicholson2018beyond,hellmann2012time}], it is worth noting that the process of mapping unoccupied/transient states often also characterizes the intrinsic thermalization timescales of electrons photo-injected into these states. In this regard, although insufficient energy resolution can sometimes limit the observation of unoccupied states, electron dynamics can nonetheless provide indirect evidence on the dispersion/gapping of the unoccupied states [see for instance \onlinecite{crepaldi2017enhanced}].

In addition, in Sec.\,\ref{TMD_map} we briefly discussed light-induced bandgap renormalizations in TMDs. We remark that other types of ultrafast modifications of electronic gaps are broadly found in TR-ARPES investigations of quantum materials, and they are often related to the transient modification of the screening following changes in electron correlations (\emph{i.e} U/W ratio, where U is the on-site interaction and W the electronic bandwidth) \cite{tancogne2018ultrafast,gatti2020light,nilforoushan2020photoinduced,beaulieu2021ultrafast,bao2023distinguishing}.

In all the systems discussed in this section electron scattering processes are slow enough to allow mapping unoccupied/transient states. When strong electron correlations are present (and thermalization times are considerably faster than the temporal resolution), accessing unoccupied states could be challenging. However, information on the unoccupied states and their dispersion can be inferred by tracking light-induced renormalizations of the electronic dispersion, as demonstrated in Bi-based cuprate superconductors where transient shifts of the chemical potential have been discussed in terms of the particle-hole asymmetry of the pseudogap \cite{miller2017particle} and transient shifts of the Fermi momentum have been attributed to the presence of small nodal pockets \cite{freutel2019optical}.

Over the next few years, we foresee improved capabilities to map unoccupied/transient states via TR-ARPES owing to technical advances in the detection of photoelectrons (\textit{i.e.}, new detectors with lower dark counts and higher dynamical range), as well as the development of ultra-stable XUV/UV and pump pulses.
Furthermore, recent advances in time-resolved momentum microscopes, which record 3D datasets $I(\text{k}_x,\text{k}_y,\omega)$ as a function of the pump-probe delay $\tau$ without the need to scan across the sample, have opened a pathway towards mapping the occupied/unoccupied electronic band structure of solids over a large momentum region and with spatial resolution of the order of a few $\mu$m \cite{kromker2008development,schonhense2015space,tusche2016multi,xian2020open,jansen2020efficient,keunecke2020time,neef2023orbital}. These technical advances may allow investigating Floquet engineering on Moir\'{e} patterned systems, an approach that promises exquisite control on the electronic and topological properties of matter \cite{topp2019topological,topp2021light,rodriguez2021low}. 

In addition to mapping of electronic dispersion, we also expect to see an increase in the number of studies applying state-of-the-art analysis of the (out-of-equilibrium) ARPES spectral features in order to disentangle population and many-body effects. Only in this way will the TR-ARPES community be able to access light-driven electronic states in a robust fashion, thus providing a pathway towards the optimized mapping/control of out-of-equilibrium phases of quantum materials.  

\begin{figure*}
\centering
\includegraphics[scale=1]{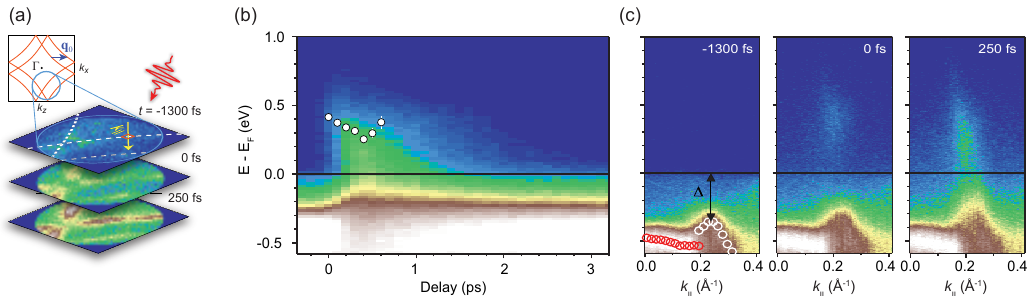}
\caption[CO1]{Ultrafast melting of the charge-order (CO) phase upon near-IR pumping in LaTe$_3$ system. 
(a) Top: tight-binding plot of the normal-state Fermi surface of LaTe$_3$ formed by the Te p-bands in the first BZ. The circle highlights the probed part of the Fermi surface and the arrow marks the CO wavevector \textbf{q$_0$}. Bottom: Fermi surface maps at different pump-probe delays across the photoinduced phase transition: the CO gap present before the pump arrival closes on a timescale of $\sim$\,100\,fs, whereas a full recovery of the long-range CO phase occurs only after $\sim$\,5.5\,ps.
(b) Detailed time evolution of the gapped region highlighted by the orange box drawn in the t \,= \,-1300\, fs slice in (a). The white dots mark the fitted positions of the upper CO band edge. 
(c) Time evolution of the band structure near the Fermi level after photoexcitation acquired along the $k_{||}$ direction highlighted in (a), yellow line. Adapted from \onlinecite{zong2019evidence}.}
\label{CO_1}
\end{figure*}

\section{Photoinduced phase transitions} \label{Phase_section}
The study of phase transitions in condensed matter is of pivotal interest for the broad scientific community not only on a fundamental level, but also as a source of future technological advances. A frequent complication encountered when attempting to model and understand phase transitions in complex materials is the involvement of several intertwined degrees of freedom, often occurring on the same energy scale. 
Moreover, with respect to second-order phase transitions, the development of long-range order may be anticipated by dynamical short-range precursors, making the identification of the energetic hierarchy challenging.
By directly accessing the time domain, time-resolved spectroscopies can unveil the dynamical nature of all degrees of freedom participating in the phase transition, thus disentangling their individual contributions \cite{giannetti2016ultrafast}. Capitalizing on its momentum- and energy- resolutions, TR-ARPES can track the dynamical behavior of specific electronic states and energy gaps and correlate them to the transient evolution of carriers or collective modes, enormously aiding the interpretation of time-resolved signals in comparison to all-optical probes. 

This section discusses a few representative cases in which TR-ARPES has offered invaluable insights into equilibrium and (dynamical) light-induced phase transitions. Section\,\ref{CO_phase} starts by reviewing TR-ARPES studies on the ultrafast melting of long-range charge order in tri-tellurides and transition metal dichalcogenides. Section\,\ref{MIT_phase} presents experimental evidence of metal-to-insulator transitions on sub-picosecond timescales, followed by a discussion of how light excitations may melt or enhance excitonic condensates in Section\,\ref{EI_phase}. Finally, Section\,\ref{SC_phase} concludes by addressing the ultrafast response of the superconducting condensate, with a specific focus on the dynamical evolution of the superconducting gap of Bi-based cuprate high-temperature superconductors. 

\begin{figure*}
\centering
\includegraphics[scale=1]{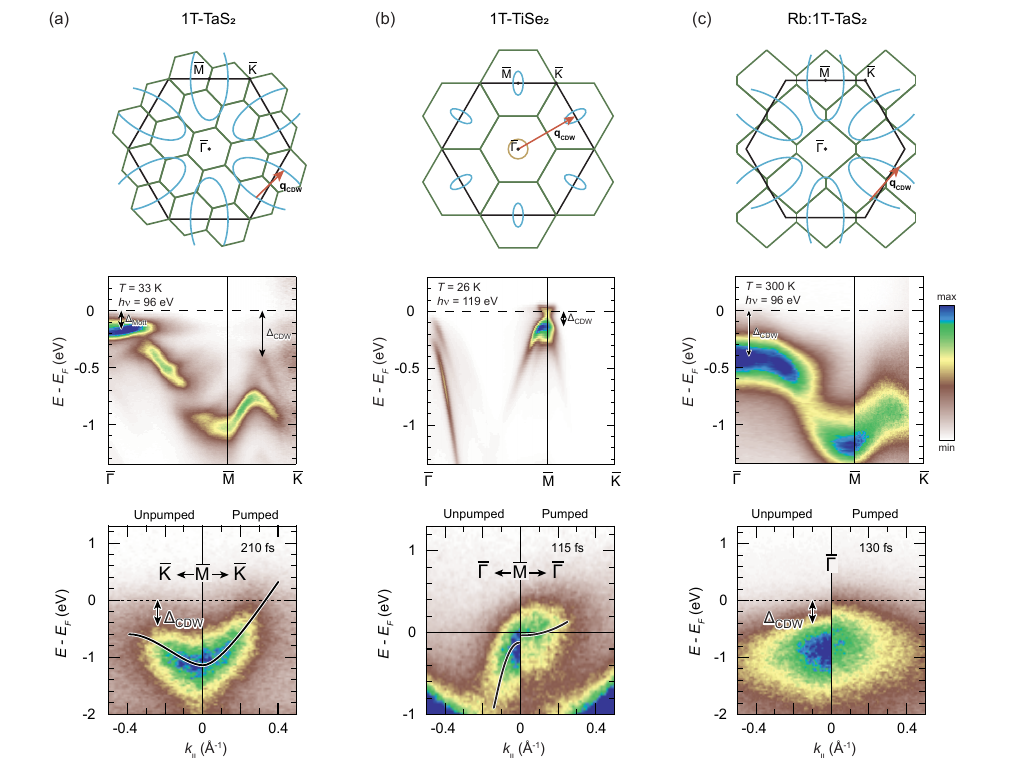}
\caption[CO2]{TR-ARPES spectroscopic signatures of the light-induced melting of the CO in various layered transition-metal dichalcogenides.
Top: Illustration of the reconstructed (green lines) and original (black) projected Brillouin zones of three CO phases in 1T-TMDs: (a) TaS$_2$, (b) 1T-TiSe$_2$, and (c) Rb:1T-TaS$_2$. Original schematic Fermi surfaces (circle and ellipses) are shown, along with CO wavevectors (arrows). Middle: Static ARPES dispersion of the three compounds, highlighting the different characteristic energy gaps in the Brillouin zone. Bottom: Unpumped/Pumped TR-ARPES spectra acquired at (a) the $\overline{\mathrm{M}}$ point of 1T-TaS$_2$, (b) the $\overline{\mathrm{M}}$ point of 1T-TiSe$_2$, and (c) the $\overline{\Gamma}$ point of Rb:1T-TaS$_2$. The absorbed pump energy density and temperature were set to 300\,J\,cm$^{-3}$ and 110\,K, respectively. 
Adapted from \onlinecite{hellmann2012time}.}
\label{CO_2}
\end{figure*}

\subsection{Melting of charge order} \label{CO_phase}
Charge order (CO), or alternatively charge density waves (CDW), are phases of matter in which the charge density self-reorganizes with a spatial periodicity different from that of the underlying lattice (either commensurate or incommensurate). CO/CDW is a recurrent feature of numerous quantum materials and it may originate from a variety of microscopic mechanisms, ranging from Fermi surface nesting (\textit{i.e.} Peierls instability or, generally, mediated by electron-phonon) to strong electron-electron interactions \cite{comin2016resonant}. Although the terms CO/CDW typically refer to slightly different phenomenology (weak/strong involvement of the lattice, respectively), the following discussion refers to the whole phenomenology as CO.
Common manifestations of CO include softening of phonons at the CO wavevector (\textit{i.e.} Kohn anomaly) and, in the case of long-range CO, folding of electronic bands and opening of a gap at the crossing points between the original and folded bands \cite{zhu2015classification}.
The amplitude of such a spectroscopic gap reflects the interaction strength underlying the CO itself. However, given the inevitable intertwinning between lattice and electronic degrees of freedom in solids, there has historically been a question of causality as to whether electron-phonon or electron-electron interactions are the main driving force of charge instabilities. 

TR-ARPES has been instrumental in addressing this question in CO systems by tracking the transient evolution of the in-gap spectral weight and the appearance/disappearance of CO-folded bands, as well as by extracting valuable dynamical information of collective phenomena, such as the CO amplitude mode. 
Rare-earth tri-tellurides ($R$Te$_3$, with $R$ indicating a rare-earth element) are a model system for investigating the CO phenomenology driven by Fermi surface nesting. Figure\,\ref{CO_1} offers an emblematic example of TR-ARPES data on LaTe$_3$ \cite{zong2019evidence}, where the Fermi surface is folded by the CO wavevector \textbf{q$_0$}.
A near-IR pump excitation of tri-tellurides results in two main observations \cite{schmitt2008transient,schmitt2011ultrafast,zong2019dynamical,zong2019evidence, maklar2021nonequilibrium, maklar2022Coherent,rettig2016persistent}: (i) coherent oscillations of electronic bands, and (ii) the suppression/filling of the CO gap (Fig.\,\ref{CO_1}c). 
In particular, while coherent oscillations of the electronic band structure may reflect specific phonon modes launched by impulsive/displacive resonant optical excitation (see Section\,\ref{CoherentPhon}), the pump-induced 2.3\,THz mode is observed solely in momentum regions in proximity of the CO gap and  disappears at high excitation fluences or above the CO onset temperature (T$_\text{CO}$). As a result of these observations, \onlinecite{schmitt2008transient} has interpreted the 2.3\,THz mode as the CO amplitude mode.
As for the dynamical evolution of the in-gap spectral weight, the ultrafast melting of the CO phase does not occur instantaneously upon pump excitation, but instead it develops on a timescale of $\sim$100\,fs, indicating the key role of the lattice in stabilizing the CO phase. Moreover, while the gap amplitude recovers within 1\,ps, the long-range coherence of the CO phase is re-established only after $\sim$5.5\,ps. \onlinecite{zong2019evidence} proposes that this long timescale arises as a consequence of the optical generation of topological defects in the CO-ordered phase.  

TMDs, such as 1T-TaS$_2$ and 1T-TiSe$_2$, are another class of CO materials that have been investigated extensively via TR-ARPES \cite{rohwer2011collapse,hellmann2012time,majchrzak2021switching,monney2016revealing,mathias2016self,zhang2022creation,sayers2023exploring}. Strong electron interactions in 1T-TMDs may prompt the emergence of Mott localization and excitonic insulator physics, challenging the study of the microscopic origin of the CO phase. The contribution of TR-ARPES is to look for dynamical signatures of the CO phase in specific momentum regions, distinguishing the different energetic contributions \cite{sohrt2014fast}. This role is clearly exemplified in the study of the CO in 1T-TiSe$_2$. While the ultrafast evolution of the spectral weight at the zone center $\Gamma$ does not directly characterize the CO phase due to the interplay of other degrees of freedom \cite{hellmann2012time,hedayat2019excitonic,perfetti2006time,perfetti2008femtosecond}, the transient disappearance of the CO-folded bands at the M point instead serves a clear signature of the light-induced melting of the CO phase \cite{rohwer2011collapse,hellmann2012time,rohde2014does,monney2016revealing,petersen2011clocking,mathias2016self,huber2022mapping}. 
\onlinecite{hellmann2012time} explored the transient response of the CO in several TMDs, as shown in Fig.\,\ref{CO_2}. Pristine 1T-TaS$_2$ shows a sub-50\,fs melting of the Mott gap at $\Gamma$ and a $\sim$225\,fs response of the CO gap at the M point, with the subsequent observation of a CO amplitude mode. The CO gap melts in approximately half the period of the CO amplitude mode, supporting a structural origin for the CO in 1T-TaS$_2$, although the presence of purely electronic contributions has also been discussed \cite{hellmann2010ultrafast,ishizaka2011femtosecond,petersen2011clocking}.

Long-range CO has also been observed in \textit{in situ} Rb intercalated 1T-TaS$_2$ (Rb:1T-TaS$_2$), with a large energy gap of 400\,meV around the BZ center. Although both the amplitude and the period of the oscillations are smaller than for the pristine case, the temporal evolution of the order parameter in Rb:1T-TaS$_2$ exhibits the same fingerprint of Peierls physics with a gap melting time of 125\,fs, corresponding to half an oscillation cycle of the amplitude mode.
In contrast, optical melting of CO in 1T-TiSe$_2$ takes place on a timescale of 30-80\,fs depending on the excitation fluence \cite{rohwer2011collapse,mathias2016self}. This time constant is too fast for a structural transition, and at the same time, slower than the conventional electron-hopping timescale \cite{hellmann2012time}. The relation between the melting of the CO gap vs. the density of transient free carriers is incompatible with a pure Peierls scenario, and it has been discussed in terms of a transient enhancement of the screening \cite{rohwer2011collapse,rohde2014does,monney2016revealing}, which is in agreement with the presence of an underlying excitonic condensate (see Section\,\ref{EI_phase} -- although one should notice that these fast electron dynamics might be related to the metallic nature of 1T-TiSe$_2$ even at low temperature). On the other hand, a recent multitechnique study combining TR-ARPES and time-resolved reflectivity has shown that electron-phonon coupling contributes to the stabilization of the CO phase and the exitonic condensate in 1T-TiSe$_2$ \cite{hedayat2019excitonic}.

\begin{figure*}
\centering
\includegraphics[scale=1]{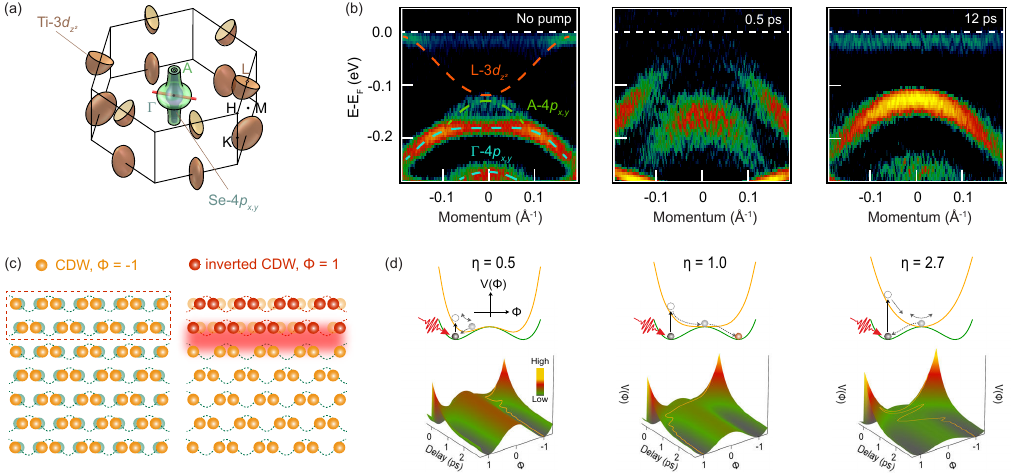}
\caption[CO3]{Probing a light-induced 2D metastable electronic state at the surface of 1T-TiSe$_2$.
(a) 3D BZ of 1T-TiSe$_2$. (b) Second-derivative ARPES spectra of the time-dependent electronic structure of 1T-TiSe$_2$ acquired along the $\Gamma$-M direction at 4\,K [red line in (a)]. Dashed blue, green and red lines mark the dispersion of the Se 4$p_{x,y}$ band at $\Gamma$, the Se 4$p_{x,y}$ band folded from point A and the Ti 3$d_{z^2}$ band folded from point L, respectively. A clear ultrafast modification of the electronic dispersion is observed, with a recovery dynamics of tens of ps. (c) Schematic of the CO inversion mechanism in real space: (left) the original 3D CO lattice has an order parameter $\phi$\,=\,$-$1; (right) upon photon excitation, phase inversion occurs at the surface leading to an order parameter $\phi$\,=\,+1. As a result, quasi-2D electronic states form between the inverted and original CO layers (red shaded area). Panels (a)-(c) adapted from \onlinecite{duan2021optical}. (d) Time-dependent Ginzburg-Landau model as a function of the order parameter for three selected pump fluences: (top row) schematics of the energy potential before and after photoexcitation and (bottom) calculated transient energy potential. Adapted from \onlinecite{duanPRL2023}.}
\label{CO_3}
\end{figure*}

TR-ARPES has both offered invaluable insights into the dynamical properties of CO in various systems and also established how light-matter interaction can drive transient/metastable states unachievable in equilibrium conditions \cite{shi2019ultrafast,duan2021optical, maklar2021nonequilibrium,zhang2022creation,zong2019evidence}. As a paradigmatic example, TR-ARPES data corroborated by MeV ultrafast electron diffraction measurements have revealed a transient metastable 2D-ordered electronic state at the surface of 1T-TiSe$_2$ \cite{duan2021optical}. As shown in Fig.\,\ref{CO_3}(b), careful analysis of the transient evolution of the spectral features has identified an ultrafast modification of the electronic dispersion at the $\Gamma$ point upon near-IR excitation [Fig.\,\ref{CO_3}(a) displays the 3D BZ], as well as a counter-intuitive sharpening of the ARPES lineshape for certain excitation fluences. 
The transient modification of the electronic dispersion and sharpening of the ARPES spectral features have been interpreted as evidence of the optically-induced inversion of the 3D CO phase at the surface of 1T-TiSe$_2$ [Fig.\,\ref{CO_3}(c)], and consequent formation of long-range 2D domain walls, realizing a new metastable phase of matter \cite{duan2021optical}.

Finally, \onlinecite{crepaldi2022optically} have studied the quasi-1D Weyl semimetal $(\mathrm{TaSe}_4)_2\mathrm{I}$, in which the emergence of CO is thought to open electronic gaps at the Weyl points leading to the formation of an axionic insulator phase. In this paper, the authors took advantage of the light-induced filling of the CO gap to restore the Weyl phase in $(\mathrm{TaSe}_4)_2\mathrm{I}$ and explore the dynamical crossover between the Weyl semimetal state and the postulated axionic insulator phase.

\subsection{Metal-to-Insulator transitions} \label{MIT_phase}
The Mott insulating phase is one of the fundamental phases that can give rise to strongly correlated materials: it emerges when the Coulomb repulsion between electrons overcomes their kinetic energy, leading to the localization of charge carriers and the development of a Mott band gap between the lower and upper Hubbard band (LHB and UHB, respectively). However, as illustrated in the previous Section\,\ref{CO_phase}, electron-electron interactions may also prompt the emergence of other quantum phases of matter, such as charge order. Therefore, although the Mott insulating phase is driven purely by electron-electron interactions, the transition into or out of such a state may be accompanied by a charge or structural re-organization, thus calling for investigative tools capable of distinguishing the different contributions (competing or coexisting) to the phase transition. 
While photoinduced Metal-to-Insulator transitions (MITs) have been reported by optical and transport studies \cite{morrison2014photoinduced,rini2007control}, TR-ARPES provides a direct probe of the temporal modifications of the electronic structure in momentum space upon the breakdown of the insulating phase, and can both track the transient renormalization and collapse of the electronic Mott gap and also disentangle the different degrees of freedom involved in the transition in the time domain.

\begin{figure}
\centering
\includegraphics[scale=1]{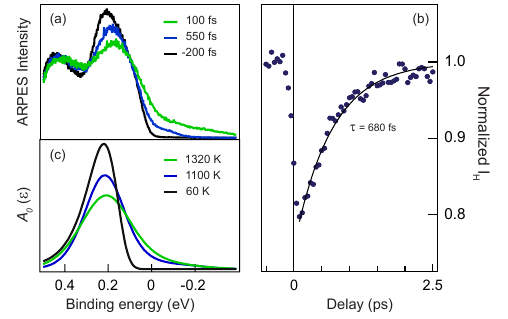}
\caption[MIT1]{Ultrafast Mott gap renormalization in 1T-TaS$_2$ upon near-IR optical excitation. (a) $\Gamma$-point EDCs at different pump-probe delays measured at 30\,K base temperature, highlighting the pump-induced transfer of spectral weight from the lower Hubbard band (LHB) to the previously gapped region. (b) Dynamic response of the Mott phase as outlined by the transient LHB intensity peak normalized to the equilibrium value. The almost-instantaneous collapse is followed by a sub-ps recovery. (c) Simulated spectral function of the curves in (a) calculated via a single-band Hubbard model in which the electronic temperature T$_e$ at different delays is the dominant term driving the phase transition. From \onlinecite{perfetti2006time}.}
\label{MIT_1}
\end{figure}

TMDs have been extensively investigated by TR-ARPES to explore their MIT \cite{perfetti2006time,perfetti2008femtosecond,hellmann2012time,ligges2018ultrafast,biswas2021ultrafast} and other ultrafast techniques, with the exemplary report of a light-induced metastable hidden phase in 1T-TaS$_2$ \cite{stojchevska2014ultrafast,stahl2020collapse}. Early TR-ARPES studies of 1T-TaS$_2$ showed that an ultrashort near-IR pump excitation leads to a nearly-instantaneous melting of the Mott insulating state at the $\Gamma$ point, as spectral weight is transferred from the LHB to the originally gapped region at the Fermi level [Fig.\,\ref{MIT_1}(a)-(c)]. 
The initial fully insulating state is then restored on a sub-picosecond timescale, as shown by the temporal evolution of the LHB peak intensity in Fig.\,\ref{MIT_1}(b). In addition to this substantial transfer of spectral weight, coherently excited phonons corresponding to the CO amplitude mode have also been detected modulating the electrons' binding energy in the TR-ARPES spectra (see also Sections\,\ref{CO_phase} and \ref{CoherentPhon}). The different relaxation time and pump fluence dependence allows dynamics of the electronic and lattice subsystems to be disentangled: while the photoinduced ionic displacement increases linearly with the employed pump fluence, the change in the LHB intensity does not, and the periodic oscillations persist for $>$\,20\,ps without affecting the Mott insulating state recovered within 1\,ps \cite{perfetti2006time,perfetti2008femtosecond}. 

\begin{figure}[b]
\centering
\includegraphics[scale=1]{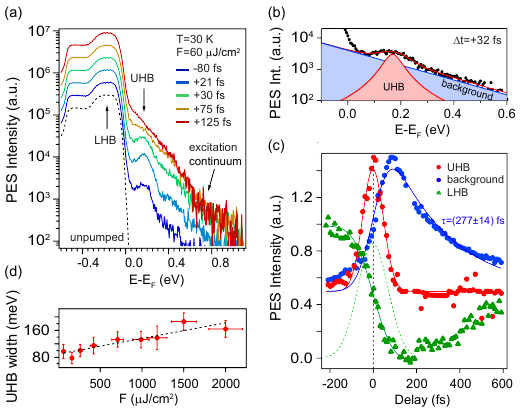}
\caption[MIT1]{Light-induced doublon dynamics in 1T-TaS$_2$. (a) Transient EDCs acquired at normal emission for selected pump-probe delay, showing the ultrafast population of the UHB. (b) Exemplary fit of a EDC above E$_F$ consisting of a linear background and a Lorentzian peak corresponding to the UHB. (c) Comparison between the ultrafast evolution of the UHB and the slower dynamics of LHB and underlying background, supporting an electronic origin of the UHB dynamics. (d) Fluence dependence of the UHB peak width. From \onlinecite{ligges2018ultrafast}.}
\label{MIT_3}
\end{figure}

These results indicate the primary role played by the electronic degrees of freedom in the photoinduced ultrafast MIT in 1T-TaS$_2$. This conclusion agrees with the report by \onlinecite{ligges2018ultrafast}, which investigated the recombination dynamics of doublons (\textit{i.e.}, doubly occupied sites). In particular, the authors demonstrated an instantaneous light-induced population of the UHB, which subsequently disappears in less than 100\,fs, as shown in Fig.\,\ref{MIT_3}(a) and (c). The doublon-hole recombination processes are estimated to occur on an analogous timescale to that of electron hopping $\hbar/J$\,$\sim$\,14\,fs, much faster than the half-period of the highest-frequency phonons in 1T-TaS$_2$ of 11.9\,THz. This finding, supported by non-equilibrium DMFT calculations, suggests that the observed doublon dynamics is purely electronic in origin, thus excluding any CO-related lattice contributions \cite{ligges2018ultrafast}.

With respect to the coexistence of Mott and CO physics in 1T-TaS$_2$, \onlinecite{wang2020band} showed that the commensurate CO phase features interlayer stacking with dimerization (\textit{i.e.}, doubling of the unit cell along the c-axis). To investigate the contributions of the out-of-plane dimerization and electronic correlations to the MIT, \onlinecite{dong2022stacking} tracked the ultrafast gap filling at $\Gamma$ of 1T-TaS$_2$ in the commensurate CO phase with different probe polarizations. Corroborated by \textit{ab-initio} calculations, they reported evidence of both light-induced melting of the CO long-range order (\textit{i.e.}, structural contribution), as well as the appearance of  mid-gap states on a timescale faster than half a period of the CO amplitude mode [a hint of electron correlations \cite{perfetti2006time,ligges2018ultrafast}]. Another TR-ARPES study tracked light-induced dynamics in an intermediate phase of 1T-TaS$_2$ that can be attained upon heating the sample from the commensurate CO to the triclinic CO phase. In this intermediate phase, the authors reported evidence of Mott physics, while the commensurate CO phase with layer dimerization behaves like a band insulator \cite{bao2023distinguishing}.
Moreover, recent studies on a similar but less studied compound, (1T-TaSe$_2$), have mapped how coherent oscillations of the CO amplitude mode ($\sim$\,2.2\,THz) modulate the energy position of the valence band at $\Gamma$, thus suggesting a direct link between the CO amplitude mode and the electronic gap in the proximity of the Fermi level \cite{sayers2020coherent}. Further exploration of ultrafast charge dynamics points towards a charge-transfer gap rather than a Mott gap at $\Gamma$ \cite{sayers2023exploring}.

A quasi-instantaneous closure of the band gap upon optical pumping has also been observed in VO$_2$, a prototypical system exhibiting a transition from a monoclinic insulating phase to a rutile metallic phase at 340\,K \cite{morrison2014photoinduced}. Interestingly, while at equilibrium this MIT transition involves a fundamental crystallographic modification, \onlinecite{wegkamp2014instantaneous} reported a transient metallic state reached by near-IR excitation that retains the monoclinic structure characteristic of the insulating state, resulting in a photoinduced state with no thermally-driven counterpart. The observed ultrafast gap renormalization in VO$_2$ is then described in terms of enhanced screening upon photoexcitation from the V valence bands, which precedes significant charge carrier relaxation (occurring within 200\,fs) and atomic rearrangement \cite{wegkamp2014instantaneous, wegkamp2015ultrafast}. A study of the sibling compound V$_2$O$_3$ investigated the non-thermal phases that develop immediately after photoexcitation not only for the paramagnetic insulator phase but also for the correlated metal phase of this prototype Mott-Hubbard compound. This TR-ARPES study showed that, in both cases, the out-of-equilibrium phases are electronic in nature and driven by transient orbital overpopulation \cite{lantz2017ultrafast}.

\begin{figure}
\centering
\includegraphics[scale=1]{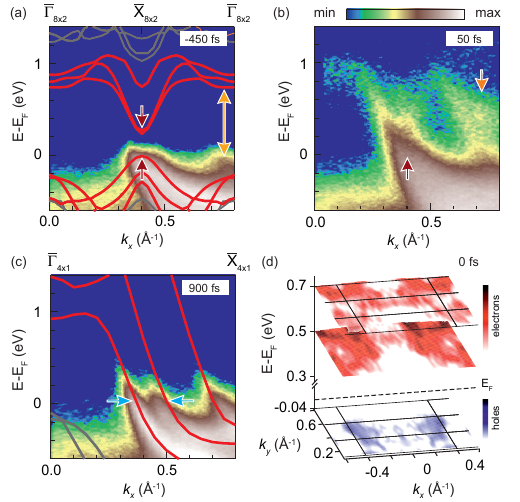}
\caption[MIT2]{Capturing the pathway of the photoinduced MIT in momentum-space, in indium nanowires on a silicon (111) surface, In/Si(111). (a)-(c) TR-ARPES spectra of In/Si(111) at 25\,K for selected time delays highlighting the transition from an insulating (8 $\times$ 2) to metallic (4 $\times$ 1) phase. Solid lines in (a)-(c) are the calculated electronic band structure in the (8 $\times$ 2) and (4 $\times$ 1) phases, respectively. The pump-induced transition occurs in three stages: closing of the gap at X$_{\text{(8 $\times$ 2)}}$ [red arrows]; downward shift of the CB at $\Gamma_{\text{(8 $\times$ 2)}}$ [yellow arrows]; separation in momentum of the now-metallic band [blue arrows] to reach the final (4 $\times$ 1) structure. (d) Experimental momentum-space distribution of excited electrons (red) and holes (blue) at the time of the pump excitation. The map is obtained by subtracting the photoemission signal measured at -1000\,fs from that measured at 0\,ps. Adapted from \onlinecite{nicholson2018beyond}.}
\label{MIT_2}
\end{figure}

The capability to unravel the full pathway of photoinduced MITs in detail is showcased in recent TR-ARPES studies of indium nanowires on a silicon (111) surface, In/Si(111) \cite{nicholson2018beyond,chavez2018band,chavez2019charge,nicholson2019excited}. This system undergoes a transition from a low-temperature CO-mediated insulating phase to a metallic state at 130\,K. While the former phase is defined by a (8 $\times$ 2) symmetry of the lattice characterized by distorted In-hexagons, the latter exhibits a (4 $\times$ 1) symmetry with arrays of In zig-zag chains leading to the emergence of three bands crossing E$_\text{F}$. When the system is irradiated in the insulating state with a near-IR pump excitation, TR-ARPES data reveal a MIT developing in three distinct stages [Fig.\,\ref{MIT_2}(a)-(c)]: first, the gap at the (8 $\times$ 2) zone boundaries closes within 200\,fs (red arrows); next, the conduction bands at $\Gamma_{\text{(8 $\times$ 2)}}$ gradually shift down in energy and cross E$_F$ after 500\,fs [yellow arrows -- due to structural transition, in agreement with time-resolved electron diffraction data \cite{frigge2017optically}]; finally, the now-metallic bands draw apart in momentum along $k_x$ (blue arrows), marking the final structural transition into the (4 $\times$ 1) metallic phase.
The map of the photoexcited carriers in momentum space shown in Fig.\,\ref{MIT_2}(d) highlights the strongly delocalized character of the photoexcited electrons across the BZ, while the photoholes are localized at the zone boundaries. Starting with the photoinduced electron and hole distributions across the BZ as input, ab-initio molecular dynamics calculations well reproduce the three stages of the MIT observed experimentally and provide insights about the dynamics of nuclei and chemical bonds. In fact, this procedure reveals the ultrafast formation of the delocalized metallic bond along the zig-zag chains on the same timescale as the gap closing. 

The work by \citeauthor{nicholson2018beyond} exemplifies how TR-ARPES can be combined with theory to achieve a comprehensive understanding of photoinduced MITs, including both momentum- and real-space descriptions. 
In addition, \onlinecite{chavez2019charge} have investigated the MIT in In/Si(111) upon sub-CO-gap excitation. They showed that mid-IR excitation with a peak electric field up to 0.9\,MV/cm ($h\nu_{\text{pump}}$\,=\,190\,meV\,$\lesssim$\,300\,meV) drives an MIT via multiphoton absorption on a timescale comparable to the one observed with near-IR excitation, thus hinting at a similar microscopic mechanism.

Finally, \onlinecite{gierster2021ultrafast} have recently reported a photoinduced ultrafast semiconductor-to-metal transition at the surface of ZnO at very low excitation fluences ($\sim$10\,$\mu$J/cm$^2$) in comparison to the other systems described in this section. This ultrafast transition is driven by the concomitant depletion of in-gap states and the emergence of a surface photovoltage (see Section\,\ref{SPV_section}) that leads to partially filling states at the Fermi level.

\subsection{Excitonic insulators} \label{EI_phase}
In semimetals or narrow-gap semiconductors, when conduction electrons and valence holes form excitons, bound states whose energy exceeds the small band gap, a macroscopic condensation may occur leading to the emergence of a novel excitonic insulating (EI) phase \cite{jerome1967excitonic}. Among the materials proposed to exhibit an EI ground state are 1T-TiSe$_2$ and Ta$_2$NiSe$_5$. Both of them have been recently investigated by TR-ARPES not only to provide supporting evidence of the excitonic nature of their insulating state, but also to pursue optical control of the electron-hole pairs for future applications.

1T-TiSe$_2$ is characterized by an indirect band gap and it undergoes a CO transition below 200\,K. The study of its spectral response in the time-domain \cite{rohwer2011collapse,hellmann2012time,monney2016revealing} revealed a characteristic sub-80fs gap melting time, which is much faster than the dynamics observed for the $M$-point Peierls-gap in 1T-TaS$_2$ (see also Section\,\ref{CO_phase}) and it exhibits a distinct pump fluence dependence. In particular, \onlinecite{rohwer2011collapse} have reported a direct relation between the gap response time in 1T-TiSe$_2$ and the inverse of $\sqrt{n}$ (where $n$ is the photoinduced carrier density), which corresponds to the timescale of the carriers' screening built-up in response to an ultrashort optical excitation. Both the ultrafast response time and the photo-enhanced screening hypothesis are consistent with a description of the insulating state in terms of EI physics [in agreement with momentum-resolved electron energy-loss spectroscopy results \cite{kogar2017signatures}], although the presence of the strong CO band folding prevents conclusively excluding the possibility that electron-lattice interactions may contribute to the gap formation in 1T-TiSe$_2$.

\begin{figure}
\centering
\includegraphics[scale=1]{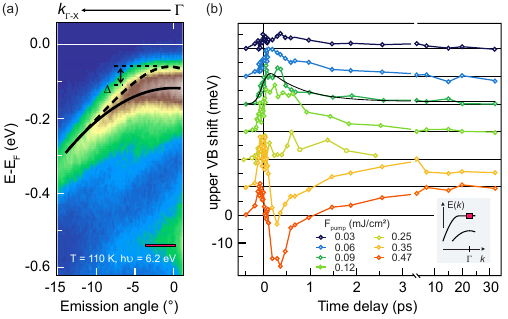}
\caption[Exc1]{Transient changes in the valence band of the excitonic insulator Ta$_2$NiSe$_5$ (TNS) upon a near-IR excitation. (a) Equilibrium ARPES spectra of TNS acquired around $\Gamma$ at 110\,K; solid (dash) black line is a guide to the eye for the VB dispersion before (after) the optical excitation. Red bar indicates the momentum integration interval for extracting EDCs to track the VB peak position. (b) Time-dependent energy shift of the upper VB peak at $\Gamma$ for different pump fluences. Two different behaviors are observed around an empirical critical fluence F$_C$\,$\sim$\,0.2\,mJ/cm$^2$: an upward shift with a 1\,ps recovery to the original binding energy at low fluences is opposed to a 200\,fs-delayed downward shift at higher fluences, which is followed by a settling into a lower energy position relative to the equilibrium case. Adapted from \onlinecite{mor2017ultrafast}.}
\label{Exc_1}
\end{figure}

In contrast with 1T-TiSe$_2$, in Ta$_2$NiSe$_5$ (TNS) the proposed excitonic insulating phase does not coexist with a CO phase, and it is defined by a direct band gap of $\sim$\,0.15\,eV at the $\Gamma$-point [Fig.\,\ref{Exc_1}(a)]. However, the characteristic flattening of the valence band below the critical temperature 326\,K is concomitant with a structural phase transition from an orthorombic to monoclinic unit cell \cite{wakisaka2012photoemission}. Whether the phase transition is excitonic or structural (\text{i.e.}, electronic or elastic) in nature remains an open question. Since the two phenomena would occur on the same energy scale and with no symmetry distinction, distinguishing them may rely on the different time dependences of the electronic and phononic instabilities. All these aspects make TNS an excellent system for investigation by low-photon-energy TR-ARPES to explore the photoinduced electronic dynamics of an EI candidate. However, despite extensive recent experimental efforts, the hierarchy of the fundamental interactions involved in the gap formation of TNS is still under debate.
Upon near-IR pump excitation, a significant loss of spectral weight occurs at the top of the upper VB in the first 100-200\,fs. The fluence dependence of the VB depletion resembles that of 1T-TiSe$_2$ (\textit{i.e.}, higher fluences lead to faster depletion) \cite{mor2017ultrafast,okazaki2018photo}, supporting the description of TNS in terms of excitonic insulator. Along with the loss of spectral weight, a photoinduced shift in binding energy of the top VB at $\Gamma$ has been reported and meticulously analyzed as a function of the pump fluence by \onlinecite{mor2017ultrafast}. As illustrated in Fig.\,\ref{Exc_1}(b), two different behaviours are observed around an empirically observed critical fluence F$_C\sim$0.2\,mJ/cm$^2$: at low fluences the VB shifts toward the Fermi level and recovers its original position within 1\,ps, whereas above F$_C$ a deviation to higher energy is detected at $\sim$\,200\,fs after the pump excitation and its relaxation process overshoot the initial equilibrium position. While the upwards shift (gap closure) is understood in terms of a gap renormalization stemming from the transiently enhanced screening (as expected in a semiconductor), the downward shift in energy (gap enhancement) has been interpreted as a signature of an increase of the excitonic condensate density. These results suggest the possibility of controlling the gap parameter in TNS by tuning the fluence of the optical pump. 

\begin{figure}
\centering
\includegraphics[scale=1]{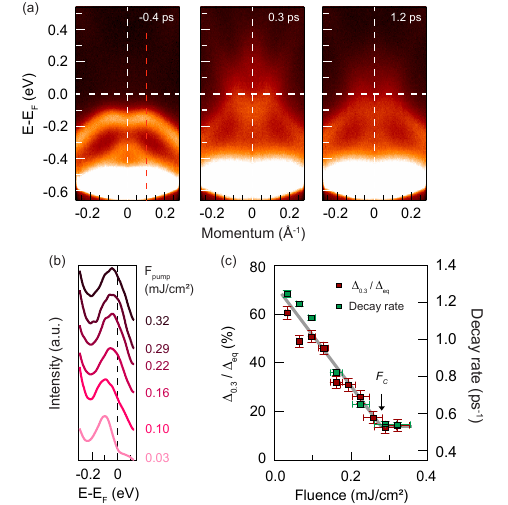}
\caption[Exc2]{Photo-melting of the excitonic gap in TNS. (a) TR-ARPES spectra of TNS acquired along the Ta and Ni chains at a base temperature of 30\,K show the ultrafast emergence of a CB electron-like pocket at $\Gamma$ upon a 1.77\,eV excitation. (b) EDCs extracted at the top of the VB [dashed red line in panel (a)] at time delay 0.3\,ps as a function of the pump fluence. (c) Fluence dependence of the photoinduced change of the energy gap extracted from the EDCs in (b) [red squares]. While a linear decrease in the gap size is reported at low fluences, a saturation point is reached at F$_C$\,=\,0.29\,mJ/cm$^2$, resulting in a residual gap even at high fluences. The same fluence dependence is observed also for the decay rate of the nonequilibrium electrons obtained in a 0.1\,eV window above E$_F$ (green squares). Adapted from \onlinecite{tang2020non}.}
\label{Exc_2}
\end{figure}

Other TR-ARPES studies of TNS have instead reported a transition from an insulating to a semimetallic state upon near-IR excitation \cite{okazaki2018photo,tang2020non,Saha2021photoinduced}, as well as transient electron dynamics in the CB \cite{mor2022ultrafast}.
In particular, a CB electron-like pocket is observed emerging at $\Gamma$ on an ultrafast timescale, as illustrated in Fig.\,\ref{Exc_2}(a). By analyzing the EDCs at the top of the VB as a function of the pump fluence, \onlinecite{tang2020non} showed that the drastic reduction of the gap amplitude upon increasing fluence reaches a saturation at F$_C$=0.29\,mJ/cm$^2$, above which a residual gap of $\sim$\,15\,$\%$ of the initial value persists [see Fig.\,\ref{Exc_2}(b)-(c)], possibly due to a forbidden band-crossing between the CB and VB. The decay rate of the nonequilibrium quasiparticles above E$_F$ exhibits the same fluence dependence. These findings demonstrate that ultrafast near-IR excitations suddenly and significantly alter the electronic distribution of TNS, though the direct band gap remains finite at all times regardless of the high electronic temperature attained. 

\begin{figure}[b]
\centering
\includegraphics[scale=1]{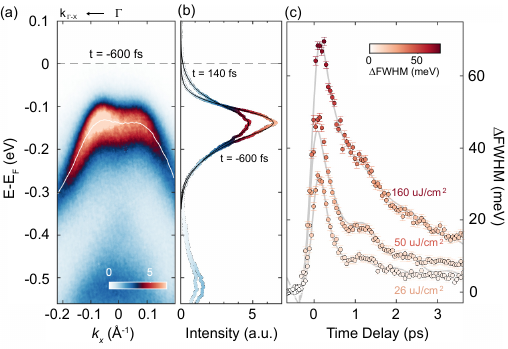}
\caption[Exc1]{Establishing the light-induced broadening of the VB lineshape as an indicator of the strength of electron-electron interactions in TNS. (a) TR-ARPES spectrum acquired along the $\Gamma$-X direction of Ta$_2$NiSe$_5$ before photoexcitation. (b) EDCs centered at $\Gamma$ before and after the optical excitation, showing a pump-induced broadening of the peak. (c) Transient variation of the spectral broadening of the VB at $\Gamma$ for three different excitation fluences. Adapted from \onlinecite{golevz2022unveiling}.}
\label{Exc_1_Sydney}
\end{figure}

In order to assess the phononic contribution to the observed gap renormalization, the pump-induced coherent oscillations of the upper VB binding energy have been analyzed in recent investigations by \onlinecite{tang2020non} and \onlinecite{baldini2023spontaneous}. Several coherent modes of A$_{\mathrm{1g}}$ and B$_{\mathrm{1g}}$ symmetry have been identified in the TR-ARPES spectra; the frequencies of these modes in the 1-5\,THz range are in agreement with previous Raman experiments on TNS \cite{werdehausen2018coherent}. Despite probing the same set of coherent modes, the two studies reach different conclusions in regard to the nature of the instability in TNS. On the one hand, \onlinecite{tang2020non} argue that the observed ultrafast phase transition is purely electronic in origin and occurs without a structural transition, owing to the pump-fluence independence of the observed coherent modes. This interpretation suggests that the system retains its low-temperature monoclinic structure across the transition, even above F$_C$. On the other hand, the combined TR-ARPES and ultrafast electron diffraction study of \onlinecite{baldini2023spontaneous} reported (i) a slow ($\sim$0.3-0.4\,ps) and a long-lasting response of the VB depletion to the optical pumping, (ii) the detection of coherent oscillations in the upper VB even during the pump-probe signal rise time, and (iii) the sudden drop followed by a slower decrease over several ps of the diffraction peaks corresponding to the monoclinic structure. Based on this evidence, the authors have excluded a dominant role for the electronic degree of freedom in the gap formation in TNS, and attribute it instead to the symmetry lowering which accompanies the structural transition.

Further contributing to the debate on the relative roles of electron-electron and electron-phonon interactions in TNS, \onlinecite{golevz2022unveiling} recently reported a significant photoinduced broadening of the valence band of TNS, while changes of the bandgap amplitude are an order of magnitude smaller. These experimental observations are summarized by the EDCs of Fig.\,\ref{Exc_1_Sydney}(b), as well as the transient evolution of the VB linewidth for three different excitation fluences in Fig.\,\ref{Exc_1_Sydney}(c).
Indeed, the response of TNS to an external optical perturbation (with $E$ and $A$ representing the electric field and vector potential, respectively) can be qualitatively modelled by two-band (CB and VB) spinless fermions in 1D coupled to dispersionless phonons of energy $\Omega_0$:
\begin{equation}
\begin{aligned}
    H=\sum_{k,\alpha \in (0,1)}[\epsilon_{k-A}]^{\alpha\alpha'}c^{\dagger}_{k,\alpha}c_{k,\alpha'}+V\sum_{i}n_{i,0}n_{i,1}+\\
    \sum_{i}[\sqrt{\lambda}X_i-E(t)]c^{\dagger}_{k,0}c_{k,1}+\sum_{i}\frac{1}{2}[X_i^2+\frac{1}{\Omega_0^2}\dot{X}_i^2]+\text{h.c.,}
\end{aligned}
\end{equation}
where $c^{\dagger}_{k,\alpha}$ is the electron creation operator, $V$ is the Coulomb interaction between the CB and VB, $\lambda$ is the electron-phonon interaction strength corresponding to a displacement $X_i$.
The ratio between $V$ and $\lambda$ determines whether an optical excitation leads to mainly a spectral feature broadening ($V$-driven) of gap closure ($\lambda$-driven). By comparing the fluence- and momentum-dependence of the ARPES linewidth with nonequilibrium many-body simulations, the authors demonstrated the dominant (but not necessarily sole) contribution of electron-electron interaction to the stabilization of the excitonic gap in TNS.
This paper further emphasizes how a detailed analysis of transient many-body interactions (encoded in the transient evolution of the spectral function, see Sec.\,\ref{quasiEq_ArpesInt}), in parallel with state-of-the-art nonequilibrium simulations, can offer an unprecedented degree of information on the microscopic origin of the order parameter in correlated systems.

\subsection{Superconducting condensate} \label{SC_phase}
\begin{figure}
\centering
\includegraphics[scale=1]{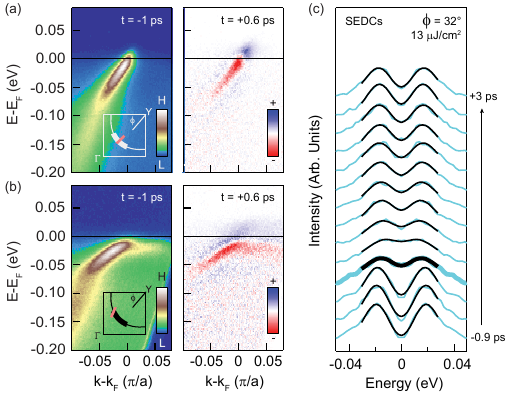}
\caption[SC1]{Tracking the transient evolution of the near-nodal superconducting gap of optimally-doped Bi$_2$Sr$_2$CaCu$_2$O$_{8+\delta}$ (Bi2212-OP91, T$_c$=91\,K). (a) TR-ARPES dispersion acquired along the gapless nodal direction [$\phi$\,=\,45$^o$] before and after near-IR optical pumping. For positive delays, the differential spectrum is shown, obtained by subtracting the ARPES map at negative delay. Insets in the left panel illustrates the location of the cut in the Fermi-Surface. (b) Same as in (a) but along the off-nodal cut [$\phi$\,=\,31$^o$], where the band bending due to the presence of the gap is clearly visible. (c) Transient evolution of the superconducting gap at $\phi$\,=\,32$^o$ mapped via symmetrized EDCs. The spectroscopic modifications observed within the originally-gapped region are described by a phenomenological model (black solid lines) that includes a single scattering term and the superconducting gap amplitude. Adapted from \onlinecite{smallwood2012tracking}.}
\label{SC_1}
\end{figure}

The normal-to-superconducting state phase transition in (quasi-)2D systems has been the subject of extensive investigations over the past four decades. Among the major classes of high-temperature superconductors, quasi-2D Bi-based copper oxides in single crystal form are particularly suitable for study by ARPES and TR-ARPES because they present a natural cleavage plane and can be doped across a wide range. 
The cuprates family is characterized by a layered perovskite structure, in which the square planar CuO$_2$ plane forms a single- or multi-layer conducting block separated by insulating charge reservoir layers. Since the Cu-O bands are the lowest-energy electronic states, the CuO$_2$ block is believed to hold a key role in the development of the macroscopic electronic properties in these materials, including high-temperature superconductivity. In particular, a band stemming from the hybridization between Cu $d_{x^2-y^2}$ and O 2$p_x$ and 2$p_y$ orbitals is observed in the vicinity of the Fermi level [see Fig.\,\ref{SC_1} (a)-(b)]. Due to the distinct $d$-wave symmetry of the superconducting order parameter in cuprates, the gap disappears along the direction diagonal to the Cu-O bond [nodal direction, panel (a)], becomes finite off-node [panel (b)], and reaches its maximum at the edge of the Brillouin zone at the antinode \cite{shen1993anomalously,damascelli2003angle}. 

Various all-optical probes have been employed over the past few decades to explore Bi-based cuprates in the time-domain \cite{giannetti2016ultrafast}. These investigations suggested that while perturbative near-IR excitations melt the superconducting condensate on sub-ps timescales in a non-thermal fashion \cite{kabanov1999quasiparticle,gedik2005abrupt}, intense mid-IR pulses may drive a transient superconducting state above the critical temperature T$_c$ \cite{liu2020pump,fausti2011light,hu2014optically}.
Only recently, the development of TR-ARPES systems with energy resolution comparable to the superconducting gap of cuprates \cite{smallwood2012ultrafast,parham2017ultrafast,cilento2016advancing} has triggered extensive studies of the transient response of the superconducting condensate in momentum space. To date, TR-ARPES studies of cuprates have almost solely relied on near-IR pumping; while this excitation range does not couple directly to the superconducting condensate, it does redistribute carriers within the charge-transfer gap, driving an ultrafast superconducting-to-normal state phase transition \cite{giannetti2016ultrafast,baldini2020electron}.

Initial TR-ARPES investigations focused on the gapless nodal direction \cite{perfetti2007ultrafast,cortes2011momentum}, where transient suppression of the coherent spectral weight at the Fermi momentum was reported and attributed to the light-induced melting of the superconducting condensate via enhancement of phase fluctuations \cite{graf2011nodal}. However, recent investigations on various Bi-based cuprates have excluded an unequivocal relation between the observed suppression of the nodal spectral weight and superconductivity, and instead discussed the phenomenon as a direct manifestation of the energy and temperature dependence of the single-particle lifetime within the Fermi Liquid formalism \cite{zonno2021ubiquitous,zonno2021time}.

A major milestone in the study of the physics of cuprates via TR-ARPES came in 2012, when \onlinecite{smallwood2012tracking} for the first time demonstrated the ability to track the evolution of the near-nodal superconducting gap in optimally-doped Bi$_2$Sr$_2$CaCu$_2$O$_{8+\delta}$ (Bi2212-OP91, T$_c$=91\,K) with sub-picosecond temporal resolution.
Figure\,\ref{SC_1}(a)-(b) shows the effect of near-IR pumping on the nodal and gapped off-nodal (gap of $\sim$\,15\,meV) electronic dispersion, respectively, via TR-ARPES differential maps. Along with the expected redistribution of photoemission intensity across the Fermi level, a transient modification of spectral weight within the originally-gapped region is observed in the off-nodal cut. In order to quantify such transient behaviour for different pump fluences and Fermi surface angles, \onlinecite{smallwood2012tracking} analyzed the symmetrized EDCs at $\mathbf{k}_\mathrm{F}$ in terms of a phenomenological model defined by the superconducting gap amplitude and a single-particle scattering rate, $\Sigma(\omega)=-i\Gamma+\Delta^2 / (\omega+i\Gamma)$ [see Fig.\,\ref{SC_1}(c)]. While for angles close to the nodal direction the pump excitation largely affects the gap response, and quasiparticle recombination rates remain higher than 20\,ps, shifting the angle away from the node reduces the pump-induced effect on the gap and makes the quasiparticle dynamics faster.

\begin{figure}
\centering
\includegraphics[scale=1]{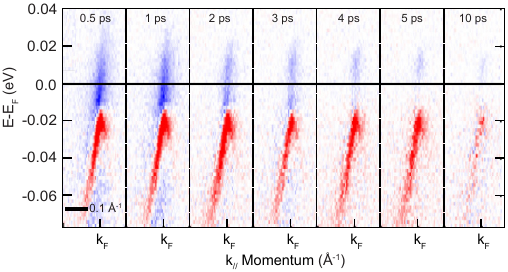}
\caption[SC2]{Visualizing the filling of the superconducting gap in optimally-doped Bi2212 upon near-IR pump excitation. The off-nodal TR-ARPES differential spectra at various pump-probe delays (0-10\,ps range) highlight the transfer of spectral weight within the gap region without a notable energy shift of the quasiparticle peak towards E$_F$. Adapted from \onlinecite{parham2017ultrafast}.}
\label{SC_2}
\end{figure}

\begin{figure*}
\centering
\includegraphics[scale=1]{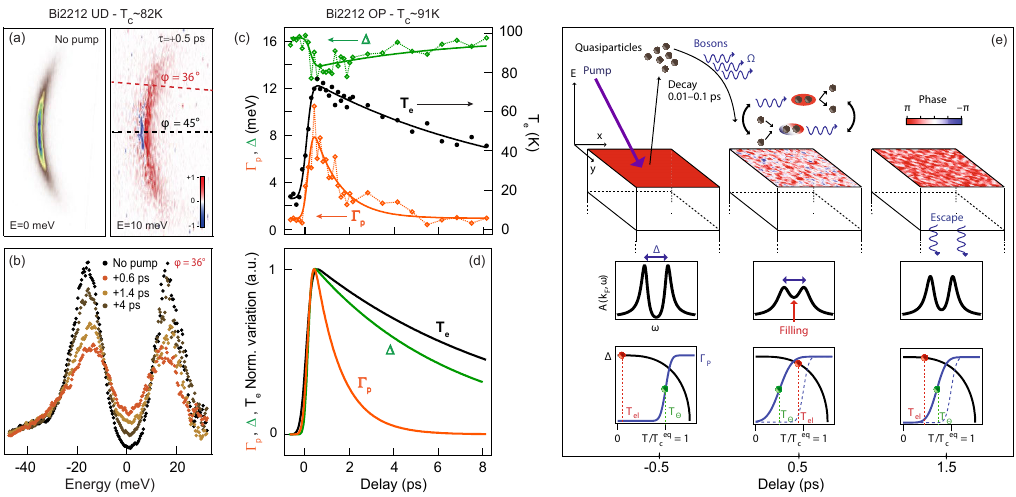}
\caption[SC3]{Establishing the key role of phase fluctuations in the transient filling of the superconducting gap in Bi-based cuprates. 
(a) Equilibrium Fermi surface and differential [\,I\,($\tau$\,=\,+0.5\,ps) -
I\,($\tau$\,=\,-0.5\,ps)\,] iso-energy contour mapping at 10\,meV above the Fermi level, for underdoped Bi2212 with T$_C$\,$\sim$\,82\,K. k$_\text{x}$ is aligned along the $\Gamma$-Y direction.
(b) Off-nodal EDCs at $\mathbf{k}_\mathrm{F}$ of underdoped Bi2212 with T$_C$\,$\sim$\,82\,K, normalized to momentum-integrated nodal EDCs ($\phi$=45$^o$) at different pump-probe delays ($\sim$8\,$\mu$J/cm$^2$ incident fluence).
EDCs have been deconvolved from the energy resolution broadening prior to the division [see \onlinecite{boschini2018collapse}].
(c) Transient evolution of the superconducting gap $\Delta$ and pair-breaking scattering rate $\Gamma_p$ extracted by fitting symmetrized EDCs (left axis), and T$_\text{e}$ extracted along the nodal direction (right axis) for optimally-doped Bi2212 with T$_C$\,$\sim$\,91\,K. Solid lines are phenomenological double-exponential decay fits to the data.
(d) Comparison between the normalized transient variation of the quantities in panel (c). The faster and disentangled dynamics characterizing $\Gamma_p$ attest to the non-thermal origin of the ultrafast enhancement of phase fluctuations. 
(e) Sketch of the transient collapse of the condensate driven by a loss of phase coherence. The top panels show a schematic diagram of the energetics of the process and the related real-space condensate phase coherence; the middle panels display the spectral function at k=k$_\text{F}$ when phase fluctuations are induced; the bottom panels show the temporal evolution of the pairing strength ($\Delta$, black line) and of $\Gamma_p$ (blue line). While the pairing is defined by T$_\text{e}$ (red spheres and dashed lines), superconductivity and the macroscopic T$_C$ are determined by the onset of phase coherence T$_\Theta$ (green spheres and dashed lines), which is proportional to the superfluid density.
Adapted from \onlinecite{boschini2018collapse} and \onlinecite{zonno2021time}.}
\label{SC_3}
\end{figure*}

This seminal work created the foundation for TR-ARPES exploration of dynamical processes in high-temperature superconductors and, in particular, the study of the hierarchy in the gap formation of cuprates, including the long-lasting investigation of the interplay between phase coherence and pairing strength in determining T$_c$ \cite{emery1995importance,caviglia2008electric}. In this regard, while early investigations reported a photoinduced gap closure in the near-nodal region \cite{smallwood2012tracking,smallwood2014time}, subsequent studies have revealed instead a filling of the superconducting gap accompanied only by a minimal modification of the gap amplitude itself upon  near-IR excitation \cite{parham2017ultrafast,zhang2017photoinduced,boschini2018collapse,zonno2021time}. 
This ultrafast filling of the superconducting gap is illustrated in Fig.\,\ref{SC_2} and Fig.\,\ref{SC_3}(b) by the temporal evolution of the TR-ARPES differential spectra of optimally doped Bi2212 and the off-nodal EDCs at $\mathbf{k}_\mathrm{F}$ normalized to the thermal broadening of underdoped Bi2212, respectively. Spectral weight is observed inside the gapped region without a significant energy shift of the quasiparticle peak towards the Fermi level. The observed transient filling of the gap has been interpreted in terms of a photoinduced loss of phase coherence, which quenches superconductivity on an ultrafast timescale. 

A quantitative analysis of this phenomenon has been conducted by introducing an additional scattering rate into the self-energy used to fit the gapped spectral function, the pair-breaking scattering rate $\Gamma_p$, which is associated with the density of phase fluctuations, $\Sigma(\omega)=-i\Gamma_s+\Delta^2 / (\omega+i\Gamma_p)$. This extended self-energy has been successfully employed to fit the transient evolution of the off-nodal EDCs at $\mathbf{k}_\mathrm{F}$ (either symmetrized or normalized to momentum-integrated EDCs along the gapless nodal direction) and track the light-induced dynamics of the pairing strength and pair-breaking scattering rate, as shown in Fig.\,\ref{SC_3}(c) \cite{boschini2018collapse,zonno2021time}. While both the dynamics of the single-particle scattering rate $\Gamma_s$ and gap amplitude $\Delta$ are locked to the evolution of the electronic occupation, \emph{i.e.} the electronic temperature T$_e$, the transient filling of the superconducting gap represented by $\Gamma_p$ exhibits much faster ($\sim$\,1\,ps) and decoupled dynamics, as shown in Fig.\,\ref{SC_3}(d). This result established the non-thermal origin of the pump-induced melting of the coherent condensate and was interpreted in terms of a light-induced non-equilibrium bosonic population that, by enhancing phase fluctuations via pair-breaking events, fills the superconducting gap and does not affect the pairing strength [Fig.\,\ref{SC_3}(e)] \cite{boschini2018collapse}. Moreover, the observation of similar gap-filling dynamics in different Bi-based compounds suggests a universal mechanism for the superconducting-to-normal state phase transition in Bi-based cuprates that relies on the loss of macroscopic phase coherence in the under-to-optimally doped regime \cite{zonno2021time}.
Note that evidence of phase fluctuations above T$_c$ has also been reported in Bi2212 via careful analysis of the nodal electron dynamics across a range of dopings and temperatures \cite{zhang2013signatures}.

Furthermore, evidence of an ultrafast filling of the SC gap has also been reported for Fe-based superconductors in a recent high energy resolution TR-ARPES study on the Fe-chalcogenide FeSe$_{0.45}$Te$_{0.55}$ superconductor \cite{nevola2023ultrafast}. However, in contrast with the cuprates, in this case the quenching of the SC phase has been associated with a metastable light-induced modification of the underlying antiferromagnetic order.

In contrast to the extensive efforts made to study the superconducting gap, the transient evolution of the pseudogap (PG), \emph{i.e.} the partial suppression of the density of states near the Fermi level even above T$_C$, has yet to receive comparable attention. This neglect is due to the larger BZ momenta defining the PG phenomenon in hole-doped cuprates, which makes such studies unattainable with the limited momentum-space accessible by low-photon energy laser systems employed to study the cuprates' near-nodal region.
To date, only a few TR-ARPES studies have focused on the transient evolution of the PG \cite{miller2017particle,cilento2018dynamics,boschini2020emergence}. 
At the antinode of the hole-doped Bi2212 cuprate, \onlinecite{cilento2018dynamics} reported the appearance of short-lived ($\sim$0.5\,ps) in-gap states in tandem with an ultrafast broadening of the O 2$p_x$ band. They attributed this finding to a resonant photoinduced transfer of electrons between Cu and O orbitals, attesting to the close relationship between low- and high-energy scales in high-temperature cuprates, as well as indicating a correlation-driven origin of the PG phenomenon \cite{cilento2018dynamics}.
Another example of TR-ARPES investigation of the PG in cuprates is the study by \onlinecite{boschini2020emergence} at the antiferromagnetic hot-spot of the optimally-doped electron-doped cuprate Nd$_{\text{2-x}}$Ce$_{\text{x}}$CuO$_{\text{4}}$, which can also be accessed at low photon energies, as discussed in Section\,\ref{SDW_spin} \cite{boschini2020emergence}.

\subsection{Summary and Outlook}
This section has summarized a few representative examples of how TR-ARPES has offered unprecedented insights into the dynamics of quantum materials, highlighting the underlying mechanisms, reaction pathways, and the ability to disentangle competing interactions of photoinduced phase transitions in quantum materials. To date, this technique has almost exclusively relied on high photon energy excitations that are not resonant with specific collective modes \cite{basov2017towards,de2021colloquium}, and thus for the most part correspond to light-induced melting of ordered phases. In fact, in all the cases discussed above, light excitations have led to the quenching of order parameters or loss of long-range coherence. 
Recent developments in the generation of tunable pump excitations in concert with high-repetition-rate XUV sources (see Section\,\ref{TR_ARPES_systems}) promise revolutionary insights into control of quantum matter by light, opening the way towards momentum-resolved surveys of novel light-induced phases of matter previously studied only via momentum-integrated techniques \cite{fausti2011light,mitrano2016possible,disa2021engineering,cavalleri2018photo,li2019terahertz}.

\section{Electron-Phonon Coupling} \label{ElPhon_section}
As discussed in Section\,\ref{TR_ARPES_section}, equilibrium ARPES is an exquisite tool to investigate the occupied electronic band structure of quantum materials but does not offer direct insights into bosonic degrees of freedom, such as the phonons' occupation and dispersion, which can instead be obtained via complementary particle or light scattering experiments. Whereas ARPES may provide hints via analysis of the electron self-energy $\Sigma(\omega,\text{k})$ as to how many-body interactions influence the electronic band dispersion and lifetime, such an approach often relies on specific working assumptions (\emph{e.g.}, knowledge of the bare electronic dispersion). This requirement makes identifying all the different electron interactions on the basis of equilibrium ARPES spectra challenging and typically impractical.
Since its first applications, the extension of ARPES into the time-domain has offered a new pathway for exploring different electron scattering processes on their intrinsic timescales, such as electron-electron (commonly within sub-100\,fs) and electron interactions with underlying collective excitations. In this regard, we note that the observation of multiple decay times is at odds with Mathiessen’s rule, which states that multiple scattering channels should lead to a single effective relaxation time \cite{kemper2018general}. In the past two decades, several methods with different levels of sophistication have been developed within the realm of TR-ARPES in an effort to provide a direct and precise estimate of relevant electron-boson couplings.

This section reviews the main TR-ARPES approaches to qualitatively or quantitatively estimate the fundamental electron-phonon coupling (EPC) phenomenon in quantum materials. Section\,\ref{TTM} discusses the use of multi-temperature models and collision integrals to describe the transient evolution of the electronic temperature and photoemission intensity, from which a qualitative estimate of the EPC is derived. Their application to two specific systems, topological insulators and graphene/graphite, is presented in detail. Section\,\ref{CoherentPhon} covers how the photoexcitation of coherent phonons may drive a transient modulation of the binding energy of electronic states, offering a direct way to estimate the EPC, as well as the application of a novel analysis procedure of TR-ARPES spectra in the Fourier frequency domain. Finally, Section\,\ref{SelfEn} focuses on how detailed tracking of the transient evolution of the electron self-energy provides insights into EPC, with particular emphasis on the case of cuprate superconductors. Finally, Section\,\ref{Ketty} reviews a recently developed TR-ARPES method for extracting the momentum- and mode-resolved electron-phonon matrix elements by detecting quantized relaxation processes of photoexcited electrons.

\subsection{Multi-temperature model and collision integrals} \label{TTM}
The multi-temperature model (MTM) is one of the most commonly used semiclassical approaches to qualitatively estimate the strength of electron scattering processes from non-equilibrium spectroscopy experiments. Going all the way back to early pump-probe studies, MTM has been implemented to extract EPC in metals and BCS superconductors \cite{anisimov1974electron,allen1987theory,kaganov1957relaxation,brorson1990femtosecond,tao2010theory}. The simplest MTM comprises just two coupled subsystems with different temperatures, and it is known as the two-temperature model (2TM) \cite{carpene2006ultrafast,bovensiepen2007coherent}. The 2TM assumes that the incoming light couples directly to the electronic bath, which subsequently transfers energy to the phononic bath (\emph{i.e.}, only excitations non-resonant to phononic modes are considered). The timescale at which such redistribution of energy occurs is dictated by the fundamental EPC strength of the material, as well as the natural frequency of the phonons involved in the process. Within this framework, tracking how the electronic temperature $T_{\text{e}}(\tau)$ evolves in a TR-ARPES experiment enables one to also infer valuable information about the phononic population and EPC in terms of the evolution of the lattice temperature $T_{\text{l}}(\tau)$.

Under the assumptions that only one Einstein phonon mode with frequency $\Omega$ is involved, and that the Eliashberg function $\alpha^2 \text{F}(\omega) \propto \delta(\omega - \Omega)$, the following rate equations hold \cite{sterzi2016smb,perfetti2007ultrafast,allen1987theory}: 
\begin{align} \label{TTM1}
    \frac{\partial T_{\text{e}}(\tau)}{\partial \tau} &= \frac{S(\tau)}{C_{e}} - \frac{3 \lambda \Omega^3 (n_e-n_{l})}{\hbar \pi k_b^2 T_{\text{e}}(\tau)},\\
    \frac{\partial T_{l}(\tau)}{\partial \tau} &= \frac{C_{e}}{C_{l}} \frac{3 \lambda \Omega^3 (n_{e}-n_{l})}{\hbar \pi k_b^2 T_{\text{e}}(\tau)}. \label{TTM2}
\end{align}
$S(\tau)$ describes the temporal profile of the optical excitation, $C_{e (l)}$ is the electronic (lattice) specific heat, $k_b$ is the Boltzmann's constant, and $n_{e,l}$ define the Bose-Einstein (BE) distribution functions for phonons calculated at temperatures $T_{\text{e}}$ and $T_{l}$, respectively.  
Here, one should note that the Eliashberg function $\alpha^2 \text{F}(\omega)$, and consequently the EPC constant $\lambda = 2 \int_0^{\infty} d\omega [\alpha^2 \text{F}(\omega)/\omega]$, depend on (i) the phononic density of states and (ii) the fundamental electron-phonon matrix element\footnote{The electron-phonon matrix element can be expressed as $g=\sqrt{\frac{\hbar}{2 M_C \Omega_{\textbf{q}}}} \bra{\textbf{k+q}}\Delta H \ket{\textbf{k}}$, where $\ket{\textbf{k}}$ and $\ket{\textbf{k+q}}$ denote the electron initial and final states connected by a phonon with momentum $\textbf{q}$, $\Delta H$ is the deformation potential, and $M_C$ is the atomic mass \cite{na2019direct,mahan2000many,grimvall1981electron,giustino2017electron}.}  \cite{mahan2000many,maklar2022Coherent,gierz2015phonon,kemper2014effect}. Note that this latter contribution, when integrated over momentum, implicitly contains the electronic density of states, \emph{i.e.} the available scattering phase space.
Therefore, any variations of the electronic structure and screening, including phase transitions accompanied by the filling and/or closure of energy gaps, may result in a transient modification of the available scattering phase space, and consequently of $\lambda$ (see for instance Sec.\,\ref{TR_ARPES_Int} and Fig.\,\ref{IntroSC}). In general, any changes of $\lambda$ across an equilibrium or out-of-equilibrium phase transition or following a modification of the electronic dispersion may result from changes in the electronic scattering phase space rather than from significant changes of the phonons’ density or fundamental electron-phonon matrix element $g$ connecting two specific $\textbf{k}$ states.

One of the main challenges of applying MTMs is determining the timescale at which temperature, which is an equilibrium property, can be properly adopted in an out-of-equilibrium context \cite{na2020establishing,mueller2013relaxation,rohde2018ultrafast,kratzer2022relaxation}.
In fact, at zero pump-probe delay when pump and probe pulses are overlapped, any optical excitation elicits a highly non-thermal state characterized by an out-of-equilibrium electronic distribution, which may exhibit a marked momentum dependence, as well as significant deviations from the equilibrium Fermi-Dirac distribution. In this respect, applying MTM in the context of TR-ARPES relies on the major assumption that the electronic bath thermalizes on ultrafast timescales via electron-electron interactions into a quasi-equilibrium state for which an effective electronic temperature can be defined. The transient evolution of $T_{\text{e}}(\tau)$ can then be usually deduced by fitting the broadening of the spectral weight at $E_F$ to a Fermi-Dirac distribution, although more advanced analysis methods can provide more accurate estimates \cite{ulstrup2014extracting}.
In addition, a light-induced shift of the chemical potential often accompanies the broadening of the Fermi edge. Although this review does not discuss this effect in detail, note that transient shifts of the chemical potential offer valuable insights into the DOS of the unoccupied states and into transient photodoping effects \cite{crepaldi2012ultrafast,miller2015photoinduced,miller2017particle,wang2012measurement,gierz2013snapshots}.

\begin{figure}
\centering
\includegraphics[scale=1]{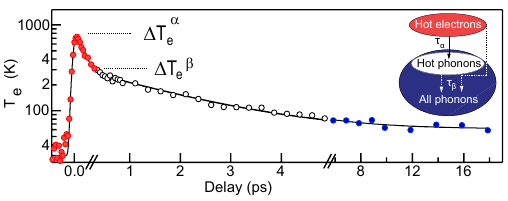}
\caption[FigTTM]{Application of the multi-temperature model (MTM) to the transient evolution of the effective electronic temperature of Bi$_2$Sr$_2$CaCu$_2$O$_{8+\delta}$ upon 1.55\,eV pump. Three cooling stages are observed, corresponding to different relaxation channels for the electronic and phononic bath following the optical excitation, as sketched in the inset. Adapted from \onlinecite{perfetti2007ultrafast}.} 
\label{FigTTM}
\end{figure}

\onlinecite{perfetti2007ultrafast} first applied an extended 2TM to TR-ARPES data of a Bi-based high-temperature cuprate superconductor by extracting $T_{\text{e}}(\tau)$ from fitting  a Fermi-Dirac distribution to the broadening of the spectral weight along the gapless nodal direction. As illustrated in Fig.\,\ref{FigTTM}, a 50\,fs thermalization time was estimated for the electronic bath, followed by further relaxation via a subset ($\sim$\,20\,$\%$) of strongly coupled phonons on a 110\,fs timescale, and then with the entire lattice on the picosecond timescale. By applying Eqs.\,\ref{TTM1} and \ref{TTM2}, the authors extracted $\lambda$\,$\sim$\,0.27 for phonons in the 40--70\,meV range, which have been proposed to be responsible for the band-structure renormalization known as the \emph{kink} phenomenon in Bi-based cuprates \cite{lanzara2001evidence}.

Similarly to MTMs, collision integrals follow a rate-equation formulation and offer a general framework for describing electron scattering processes \cite{ashcroft1976solid,yang2015inequivalence,del2000nonequilibrium,sjakste2018hot}:
\begin{equation} \label{CollInt}
\begin{aligned}
    \frac{\partial f(\epsilon_{\textbf{k}})}{\partial \tau} &= - \int \frac{d\textbf{k}'}{(2 \pi)^3}W_{\textbf{k},\textbf{k}'}f(\epsilon_{\textbf{k}})[1-f(\epsilon_{\textbf{k}'})] \\
    &+ \int \frac{d\textbf{k}'}{(2 \pi)^3} W_{\textbf{k}',\textbf{k}} f(\epsilon_{\textbf{k}'}) [1-f(\epsilon_{\textbf{k}})],
\end{aligned}
\end{equation}
where $\epsilon_{\textbf{k}}$ is a specific electronic state at momentum $\textbf{k}$, $f(\epsilon_{\textbf{k}})$ is the occupation of the $\epsilon_{\textbf{k}}$ state, and $W_{\textbf{k},\textbf{k}'}$ is the probability of scattering from $\textbf{k}$ to $\textbf{k}'$. Collision integrals, and rate-equation-based models in general, can describe the transient evolution of the photoemission intensity $I(\textbf{k},\omega,\tau)$ and estimate intrinsic electron relaxation times of the system.
However, \onlinecite{yang2015inequivalence} highlighted that $I(\textbf{k},\omega,\tau)$ maps the population dynamics and it is not directly related to the single-particle lifetime (which is inversely proportional to the imaginary part of the electron self-energy) and, consequently, is not directly related to the momentum-averaged fundamental electron coupling constants. Since the relaxation times of the population dynamics are strongly affected by the momentum- and energy-dependence of the out-of-equilibrium electronic distribution and light-induced modifications of the scattering phase space \cite{maklar2021nonequilibrium}, relying solely on fitting of $T_{\text{e}}(\tau)$ and $I(\textbf{k},\omega,\tau)$ provides only phenomenological/qualitative estimates of the underlying scattering times and electron coupling constants.

Over the past two decades, MTMs and collision integrals (and, more generally, other phenomenological multi-exponential decay models) have then been applied to a variety of quantum materials, ranging from topological insulators \cite{crepaldi2012ultrafast,sobota2012ultrafast,crepaldi2013evidence,sobota2014ultrafast,sterzi2017bulk,sanchez2017subpicosecond,reimann2014spectroscopy,zhu2015ultrafast,kuroda2016generation} to graphene/graphite \cite{ishida2011non,gierz2013snapshots,johannsen2013direct,johannsen2015tunable,stange2015hot,ulstrup2015ultrafast,yang2017novel,rohde2018ultrafast}, as well as superconductors \cite{perfetti2007ultrafast,cortes2011momentum,dakovski2015quasiparticle,piovera2015quasiparticle,smallwood2015influence,fanfarillo2021photoinduced,konstantinova2018nonequilibrium,zhang2016stimulated}, semicondutors/semimetals \cite{tanimura2015ultrafast,sterzi2016smb,tanimura2016formation,chen2020ultrafast,crepaldi2017enhanced,caputo2018dynamics,chen2018ultrafast,bao2022population}, and half-metals \cite{battiato2018distinctive}. Before continuing on to closely examine the application of MTMs and collision integrals in two classes of materials that have been extensively studied by the TR-ARPES community (topological insulators and carbon-based layered compounds such as graphene and graphite), we remark that these measurements mainly access local effects (\emph{i.e.}, relaxation processes) and provide little or at best indirect insights into nonlocal effects (\emph{i.e.}, diffusive transport). \onlinecite{brorson1987femtosecond} proposed an experimental strategy to distinguish local and nonlocal effects in pump-probe studies by optically exciting the sample on both sides (front and back) while probing its out-of-equilibrium response only on one side (front). This method was successfully applied to a Au/Fe/MgO(001) heterostructure \cite{beyazit2020local,beyazit2023ultrafast} and, by applying a TTM accounting for diffusive electronic transport, it revealed the transition from superdiffusive to diffusive electronic transport in the Au layer as a function of its thickness  \cite{kuhne2022ultrafast}.

\subsubsection{Topological Insulators} \label{TTM_TI}
\begin{figure}
\centering
\includegraphics[scale=1]{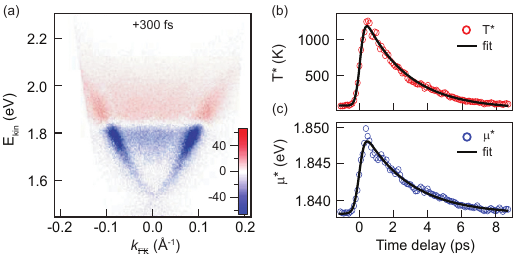}
\caption[EPCrepaldi]{Transient evolution of the electronic temperature and chemical potential in the topological surface state of Bi$_2$Se$_3$ after near-IR pump excitation. (a) Pump-probe TR-ARPES spectrum displayed as the difference between the signal at +\,300\,fs and the signal at -\,500\,fs. Temporal evolution of (b) the effective temperature T$^{*}$ and (c) the effective chemical potential $\mu^{*}$ obtained from fitting a sequence of EDC curves at k$_{\text{F}}$. In each case, a single decay exponential function captures the observed temporal dynamics, with characteristic relaxation times equal to $\sim$\,2.5\,ps and $\sim$2.7\,ps, respectively. A simple thermal model based on the extracted T$^{*}$($\tau$) and $\mu^{*}$($\tau$) reproduces the ultrafast evolution of the nonequilibrium charge population in the topological surface state at different binding energies. Adapted from \onlinecite{crepaldi2012ultrafast}.}
\label{EP_crepaldi}
\end{figure}

Since the first observation of the topological surface state in TIs \cite{hasan2010colloquium}, the ARPES community has sought to estimate the strength of EPC for Dirac electrons, with contradictory results \cite{hatch2011stability,pan2012measurement}.
In an effort to resolve this controversy, MTM together with general considerations about the decay rates of the photoinduced electronic population have been applied in a qualitative fashion to various TIs to gain insights into electron scattering processes within the bulk bands and the topological surface state. 
As discussed in Section \ref{TIs}, near-IR optical excitation of TIs induces a non-thermal redistribution of carriers within both valence and conduction bulk bands and surface states.
Due to the limited scattering phase space of the surface state, relaxation timescales can extend up to hundreds of picoseconds for Dirac electrons in systems where the chemical potential falls inside the bulk band gap, and a few picoseconds for bulk electrons in systems where the chemical potential lies outside the bulk band gap. Several intertwined relaxation mechanisms have been proposed to account for these relaxation timescales, ranging from inefficient electron-phonon scattering to surface-to-bulk diffusion  \cite{sobota2012ultrafast,hajlaoui2012ultrafast,wang2012measurement,crepaldi2012ultrafast,crepaldi2013evidence,sobota2014distinguishing,reimann2014spectroscopy,sanchez2017subpicosecond,sterzi2017bulk,freyse2018impact,khalil2019bulk,zhong2021light}, as well as bottleneck effects at the Dirac point and surface photovoltage effects (SPV, see section\,\ref{SPV_section}) \cite{zhu2015ultrafast,sanchez2016ultrafast,sumida2017prolonged,hedayat2018surface,sumida2019inverted,hajlaoui2014tuning,neupane2015gigantic,papalazarou2018unraveling,ciocys2020manipulating}.
With particular emphasis on electron-phonon coupling in TIs, pioneering studies by \onlinecite{sobota2012ultrafast}, \onlinecite{crepaldi2012ultrafast,crepaldi2013evidence}, and \onlinecite{wang2012measurement} have extracted the transient electronic temperature for the unoccupied conduction band of Bi$_2$Se$_3$ and Bi$_2$Te$_3$, and associated the observed decay of $\sim$0.7--2.5\,ps to inefficient electron-phonon coupling. Figure\,\ref{EP_crepaldi} showcases the transient evolution of the effective electronic temperature and chemical potential -- with characteristic relaxation times of 2.5\,ps and 2.7\,ps, respectively -- for the topological surface state of Bi$_2$Se$_3$ upon near-IR excitation. While the former is ascribed to electron-phonon relaxation processes (via TTM considerations), the latter is attributed to the relaxation of the excess charge in the CB via diffusion \cite{crepaldi2012ultrafast}. Moreover, \onlinecite{sobota2014ultrafast} reported a counter-intuitive temperature dependence of the decay-rate of the conduction band of Bi$_2$Se$_3$ (lower decay rates with higher temperatures), and they showed via collision integrals that it originates directly from electron-phonon scattering processes .

\begin{figure*}
\centering
\includegraphics[scale=1]{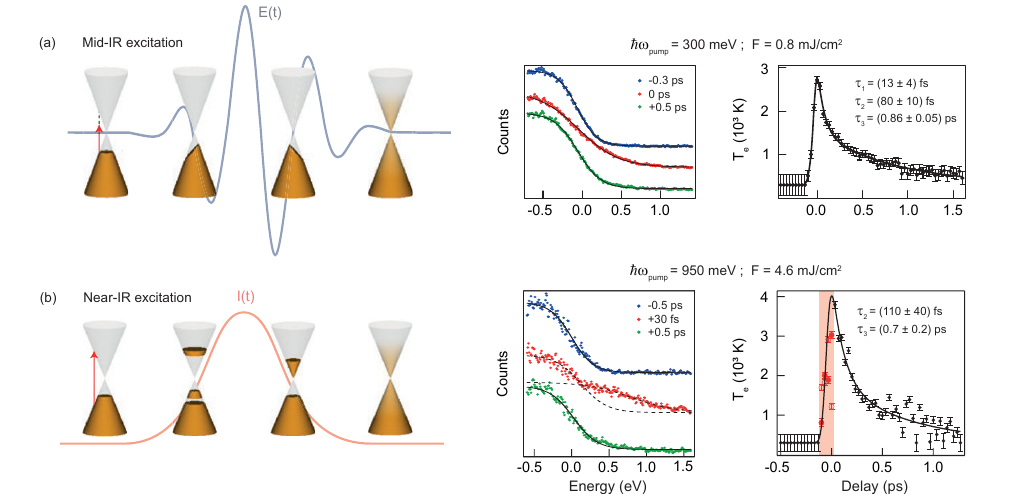}
\caption[EP1]{MTM analysis of TR-ARPES data of graphene upon (a) mid-IR and (b) near-IR excitations. Left: Schematics of the two different excitation schemes, illustrating how only the near-IR pump pulse accesses direct optical transitions. The light grey line represents the electric field of the 300\,meV pump pulse and the light red line represents the intensity of the 950 meV pump pulse.
Middle: Momentum-integrated EDCs around the K-point for selected pump-probe delays, along with Fermi-Dirac distribution fits. Right: Transient electronic temperature with multi-exponential decay fits. Note that while in (a) the pump-induced thermal broadening follows the conventional Fermi-Dirac distribution for all delays, in (b) EDCs at times close to zero are best described by two different Fermi-Dirac fits for electrons and holes (dashed black lines). These time delays are indicated by red data points and the shaded area in the transient evolution of T$_e$. Adapted from \onlinecite{gierz2013snapshots}.}
\label{EP1_MTM}
\end{figure*}

The weak electron-phonon coupling of the topological surface state has been further qualitatively assessed in studies that (i) observed a power-law dependence of the surface temperature decay-rate on carrier density \cite{wang2012measurement}, (ii) used XUV light (17.5\,eV) to increase the surface sensitivity, which revealed a saturated electronic temperature at the surface \cite{crepaldi2013evidence}, and (iii) showed the vanishing of transient pump-induced spin-polarization in the topological surface state after $\sim$1.2\,ps, a timescale compatible with electron-phonon scattering \cite{sanchez2016ultrafast} [see also Sec.\,\ref{TI_spin}].

\subsubsection{Graphene/Graphite}
Graphene and its three-dimensional allotropic compounds, such as graphite, are characterized by Dirac-like dispersive bands at the corners of the hexagonal Brillouin zone, as well as strong coupling of electrons to optical phonons with momentum Q=0 ($\Gamma$-phonons) and Q=K (K-phonons). Efficient electron-phonon scattering is one of the possible limiting factors for the transport properties of graphene-based devices, hindering their use for energy harvesting purposes \cite{bonaccorso2010graphene,tielrooij2013photoexcitation,johannsen2015tunable,scheuch2011temperature}.

MTMs have been extensively applied in TR-ARPES investigations of graphene/graphite to gain valuable insights into the intrinsic timescale of electron-phonon scattering.
Although no electronic states are expected in the proximity of E$_F$ away from the BZ corners, an early 6-eV TR-ARPES investigation of graphite reported a transient modification of the photoemission intensity at the $\Gamma$ point \cite{ishida2011non}. Such incoherent spectral weight in the BZ center stems from scattering of K/K'-electrons into $\Gamma$ mediated by the emission of an optical phonon with momentum Q=K, and is therefore a manifestation of the strong electron-phonon coupling characteristic of graphite \cite{liu2010phonon}. By inspecting the transient evolution of the spectral broadening at $\Gamma$, \onlinecite{ishida2011non} reported the emergence of a non-thermal electronic distribution in the first $\sim$\,200\,fs, and suggested that generation of strongly coupled optical phonons occurs mainly during this ultrafast timescale. Moreover, given the impossibility of reliably extracting the transient $T_{\text{e}}(\tau)$ at these early time delays, the author proposed that the 2TM is applicable in graphite only after approximately $\sim$200\,fs.

Subsequent TR-ARPES studies employed XUV probe light to explore the ultrafast electrodynamics at the K point of graphene/graphite. Among these, \onlinecite{gierz2013snapshots} investigated the evolution of $T_{\text{e}}(\tau)$ of 200-meV hole-doped graphene upon both mid-IR (0.3\,eV) and near-IR (0.95\,eV) excitations, as shown in Fig.\,\ref{EP1_MTM}. While the mid-IR photon energy is too low to drive any direct optical transitions and leads to a spectral broadening well-captured by a Fermi-Dirac distribution [panel (a)], a non-thermal distribution is observed around time zero upon near-IR pump light [panel (b)]. In the latter case, the photoemission signal can be described as the sum of two distinct Fermi-Dirac distributions centred above/below the Dirac point \cite{ulstrup2014ultrafast,gierz2015tracking,gierz2017probing,gierz2015population}. Moreover, in both excitation schemes of Fig.\,\ref{EP1_MTM}, $T_{\text{e}}(\tau)$ exhibits a $\sim$\,100\,fs decay associated with the fast scattering of electrons off optical phonons, and a slower picosecond component stemming from the anharmonic decay of optical phonons into acoustic phonons and from electron-acoustic phonon scattering.
Other XUV-probe TR-ARPES studies on graphene \cite{johannsen2013direct,johannsen2015tunable,stange2015hot,ulstrup2015ultrafast} also reported similar decay times. 

Extended MTMs have been applied to TR-ARPES data also in an effort to evaluate the impact of supercollisions [\emph{i.e.}, the coupling between electrons and acoustic phonons mediated by disorder, which acts as a source of momentum \cite{alencar2014defect}] on the electron relaxation processes. These studies have led to apparently contradictory results: whereas the inclusion of supercollisions appears necessary to describe the evolution of $T_{\text{e}}(\tau)$ in graphene \cite{johannsen2013direct,someya2017suppression}, a basic strongly-coupled phonon model, which neglects any coupling between electrons and acoustic phonons, succeeds in reproducing the experimental data of graphite \cite{stange2015hot}. This apparent disagreement may simply originate from the different impurity levels of graphite and exfoliated graphene on different substrates.

In addition, \onlinecite{pomarico2017enhanced} first investigated how the EPC of bilayer graphene on SiC(0001) may be modulated via resonant excitation of the in-plane E$_{1u}$ bond-stretching phonon at 200\,meV. By combining THz time-domain spectroscopy and analysis of the decay time of the transient electronic temperature extracted from TR-ARPES data, they reported a transient threefold enhancement of the EPC constant.

Despite the extensive adoption of MTMs in studying graphene/graphite, two general limitations remain in such an approach to analyzing TR-ARPES data: (i) MTMs cannot identify specific optical phonons involved in the ultrafast thermalization processes; (ii) MTMs are not strictly valid for short pump-probe delays when an electronic temperature cannot be defined unambiguously. 
The first obstacle can be overcome either by combining TR-ARPES with complementary time-resolved techniques, such as time-resolved Raman spectroscopy \cite{yang2017novel} or ultrafast electron diffuse scattering spectroscopy \cite{stern2018mapping}, or by applying more advanced analysis procedures to the TR-ARPES signal \cite{na2019direct} (see Sec.\,\ref{Ketty}). Exploiting these methods, the aforementioned experimental investigations established that the optical K-phonons are the main drivers of the observed ultrafast sub-100\,fs relaxation processes. 
\onlinecite{rohde2018ultrafast} addressed the second obstacle by showing that the non-thermal electronic distribution converges to a single Fermi-Dirac distribution within $\sim$\,50\,fs (incident fluence $>$\,0.9\,mJ/cm$^2$) as a result of electron-electron and electron-phonon scattering. 

However, the duration of the non-thermal window depends on the excitation density and may lengthen in low-fluence regimes \cite{na2020establishing,kratzer2022relaxation}. We emphasize that semiclassical models of the TR-ARPES data based on collision integrals do not strictly require equilibrium thermodynamic variables such as temperature to be well defined, thus allowing a description of the overall transient evolution of the photoemission intensity even in highly non-thermal regimes \cite{na2020establishing}.

\begin{figure*}
\centering
\includegraphics[scale=1]{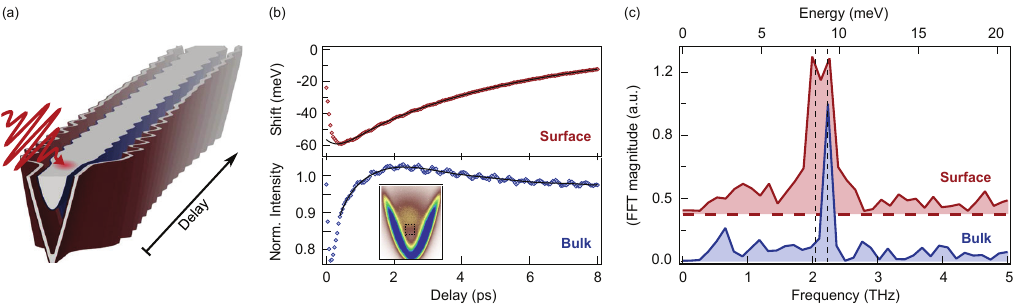}
\caption[EP2]{Disentangling the surface and bulk contributions to the light-induced coherent oscillations in Bi$_2$Se$_3$. (a) Cartoon of the optical excitation and relaxation of coherent phonons in the bulk and surface bands. (b) Time-dependent shift of the TSS and CB binding energies, along with the respective smooth curve fits (black solid line, tenth-order polynomial). (c) Fast Fourier Transform (FFT) of the residuals from the fit for the TSS and CB, showing the presence of distinct frequencies: while the oscillations of the CB can be captured by a single undamped cosine, the oscillations of the TSS are best fit by a beat pattern represented by the sum of two cosine functions.
 Adapted from \onlinecite{sobota2014distinguishing}.}
\label{EP2_Sobota}
\end{figure*}

\subsection{Light-driven Coherent Phonons} \label{CoherentPhon}
When performing TR-ARPES experiments, it is not uncommon to observe coherent oscillations in the transient photoemission intensity or the binding energy of the probed electronic states. Light-inducing coherent phononic modes provide advantageous approaches -- via resonant phonon pumping \cite{forst2011nonlinear}, non-resonant excitation \cite{dhar1994time,zeiger1992theory}, or coherent control \cite{rettig2014coherent} -- to investigate how the electronic bath couples to specific phonon modes. While TR-ARPES studies with light excitations resonant to specific phonon modes are still rare \cite{gierz2015phonon,pomarico2017enhanced}, the use of non-resonant visible and near-IR pump pulses has led to the observation of phonon-mediated coherent oscillations in the photoemission signal. This phenomenology has been reported for different classes of materials, ranging from metals \cite{papalazarou2012coherent,sakamoto2022connection} and superconductors \cite{gerber2017femtosecond,yang2019mode,yang2021experimental,okazaki2018antiphase,suzuki2021hhg,yang2014ultrafast} to topological insulators \cite{sobota2014distinguishing,golias2016observation,sobota2023influence} and chalcogenides \cite{perfetti2006time,hein2020mode,suzuki2021detecting,tang2020non,baldini2023spontaneous}.  
Non-resonant high-photon energy light pulses can drive coherent phonons through two main mechanisms \cite{garrett1996coherent}: (i) impulsive stimulated Raman processes, usually occurring in transparent compounds where the pump photon energy is smaller than the optical gap and prompting sine-like oscillations \cite{zeiger1992theory}; (ii) displacive processes, commonly observed in opaque materials as a consequence of an abrupt modification of the electronic density, which in turn triggers cosine-like displacive motions of ions around new coordinates \cite{dhar1994time}.

Generally speaking, any shift in the binding energy $\Delta_{k}$ of the Bloch state $\ket{k}$ can be related to the atomic displacement $u$ driven by a particular mode and the electron-phonon matrix element $g_k$ for that particular electronic state and phonon mode, $\Delta_{k}=g_k u$ \cite{grimvall1981electron,giustino2017electron}. The direct relation between $\Delta_{k}$ and $g_k$ enables valuable insights into the coupling of the electronic states probed via TR-ARPES with coherent phonon modes by tracking the coherent oscillations of the electronic band structure at the frequency of the light-induced phonon mode.

\begin{figure*}
\centering
\includegraphics[scale=1]{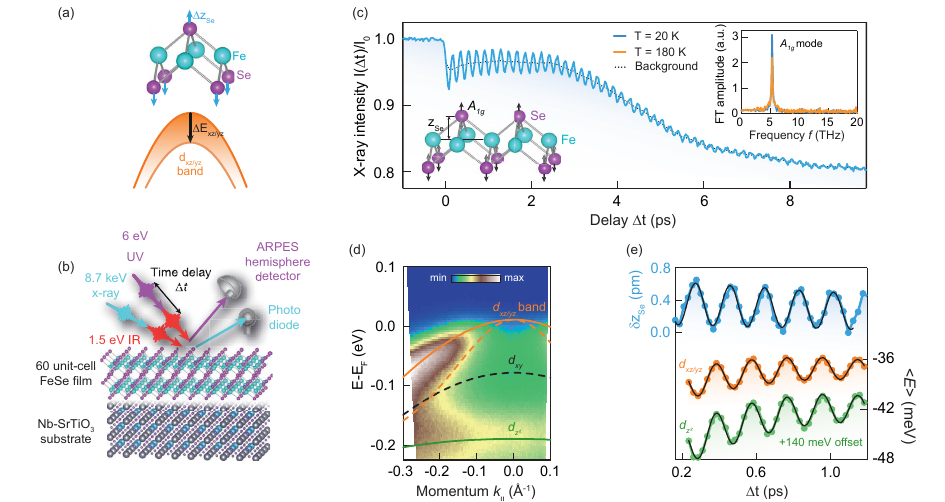}
\caption[EP2]{Tracking the response to optical excitations of the A$_{\text{1g}}$ phonon mode in FeSe/SrTiO$_3$ films via combined time-resolved x-ray diffraction and TR-ARPES measurements. (a) Schematic of the A$_{\text{1g}}$ phonon mode, which periodically modulates the electronic band energies. (b) Sketch of the experimental strategy. (c) Temporal evolution of the x-ray intensity of the (004) Bragg peak upon near-IR pumping, which excites a coherent A$_{1g}$ phonon mode (see inset). (d) Equilibrium ARPES spectrum of FeSe along the $\Gamma$-X direction acquired at 20\,K. Renormalized density functional theory calculations for the electronic band dispersions are represented by superimposed lines (solid for ARPES-detected bands and dashed for unresolved bands). The orbital characters of the bands are also indicated. (e) Relationship between (blue) the oscillations of the Se displacement $\delta$z$_{Se}$ probed by time-resolved x-ray scattering and (orange and green) the momentum-averaged binding energy shift of the electronic bands obtained by TR-ARPES.  Adapted from \onlinecite{gerber2017femtosecond}.}
\label{EP2_CohPh}
\end{figure*}

\subsubsection{Coherent oscillations in the TR-ARPES signal}
\label{CoherentOscill}

This section focuses on two emblematic TR-ARPES studies in which detection of coherent oscillations allowed quantitative information on the underlying EPC to be extracted.

In the work by \onlinecite{sobota2014distinguishing}, out-of-plane A$_{1g}$ optical phonons were coherently driven by near-IR light in the topological insulator Bi$_2$Se$_3$. As a result, both the topological surface state and the bulk conduction band displayed oscillations in their binding energies [see Fig.\,\ref{EP2_Sobota}(a) and (b)]. Interestingly, the frequency of the coherent modulation of the surface's energy shift was $\sim$10$\%$ lower than the one observed for the bulk band [2.05\,THz and 2.23\,THz, respectively, as depicted in Fig.\,\ref{EP2_Sobota}(c)]. This work showcased the capability of TR-ARPES to reveal how coherent phonons may couple differently to surface and bulk states, and crucially relied on high frequency sensitivity to resolve the softening of the phonon mode at the surface.

In their seminal work, \onlinecite{gerber2017femtosecond} combined time-resolved x-ray diffraction and TR-ARPES to track the orbital-resolved deformation potential for the A$_{1g}$ mode in bulk FeSe, an unconventional superconductor in which strong electron correlations are expected to renormalize the EPC. Figures\,\ref{EP2_CohPh}(a) and (b) illustrate the motivating concept of this work: relate lattice changes (probed via time-resolved x-ray diffraction) to energy shifts of electronic states (accessed via TR-ARPES). 
Figure\,\ref{EP2_CohPh}(c) shows the coherent oscillations in the x-ray intensity induced by the excitation of the displacive 5.3\,THz A$_{1g}$ optical phonon mode, which triggers a periodic variation of the anion height $\delta$z$_{Se}$ (see inset). This atomic variation was related to the pump-induced energy modulation of two electronic bands observed by TR-ARPES at the $\Gamma$ point; the two bands exhibit different orbital characters ($d_{xz/yz}$ and $d_{z^2}$), as illustrated in Fig.\,\ref{EP2_CohPh}(d)-(e). By combining the lattice and electronic experimental values, \onlinecite{gerber2017femtosecond} extracted the value of the deformation potential for the two bands and compared it to the theoretical value obtained by DFT and self-consistent DFT–dynamical mean field theory calculations. Only the latter, which accounts for electron correlations, was able to reproduce the experimental results, thus demonstrating the key role of electron correlations in renormalizing the EPC in FeSe.
A recent TR-ARPES study, by tracking light-induced shifts of the $d$-orbitals around the Brillouin zone center, has determined that the tetragonal-to-orthorhombic structural transition plays a key role in the emergence of the electronic nematic phase of FeSe \cite{yang2022anomalous}.

To conclude, two recent works investigated light-induced coherent phonons in topological insulators by combining TR-ARPES and time-resolved x-ray diffraction. In particular, they demonstrated how the topological surface state of Bi$_2$Te$_3$ couples to different phonon modes \cite{sobota2023influence} and extracted the coupling between the topological surface states of Bi$_2$Se$_3$ and Bi$_2$Te$_3$ and light-induced A$_{1g}$ phonon modes by evaluating mode- and band-resolved deformation potentials \cite{huang2023ultrafast}.

\subsubsection{Frequency-Domain ARPES} \label{FDARPES}
The detection of coherent oscillations of the band structure requires the use of a probe pulse with (i) a bandwidth sufficiently narrow to resolve any energy shift of the band (commonly of the order of few meV) and (ii) a temporal duration smaller than the phonon cycle. These two conditions have limited the detection of coherent modulations of the electronic dispersion in the TR-ARPES signal to \emph{slow} phonons ($<$\,3--4\,THz).
A novel approach known as Frequency-Domain ARPES (FD-ARPES) promises to circumvent these difficulties in detecting high-frequency phonons by offering a direct estimate of EPC via Fourier-like analysis of the TR-ARPES signal. FD-ARPES has been recently discussed theoretically \cite{de2020direct} and demonstrated experimentally \cite{hein2020mode,suzuki2021detecting,lee2023layer,ren2023phase}. 
The key advantage of FD-ARPES relies on the fact that, even if the probe bandwidth greatly exceeds the band shifts in energy $\Delta_k$, the Fourier-transformed photoemission signal grants access to the diagonal and off-diagonal electron-phonon matrix elements.

Under the assumption that a single phonon mode $\Omega_0$ [whose displacive motion is parameterized by $u(\tau)$\,=\,$u_0$\,$sin(\Omega_0 \tau$)] induces variations in both the band energy and orbital character, the Fourier transform of the TR-ARPES signal (see Eq.\,\ref{TRARPES_int_EQ} in Section\,\ref{TR_ARPES_Int}) evaluated at $\Omega_0$ can be expressed as \cite{de2020direct}: 
\begin{equation} \label{EqFDARPES}
    \tilde{I}(\omega,\Omega_0) \propto \frac{i}{2} M \Delta_k \frac{\delta A(\omega)}{\delta \omega} +\frac{1}{2} \frac{\delta M[u(\tau)]}{\delta u(\tau)} \Omega_0 A(\omega).
\end{equation}
Equation\,\ref{EqFDARPES} is valid when (i) the band shift $\Delta_k$ is smaller than the probe bandwidth, (ii) $u_0$ is sufficiently small, and (iii) the electron-phonon coupling is adiabatic. 
The first term of $\tilde{I}(\omega,\Omega_0)$ depends on the ARPES matrix elements and the derivative of the spectral lineshape, and is proportional to the diagonal electron-phonon matrix element $g_k$\,=\,$\frac{\Delta_k}{u_0}$. The second term of Eq.\,\ref{EqFDARPES} stems from the transient modulation of the photoemission matrix-elements weighted by the ARPES lineshape, and it is proportional to interband electron-phonon matrix elements \cite{de2020direct}.
In single-band systems with a well-defined orbital character only the first term of Eq.\,\ref{EqFDARPES} survives and the FD-ARPES signal directly probes the EPC. Multi-band systems in which interband electron-phonon coupling may have non-negligible contributions can be more challenging to unravel. In addition, the pump-induced modification of the photoemission matrix elements, recently described theoretically \cite{freericks2016constant} and reported experimentally \cite{boschini2020role}, may further complicate a quantitative analysis of the FD-ARPES signal. 

\begin{figure}
\centering
\includegraphics[scale=1]{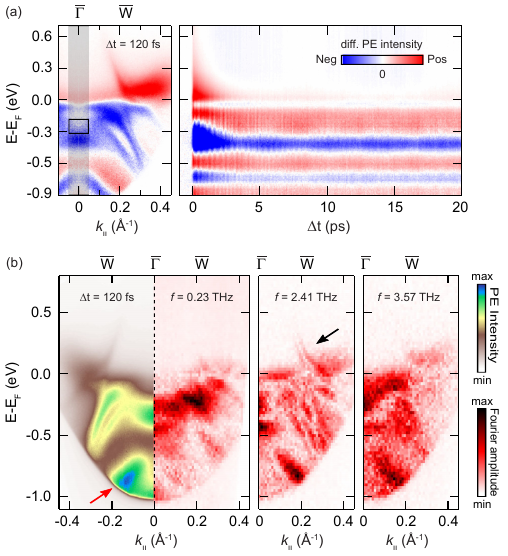}
\caption[EP3]{Momentum-space visualization of the band selectivity of the phonon coupling in T$_d$-WTe$_2$ via Frequency-Domain ARPES.
(a) Left: differential ARPES spectrum of T$_d$-WTe$_2$ along $\overline{\Gamma}$-$\overline{\mathrm{W}}$ acquired 120\,fs after near-IR pump excitation. Right: transient evolution of the differential ARPES intensity around $\Gamma$, that shows clear oscillations of the binding energies modulated by multiple frequencies. 
(b) Frequency-Domain ARPES maps (\emph{i.e.}, Fourier-transformed photoemission signals) for selected frequencies. The analysis highlights the band selectivity of the phonon excitations (see, for example, red arrow) and also applies to excited states above E$_F$ (see black arrow). From \onlinecite{hein2020mode}.}
\label{EP3_FD}
\end{figure}

Nevertheless, \onlinecite{hein2020mode} first demonstrated the capabilities of FD-ARPES on T$_d$-WTe$_2$, a type-II Weyl semimetal. The non-centrosymmetric T$_d$ crystal structure is a pivotal requirement for the emergence of Weyl points in non-magnetic materials \cite{soluyanov2015type}, and a light-induced structural phase transition into a centrosymmetric phase is predicted to annihilate Weyl points. Such a structural phase transition is expected to occur upon the excitation of the 0.24\,THz interlayer shear mode \cite{sie2019ultrafast}. In the work of \onlinecite{hein2020mode} five A$_1$ coherent phonon modes were excited via near-IR optical pump, including the 0.24\,THz shear mode [Fig.\,\ref{EP3_FD}(a) shows the presence of coherent phonon modes via conventional analysis of the differential ARPES image, as discussed in Sec.\,\ref{CoherentOscill}]. Figure\,\ref{EP3_FD}(b) compares the TR-ARPES map acquired along the $\overline{\Gamma}$-$\overline{W}$ direction immediately after the pump-excitation (far left panel) with the corresponding FD-ARPES maps obtained at three different frequencies (red coloured maps). The data illustrate how the different pump-induced phonons selectively couple to specific electronic bands, as exemplified by the dispersion indicated by the red arrow, where spectral weight is nearly null in the 0.23\,THz FD-ARPES map but very intense at 2.41\,THz. Due to the nature of TR-ARPES, this method can resolve the effect of the electron-phonon coupling on individual bands even above the Fermi level (see black arrow in the middle panel).   
By specifically focusing on the FD-ARPES map at 0.23\,THz, \onlinecite{hein2020mode} showed how the shear mode modifies the electronic structure of T$_d$-WTe$_2$ and modulates the spin-splitting of bands, a spectral signature which is linked to the broken inversion symmetry of the crystal. 

\subsection{Transient evolution of the Self Energy} \label{SelfEn}
As discussed in Section\,\ref{TR_ARPES_Int}, since the photoemission intensity is proportional to the one-electron removal spectral function, TR-ARPES can access the transient evolution of many-body interactions via the electron self-energy $\Sigma=\Sigma'+i\Sigma''$, where the real and imaginary parts of $\Sigma$ are related by Kramers-Kronig relations. In particular, careful analysis of the experimentally-determined ARPES band structure renormalizations and spectral broadening can evaluate $\Sigma'$ and $\Sigma''$ \cite{damascelli2003angle,sobota2021angle}.
However, when a material is perturbed by an optical excitation, the electron self-energy is expected to vary \cite{sentef2013examining,kemper2013mapping,kemper2014effect,kemper2017review,kemper2018general}. Even for an interacting electron gas without any couplings to bosonic modes, a sudden increase in the electronic temperature is accompanied by a broadening of the spectral lineshape [in the Fermi-Liquid scenario $\Sigma'' _{\text{FL}} (\omega, \text{T}_e (\tau))  \propto \beta\, (\omega^2 + \pi^2 k_B^2 \text{T}_e (\tau)^2)$, \cite{zonno2021ubiquitous}].

\begin{figure}
\centering
\includegraphics[scale=1]{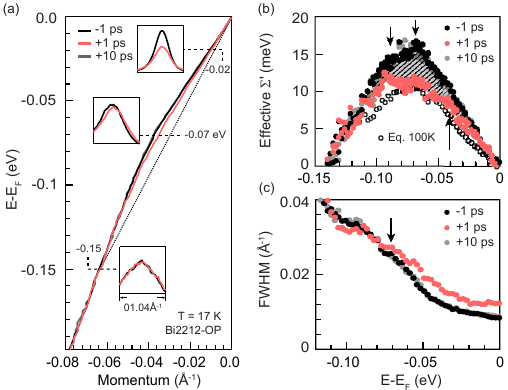}
\caption[EP4]{Evaluation of the electron-phonon coupling response in optimally-doped Bi2212 cuprate via analysis of the transient electron self-energy. (a) Nodal dispersion of optimally-doped Bi2212 for three distinct pump-probe delays, showing the softening of the characteristic \textit{kink} at 70\,meV at +1\,ps. The dashed black line corresponds to the linear bare band dispersion. The insets compare MDCs before/after near-IR pumping around the kink energy. (b) Real part of the electron self-energy extracted at different delays (full circles) and the equilibrium values measured at 100\,K (open circles). Black arrows mark the suppression of different peaks. (c) Corresponding MDC's full-width at half maximum (FWHM), which is proportional to the imaginary part of the self-energy. A clear pumped-induced modification of both the real and imaginary parts of $\Sigma$ is reported at the kink energy. Adapted from \onlinecite{zhang2014ultrafast}.}
\label{EP4_SelfEn}
\end{figure}

\begin{figure*}
\centering
\includegraphics[scale=1]{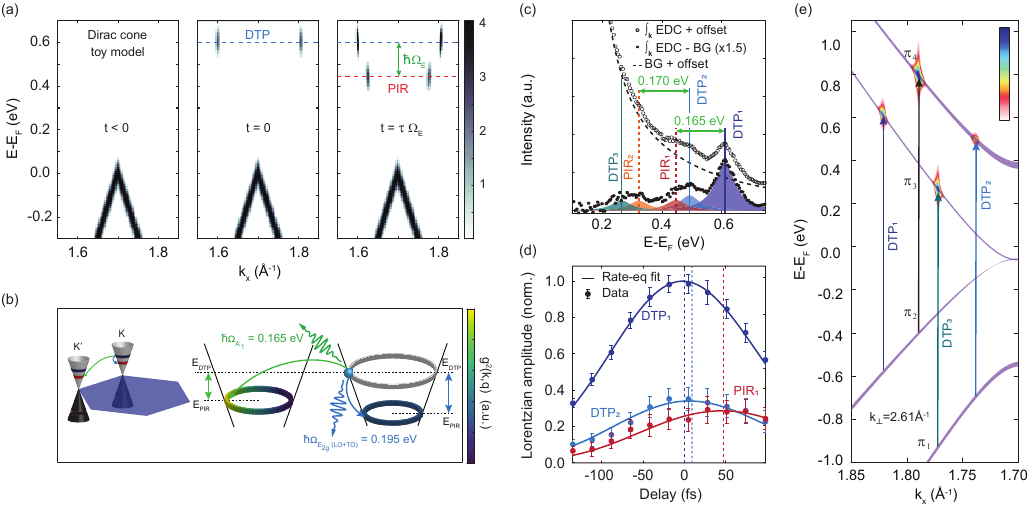}
\caption[EP5]{Momentum-resolved analysis of the quantized electron-phonon decay processes in graphite, allowing the extraction of the mode-projected electron-phonon matrix element $g_{\bar{k}}$. 
(a) Simulation of the transient TR-ARPES intensity for a Dirac cone pumped with 1.2\,eV photons, including a retarded e-ph interaction with a phonon of energy $\hbar\Omega_0$. At time t\,=\,0, the direct transition peak (DTP) feature is observed at E$_{\text{DTP}}$\,=\,0.6\,eV; at t\,=\,$\tau_{\Omega_0}$, the phonon-induced replica (PIR) is observed at E$_{\text{DTP}}$-$\hbar\omega_0$. 
(b) Calculation of $g_{\bar{k}}$ for both an electron in K scattering with an A'$_1$ mode with momentum $\sim$K (green) into a final state in K', as well as an electron in K scattering with an E$^2_g$ mode with momentum $\sim$\,0 (blue) into a final state in K.
(c) Momentum-integrated EDC around the K point immediately after the pump excitation. Multiple DTPs and PIRs can be identified on top of the hot-electron background based on the momentum-resolved optical joint DOS of panel (e).
(d) Transient evolution of the amplitude of the most prominent peaks: DTP$_1$ (DTP$_2$) in dark (light) blue and PIR$_1$ in red. Dashed lines indicate the
peak delays: DTP$_2$ (PIR$_1$) is delayed 9\,fs (47\,fs) with respect to DTP1. Solid curves represent the electronic occupation of the specified states derived from a rate-equation model fit. The transfer of spectral weight
from DTP$_1$ to PIR$_1$ is associated with an e-ph scattering time constant $\tau_{\Omega_0}$\,=\,174\,$\pm$\,35\,fs. Adapted from \onlinecite{na2019direct}.}
\label{EP5_PIR}
\end{figure*}

In the specific case of electrons interacting with a bosonic mode of energy $\Omega_0$, the band dispersion deviates from the non-interacting case and a \emph{kink} appears in the dispersion. Paradigmatic examples of the observation and analysis of an electron-boson \emph{kink} in the ARPES spectrum range from metals \cite{valla1999many}, to cuprates (at $\Omega_0 \sim$70\,meV) \cite{lanzara2001evidence,johnson2001doping}, and graphene (at $\Omega_0 \sim$200\,meV) \cite{bostwick2007quasiparticle}.
In equilibrium, the imaginary part of a generic electron-phonon self-energy has the form 
\begin{equation} \label{EqSigmaEPh}
    \Sigma'' _{\text{e-ph}}(\textbf{k},\omega)= \int d\omega' \alpha^2 F(\textbf{k},\omega') K(\omega,\omega'),
\end{equation}
where the kernel $K(\omega,\omega')$ depends on the electronic and bosonic distributions, as well as the electronic DOS \cite{mahan2000many}. 
In the illustrative scenario wherein a single Einstein mode $\Omega_0$ is present [\emph{i.e.} $\alpha^2 F(\Omega)$\,=\,$\delta(\Omega-\Omega_0)$], the real part of the electron-phonon self-energy peaks at $|\omega|$\,=\,$\Omega_0$, while the imaginary part of the self-energy becomes non-zero for $|\omega|$\,$>$\,$\Omega_0$.
By extending Eq.\,\ref{EqSigmaEPh} into the time domain, it becomes clear that any light-induced changes of the electronic distribution, electron-phonon coupling or electronic DOS may lead to a transient modification of the self-energy $\Sigma _{\text{e-ph}}(\tau)$.  

Different experimental methods have been employed for extracting $\Sigma$ from TR-ARPES data. Given the inverse relation between $\Sigma''$ and the quasiparticle lifetime, qualitative estimates of underlying many-body interactions have been offered by tracking the energy- and momentum-resolved population dynamics. Although population dynamics do not correspond to the single-particle lifetime \cite{yang2015inequivalence}, the energy-dependence of the population dynamics may qualitatively mimic $\Sigma''(\omega)$ \cite{yang2015inequivalence,smallwood2015influence}. This approach, which establishes the basis for extracting the decay times of the transient photoemission intensity, has been successfully applied in Bi-based cuprates where the population relaxation rates change abruptly outside the boson window $|\omega|>$\,70\,meV \cite{rameau2016energy}. 

The development of TR-ARPES systems with an energy resolution better than the electron-phonon energy window has paved the way to extract the temporal evolution of the many-body self-energy $\Sigma(\tau)$ via analysis of the transient spectral function \cite{ishida2016quasi,zhang2014ultrafast,rameau2014photoinduced,miller2018interplay,duvel2022far,hwang2019ultrafast,zhang2022self}. For example, \onlinecite{zhang2014ultrafast} reported transient changes of the electron-boson \emph{kink} in Bi$_2$Sr$_2$CaCu$_2$O$_{8+\delta}$ (Bi2212) cuprate superconductors. As depicted in Fig.\,\ref{EP4_SelfEn}, near-IR excitation leads to an ultrafast softening of the $\sim$\,70\,meV band structure renormalization, which manifests as transient modifications of both the real and imaginary part of the electron self-energy at the phonon energy [panels (b) and (c), respectively]. The authors also reported a close link between the extracted $\Sigma(\tau)$ and the superconducting gap, thus suggesting that the electron-boson interaction responsible for the $\sim$\,70\,meV kink also plays an important role in the formation of the pairing gap.
Nonetheless, Eq.\,\ref{EqSigmaEPh} shows that transient changes in the self-energy are not unequivocally driven by the modification of the EPC, \emph{i.e.} $\alpha^2 F(\omega)$, but may also stem from pump-induced modifications to the electronic DOS, which has indeed been reported to occour on an ultrafast timescale in cuprates \cite{boschini2018collapse,smallwood2012tracking,zhang2017photoinduced,parham2017ultrafast}, although TR-ARPES data on metallic Bi2212 (non-superconducting) still shows signature of transient changes of the EPC \cite{hwang2019ultrafast}.

Although the extraction of $\Sigma(\tau)$ from TR-ARPES data has so far followed equilibrium analysis procedures, one should note that this procedure is not formally correct since the optical creation of a non-thermal electronic distribution may significantly impact the functional form of the self-energy (at least at early pump-probe delays) \cite{kemper2017review,kemper2018general}. Therefore, achieving a new robust understanding of light-induced changes of electron interactions will require meticulous analysis of the TR-ARPES spectra, accompanied by a full-non-equilibrium treatment of the TR-ARPES signal. 

\subsection{Mapping quantized electron-phonon decay processes} \label{Ketty}
When escaping from a solid, photoelectrons can couple to gapped collective modes and exchange a quantized amount of energy, leading to the emergence of shake-off replica bands in the equilibrium ARPES spectra \cite{li2018electron}. A similar phenomenon has recently been observed in the time domain via TR-ARPES, offering a new method for quantitatively extracting the momentum- and mode-resolved electron-phonon matrix elements \cite{na2019direct}. 

The experimental strategy consists of selectively injecting electrons into specific unoccupied states by optical pumping and then carefully tracking the transfer of spectral weight from those photoexcited states as they begin to relax to lower-energy states via emission of a phonon with energy $\hbar \Omega_0$, as sketched in the toy model of Fig.\,\ref{EP5_PIR}(a). The time constant associated with this transfer of spectral weight ($\tau_{\Omega_0}$) is directly related to the electron-phonon contribution to $\Sigma$ for the specific phonon involved \cite{na2019direct}:
\begin{equation}
    \frac{1}{\tau_{\Omega_0}}=2 \pi \hbar g_{\bar{k}} ^2 D(\omega_p-\Omega_0).
\end{equation}
Here, $\omega_p$ is the energy of the direct optical transition, $g_{\bar{k}} ^2$ is the square of the mode-projected electron-phonon matrix element averaged over the states $\bar{k}$ populated by the optical excitation, and $D$ is the electronic DOS. Note that the time constant $\tau_{\Omega_0}$ defines the contribution to the single-particle lifetime of an electron photoexcited in $\omega_p$ and then scattering off the specific phonon $\Omega_0$, and it is not directly related to the population dynamics extracted by tracking the transient evolution of the photoemission intensity (discussed in Section\,\ref{TTM}). 
\onlinecite{na2019direct} observed that the relaxation of non-thermal carriers at the K-point of graphite occurs in quantized energy steps corresponding to the emission of strongly coupled optical K-phonons. As seen in Fig.\,\ref{Intro_systems}(b) and Fig.\,\ref{EP5_PIR}(c) a series of peaks emerge in the momentum-integrated EDC of the unoccupied states; each peak represents either a direct transition peak (DTP) or a phonon induced replica (PIR). By tracking the transient evolution of DTP and PIR pairs [Fig.\,\ref{EP5_PIR}(d)], as well as by solving a system of rate equations, the authors were able to model the ultrafast timescales of the transfer of spectral weight and also to provide a quantitative estimate for the specific EPC involved in the process. The calculated $g_{\bar{k}}^2=0.050 \pm 0.011$\,eV$^2$ for the coupling of photoexcited electrons $\sim$\,0.6\,eV above the E$_F$ to K-optical A$_1$ phonons agreed with theoretical estimates [Fig.\,\ref{EP5_PIR}(b)], as well as with the experimental value obtained from ultrafast electron diffuse scattering \cite{stern2018mapping}.

In principle, this approach may find applications to a wide variety of quantum materials in which electrons are strongly coupled to a small number of bosonic modes and specific optical transitions are available. By monitoring quantized decays processes across the whole momentum space, this newly proposed method may allow the investigation of the microscopic origin and momentum dependence of the electron-boson coupling in quantum materials.

\subsection{Summary and Outlook}
This section reviewed multiple TR-ARPES approaches for extracting the electron-phonon coupling strength, and more generally electron-boson coupling, in a variety of quantum materials. 
Although many of these methods could in principle provide quantitative estimates of electron-boson coupling, in practice interpreting the TR-ARPES signal unambiguously is often challenging (see Sec.\,\ref{TR_ARPES_Int}), hindering a precise estimate of electron-boson scattering processes. The development of advanced TR-ARPES experimental and analytical approaches, in combination with other ultrafast techniques, will be crucial to disentangle electron-boson scattering events by removing spurious contributions (such as transient changes of the scattering phase space, screening and electronic distribution). 
We note that nearly all the TR-ARPES studies reviewed above rely on light excitations non-resonant to bosonic modes. That is, estimates of electron-boson coupling strength are obtained following extensive redistribution of the electronic density, which potentially changes the materials’ state, such as its polarizability and screening. There is a clear opportunity for new studies at the forefront of the field to more quantitatively assess  electron-boson processes in quantum materials by combining the analysis procedures discussed above with selective excitation of collective modes.

\begin{figure*}
\centering
\includegraphics[scale=1.02]{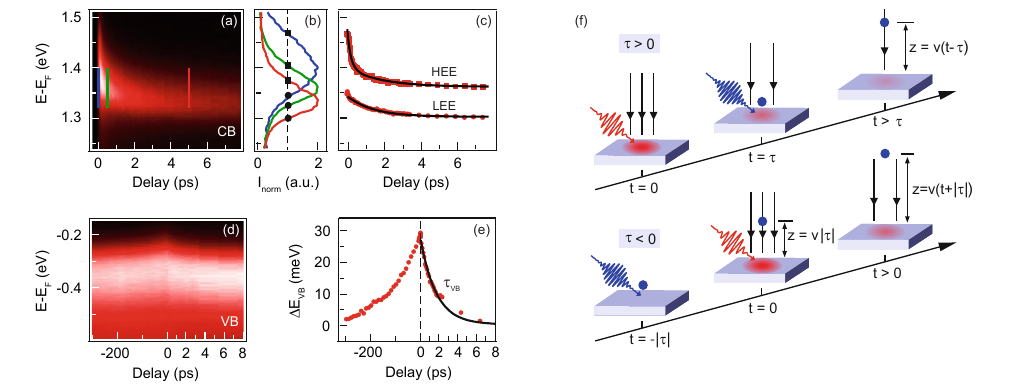}
\caption[SPV1]{Emergence of surface photovoltage (SPV) in GaAs(110). (a) Conduction band (CB) dynamics around the $\overline{\Gamma}$ point of p-type GaAs(110) upon ultrashort mid-IR optical excitation. (b) Normalized CB EDCs at 0\,ps (blue), 0.5\,ps (green) and 5\,ps (red). (c) Temporal evolution of the half-maximum point of the high-energy edge (HEE, squares) and low-energy edge (LEE, circles) of the EDC peak. (d) Valence band (VB) $\overline{\Gamma}$-EDC transient map and (e) energy shift dynamics. For both CB and VB the energy shift observed at positive delays can be described by double- or single-exponential fits: $\tau_{\mathrm{H1}}$\,=\,200\,$\pm$\,20\,fs and $\tau_{\mathrm{H2}}$\,=\,1.8\,$\pm$\,0.3\,ps for HEE edge; $\tau_{\mathrm{L}}$\,=\,1.4\,$\pm$\,0.1\,ps for LEE edge; $\tau_{\mathrm{VB}}$\,=\,1.7\,$\pm$\,0.1\,ps for VB. The shared $\sim$\,1.5\,ps dynamics stem from the emergence of the pump-induced SPV, which also causes the additional upwards shift of the VB at \textit{negative} delays. (f) Schematic interpretation of the positive- and negative-delay dynamics in the presence of SPV, where the black lines perpendicular to the sample surface represent the SPV-induced electric field.
Adapted from \onlinecite{yang2014electron}.}
\label{SPV_GaAs}
\end{figure*}

\section{Surface Photovoltage} \label{SPV_section}
As mentioned in Section \ref{TR_ARPES_Int}, TR-ARPES provides access to the transient electronic spectral function as a function of energy, momentum and time. For negative pump-probe delays, \textit{i.e.} before the arrival of the pump excitation at the sample surface, the system is usually postulated to be in thermal equilibrium, so no dynamics are expected. According to this picture, the energy and momentum of the photoelectrons are fully determined at the moment of photoemission from the material, and any electronic dynamics in the sample left behind do not affect their propagation in vacuum. However, these assumptions do not remain valid if the arrival of the pump pulse on the material's surface induces a time-dependent electric field, as may occur in systems with spatially separated electron-hole pairs, such as semiconductors. This surface photovoltage (SPV) plays a crucial role in governing the efficiency of photoelectric devices. In a simplified description, the SPV can be understood in terms of a light-induced variation of the surface potential driven by the redistribution of photo-excited carriers affecting the intrinsic band bending. 
To the first approximation, the SPV manifests in the TR-ARPES data as a time-dependent rigid shift in the binding energy of the electronic band structure, whose direction depends on the direction of the light-induced surface electric field. Of course, transient modifications of the surface potential may also result in changes in the confinement of surface states.

While SPV in conventional semiconductors had been extensively studied and exploited for many years \cite{kronik1999surface,schroder2001surface,widdra2003time,monch2013semiconductor,marsi2000transient,marsi1997surface}, the role of SPV in the context of TR-ARPES measurements was not addressed theoretically until a decade ago by \onlinecite{tanaka2012utility}. By systematically solving the classical equations of motion of photoelectrons from the sample's surface to the analyzer, they analyzed the detection by pump-probe photoemission of both the onset and the decay (\textit{i.e.} at negative and positive delays) of the transient SPV, establishing the importance of the proper consideration of the electron propagation.
Following this theoretical work, various experimental TR-ARPES studies have addressed the SPV phenomenon and factored its effect into both data acquisition and analysis \cite{yang2014electron,ciocys2019tracking,rettig2012ultrafastQWSPb}. 

\begin{figure*}
\centering
\includegraphics[scale=1]{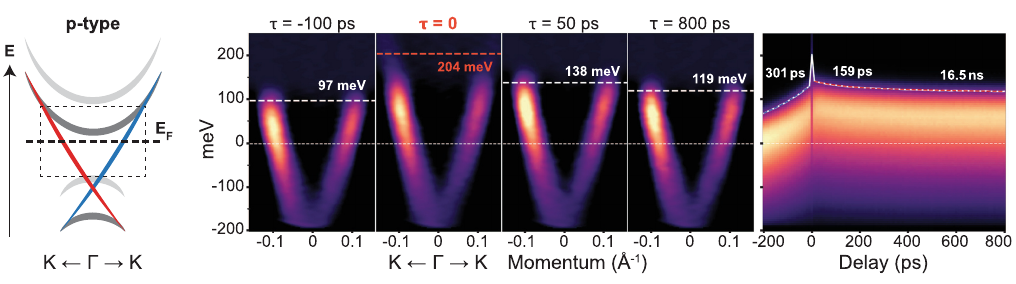}
\caption[SPV2]{Visualization of the effects of light-induced SPV in TR-ARPES data of p-doped Bi$_2$Se$_3$. The sketch on the left illustrates the topological surface state (red and blue) and bulk bands (gray) of p-doped Bi$_2$Se$_3$, where the Fermi level E$_{\text{F}}$ lies within the bulk band. The ARPES spectra at different pump-probe delays highlight the shift of the chemical potential observed before and after the optical excitation due to the SPV effect. The far right panel shows the full dynamics of the momentum-integrated intensity along K--$\Gamma$--K. Adapted from\,\onlinecite{ciocys2020manipulating}.}
\label{SPV_Lanzara}
\end{figure*}

This section discusses the signatures of emerging SPV in TR-ARPES experiments, its implications for interpreting the data, and how SPV can ultimately be exploited to transiently manipulate macroscopic properties of quantum materials, focusing on two classes of materials: semiconductors and semimetals in Section\,\ref{Semi_SPV} and 3D topological insulator in Section\,\ref{TI_SPV}.

\subsection{Semiconductors and semimetals} \label{Semi_SPV}

Due to their poor electronic screening, semiconductors are particularly sensitive to surface photovoltage. Focusing on p-type GaAs(110) as the typical semiconductor model system, \onlinecite{yang2014electron} first explored how SPV may manifest in a TR-ARPES experiment and outlined a basic framework to disentangle the system's intrinsic electronic dynamics from the transient field-induced features in the photoemission signal.
Upon 1.5\,eV pump excitation, both the valence band and the optically occupied conduction band exhibit a transient energy shift at positive delays with similar relaxation times of $\sim$\,1.5\,ps. Additional ultrafast dynamics are instead observed solely for the CB in the first 0.5\,ps accompanied by a change of lineshape [see Fig.\,\ref{SPV_GaAs}(a)-(c)]. While the latter timescale is attributed to the intraband CB electron-phonon scattering, the longer timescale dynamics shared by both CB and VB are associated with the SPV driven by the downward band bending in a p-type semiconductor, which leads to a rigid shift of the whole band structure. Furthermore, in contrast with what may be expected in a system at equilibrium, an energy shift of the VB is observed also at negative delays (\textit{i.e.} before pump excitation) lasting for hundreds of ps, as shown in Fig.\,\ref{SPV_GaAs}(d)-(f). This kind of long-lasting pre-excitation dynamics represents one of the main signatures of SPV in a TR-ARPES experiment, and it has been observed in different systems, such as the Kondo insulator SmB$_6$ \cite{ishida2015emergent}, graphene and black phosphorous \cite{ulstrup2015ramifications, hedayat2021non}, and various topological insulators (see Section\,\ref{TI_SPV}). As a technical note, it is worth mentioning that the SPV-induced negative-delay dynamics extending to hundreds of ps may serve as a convenient tool to facilitate the routine operations of spatially and temporally overlapping pump and probe beams on the sample surface -- a crucial requirement for any TR-ARPES experiment.

In order to disentangle the contributions of different physical processes, \onlinecite{yang2014electron} specifically included the propagation of photoelectrons through the SPV field in their interpretation of TR-ARPES data of GaAs(110), in line with the theoretical work of \onlinecite{tanaka2012utility}. While for positive delays the pump-induced field is generated on the sample's surface before the electrons are photoemitted, thus affecting the entire length of their propagation in vacuum, for negative delays the photoelectrons travel a finite distance before the SPV field develops, leading to a signal that maps the spatial profile of the field. 

Owing to its high carrier mobility and tunable band gap, black phosphorus represents a promising candidate for semiconducting devices and its photovoltaic response has recently been studied by TR-ARPES under chemical gating. In particular, \onlinecite{hedayat2021non, kremer2021ultrafast} have shown that the strong surface electric dipole, and associated downward band bending, induced by deposition of alkali atoms (such as K or Cs) can be fully compensated via illumination of the surface due to the emergence of SPV lasting on a timescale of ns. Two main SPV-induced effects are observed in the electronic structure at positive pump-probe delays: (i) a rigid energy shift affecting both the CB and surface VB; (ii) a transient modification of the VB spectral lineshape. The latter is caused by the SPV neutralization of the intrinsic band bending, which leads to bulk and surface VB contributions moving to similar binding energies. Therefore, in this scenario the dynamics of the VB bandwidth reflect the temporal evolution of the SPV as probed by TR-ARPES: the linewidth narrows in the first ps, remains roughly constant over the next 60\,ps then slowly returns to its equilibrium value over $\sim$1\,ns.

Finally, the observation of SPV may reflect additional intrinsic phenomena and interactions of quantum materials. For example, in SmB$_6$, where topological surface states are predicted to emerge upon formation of a Kondo hybridization bulk gap, an SPV signal lasting $>$\,200\,$\mu$s was detected only below the hybridization temperature 90\,K, indicating the development of the surface band bending associated with the evolution of the hybridization band itself \cite{ishida2015emergent}.

\begin{figure*}
\centering
\includegraphics[scale=1]{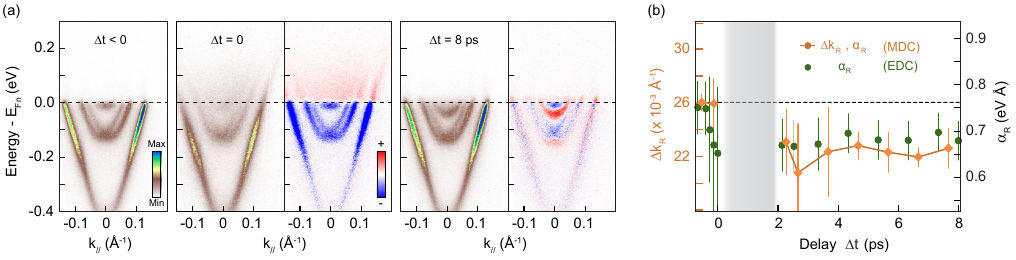}
\caption[SPV3]{SPV as a tuning knob of the Rashba spin-orbit coupling of a 2D electron gas at the surface of Bi$_2$Se$_3$. (a) ARPES dispersion at negative time delay (before pump arrival), at time zero (pump and probe fully overlapped), and at 8\,ps delay after the excitation. The differential spectra (obtained by subtracting the spectrum acquired at –0.5\,ps) are also shown for the zero and positive delays. An observable time-dependent energy shifts of the quantum well states (QWSs) at the Brillouin zone center is induced by the emergence of SPV. (b) Transient evolution of the Rashba spin-orbit coupling strength $\alpha_R$ in the first QWS as extracted by fitting momentum (orange) and energy (green) splitting at several time delays. Adapted from\,\onlinecite{michiardi2022optical}.}
\label{SPV_QMWs}
\end{figure*}

\subsection{Topological insulators} \label{TI_SPV}
As discussed in Section \ref{TIs}, TR-ARPES has been exploited extensively to study the electronic response to ultrafast optical excitations of both bulk and surface states of 3D topological insulators. SPV is one of the phenomena explored in relation to TIs, primarily in respect to electronic relaxation dynamics in bulk-insulating TIs. By comparing various $p$- and $n$-type Bi-chalcogenide TIs, \onlinecite{hajlaoui2014tuning} showed how different surface band bendings spatially separate photoexcited electrons and holes and affect the interplay between surface and bulk states dynamics. Upon near-IR excitation, a large charge asymmetry is observed after a few ps with an excess of holes in the surface states of the $n$-type sample due to a small upward band bending. In contrast, in the $p$-type Bi$_{2.2}$Te$_3$ the pre-existing downward band bending leads to an even greater excess of hot electrons in the Dirac cone. This charge disequilibrium has been suggested to be the main cause of the long carrier relaxation time of $>$\,50\,ps in Bi$_{2.2}$Te$_3$ and of the simultaneous transient shift of the chemical potential for the 2D Dirac states as large as $\sim$\,100\,meV, which corresponds to half the equilibrium gap size \cite{hajlaoui2014tuning}. Similarly, \onlinecite{ciocys2020manipulating} observed a long-lasting ($\sim$10\,ns) SPV in n-doped Bi$_{2}$Te$_3$ and p-doped Bi$_2$Se$_3$ compounds upon near-IR excitation. In this study the authors employed a large pump-probe delay range (for both negative and positive delay values) to reveal the full intrinsic timescale of SPV in TIs. As shown in Fig.\,\ref{SPV_Lanzara}, TR-ARPES mapping of p-doped Bi$_2$Se$_3$ reveals a shift of the chemical potential at negative delays with a $\sim$\,300\,ps timescale, induced by the interaction between photoelectrons and the SPV-driven electric field at the sample's surface (as discussed in the previous section also for semiconductors), whereas at positive delays this effect persists on a ns timescale. These results illustrate that examining a large pump-probe delay range is essential to extract the intrinsic formation and decay timescales of the SPV. A narrow time range may preclude detection of the full negative delay dynamics and incorrectly suggest that the negative delay shift of the chemical potential arises primarily as a build-up effect from previous pump pulses, thus leading to inaccurate estimates of the SPV lifetime.

A significant rigid shift of the chemical potential $\mu$ has been observed also for bulk-insulating ternary TIs driven by the emergence of large SPV \cite{yoshikawa2019bidirectional,papalazarou2018unraveling,neupane2015gigantic,sanchez2017laser,sumida2017prolonged,yoshikawa2018enhanced}. In Bi$_2$Te$_2$Se the shift of $\mu$ due to the SPV reaches a quasistatic value of $\sim$\,100\,meV after 100\,ps and subsequently relaxes on a very long timescale far exceeding the 4\,$\mu$s probed in the experiment \cite{neupane2015gigantic}. The same timescale characterizes the SPV observed at both room and low temperature for the $x$=0.55 compound of the Bi$_{1-x}$Sb$_{x}$Te$_3$ series \cite{sanchez2017laser,sumida2017prolonged}. In both cases this long-lasting SPV is linked to the much longer relaxation lifetime of the Dirac surface states with respect to the bulk bands, as both phenomena are affected by the different diffusion rates of hot carriers in the surface and bulk bands in the case of a truly bulk-insulating TI. In this respect, \onlinecite{papalazarou2018unraveling} have discussed the role of surface defects, rather than laser-induced band bending, in creating charge carrier separation after reporting a temperature-dependent change of sign of the SPV in Bi$_2$Te$_2$Se. According to this interpretation, special defects on the cleaved surface (such as donor-type vacancies) may change the charge density in the topmost layers leading to a near-surface electrostatic potential which confines photoexcited holes into the bulk. The observed SPV would then be induced by impurity states lying within the bulk energy gap.

Very recently, a few TR-ARPES studies on the emergence of SPV have addressed the exciting possibility of optically generating spin-polarized currents on the surface of a TI, owing to the helical texture of the topological surface state. Such an avenue is very promising in the context of potential spintronics applications, where active control of the electrons' degree of freedom is required. \onlinecite{michiardi2022optical} and \onlinecite{ciocys2022driving} demonstrated how light-induced surface photovoltage and charge redistribution can be utilized to manipulate the intrinsic material's spin properties. 
\onlinecite{michiardi2022optical} in particular analyzed the spectroscopic response to optical pumping of the Rashba-split quantum well states (QWSs) generated at the surface of Bi$_2$Se$_3$ via deposition of alkali atoms. As a typical fingerprint of the Rashba effect, these states consist of spin-polarized electronic bands that are offset in momentum and whose splitting is directly related to the strength of the Rashba spin-orbit coupling in the system $\alpha_R$. 
Upon near-IR excitation, a transient modification of the QWSs binding energy was observed on a picosecond timescale, whereas the topological surface state did not exhibit any significant change, as shown in Fig.\,\ref{SPV_QMWs}(a). This distinctive long-lasting response of the QWSs translates into a transient decrease of the splitting in both energy and momentum of the Rashba subbands and consequently of the intrinsic spin-orbit coupling strength $\alpha_R$. The latter was found to decrease by as much as 15\% over hundreds of picoseconds [Fig.\,\ref{SPV_QMWs}(b)]. \onlinecite{michiardi2022optical} ascribed the observed phenomena to a pump-induced SPV: following the optical excitation, excess negative charge accumulates at the surface, softening the original downwards band bending, which in turn significantly affects the dispersion of the QWSs [similar to what has been observed in thin epitaxial Pb films on Si(111) \cite{rettig2012ultrafastQWSPb}], while shifting the more surface-localized topological surface state in a more rigid fashion. 

Another recent TR-ARPES study on $\alpha$-GeTe(111) has shown an ultrafast enhancement of the Rashba coupling mediated by SPV in tandem with ion displacement, deduced by tracking the transient energy shift of the bulk bands, see Sec.\,\ref{CoherentPhon} \cite{kremer2022field}. The emergence of SPV prompts the excitation of displacive coherent phonons, thus ultimately controlling the ferroelectric properties of the $\alpha$-GeTe(111) system. 
These results attest to the potential of the TR-ARPES technique not only to test and characterize the SPV effect in quantum materials, but also to exploit this phenomenon to optically manipulate their physical properties, including the spin character, for future optically-controlled spintronics devices.

\subsection{Summary and Outlook}
Research conducted over the last decade has provided a robust description of the signatures of SPV in the context of TR-ARPES and we believe that dynamic SPV effects are now well characterized and understood. Design and analysis of TR-ARPES measurements require thoughtful consideration of SPV effects as they can be an important factor in understanding TR-ARPES data. Moreover, SPV also represents an opportunity to expand our optical control approaches to quantum materials, as very recent TR-ARPES experiments have demonstrated the possibility of exploiting it as a means to transiently manipulate the electronic and magnetic properties of topological insulators. Ongoing technical improvements to the laser sources and ARPES systems will enable detection of smaller SPV effects, enabling this methodology to evolve into a fine tuning knob supporting novel technological applications.

\section{Spin textures and their dynamics} \label{Spin_section}
Electron spin and its interactions with other degrees of freedom (\textit{e.g.}, spin-orbit coupling) often play a significant role in determining a material's electronic structure, from topological insulators \cite{cao2013mapping,zhu2013layer} to high-temperature superconductors \cite{gotlieb2018revealing}. Experimentally accessing these interactions is essential to reach a deeper understanding of the physical properties of quantum materials, as well as to advance the growing field of spintronics technology. The development and inclusion of spin polarimeters (either Mott or VLEED scattering detectors) in conventional ARPES systems, in concert with the use of the optical selection rules of helical light to excite and probe spin-orbital entangled states (dichroism), have created the foundation to study the spin texture of the electronic bands directly in momentum space. Accessing the spin-degree of freedom via ARPES is far from straightforward -- for instance, spin-resolved ARPES has been challenged by the limited efficiency of spin polarimeters, and the link between spin order and dichroic signal is not at all direct. Nevertheless, interest in studying spin interactions has driven efforts to develop combined spin-resolved and dichroism-based time-resolved ARPES approaches able to access not only the spin character of the electronic bands, but also the dynamical role played by the spin degree of freedom in the electronic relaxation processes following an optical excitation.

\begin{figure*}
\centering
\includegraphics[scale=1]{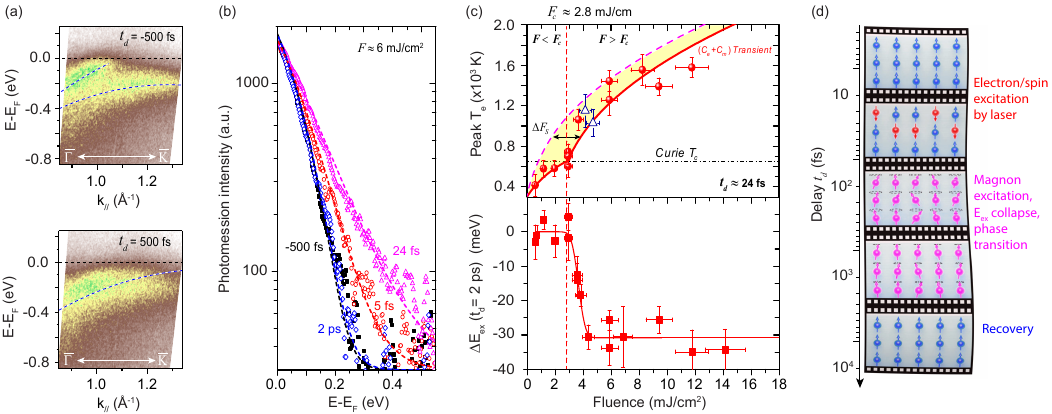}
\caption[Spin1]{Critical phenomena connecting the ultrafast demagnetization in ferromagnetic Ni to the equilibrium magnetic phase transition. 
(a) TR-ARPES spectra of Ni(111) along the $\Gamma$--K direction before (-500\,fs) and after (500\,fs) laser excitation, highlighting the transient reduction of the exchange splitting $\Delta$E$_{\text{ex}}$. 
(b) Momentum-integrated EDCs on a logarithmic scale at different pump-probe delays at $\sim$6\,mJ/cm$^2$ pump fluence. The electronic temperature, extracted by fitting momentum-integrated EDCs to a Fermi-Dirac distribution, reaches its maximum within $\sim$24\,fs after the optical excitation.
(c) Top: maximum electronic temperature at 24\,fs pump-probe delay as a function of the pump fluence. The yellow-colored region highlights the energy transferred to the spin bath ($\Delta F_S$) within $\sim$20\,fs. Bottom: $\Delta$E$_{\text{ex}}$ at 2\,ps pump-probe delay as a function of the pump fluence. Both panels report a similar critical fluence $F_c \approx$ 2.8\,mJ/cm$^2$.
(d) Schematic of the light-induced demagnetization process of Ni: upon a femtosecond pump excitation, the transient electronic temperature approaches and can surpass the Curie temperature within 20\,fs, which induces high-energy spin excitations, which store the magnetic energy. Demagnetization occurs later, in $\sim$\,175\,fs, driven by relaxation of nonequilibrium spins and the likely excitation of low-energy magnons. Full recovery of the spin system occurs within $\sim$\,0.5--100\,ps, depending on the laser fluence.
Adapted from\,\onlinecite{tengdin2018critical}.}
\label{FM_Ni}
\end{figure*}

This section reviews how TR-ARPES has been adapted to gain insights into the spin degree of freedom of various quantum materials, beginning with the achievements of the spin-resolved TR-ARPES technique in Section\,\ref{spin_TRARPES}, focusing on research into ferromagnetic materials and topologically protected systems. Next, Section\,\ref{dichroic_TRARPES} highlights how dynamical information about the spin may be obtained via analysis of the transient dichroic signal generated by utilizing different pump or probe polarizations. Finally, Section\,\ref{SDW_spin} briefly reviews how the capabilities of TR-ARPES can be instrumental in furthering our understanding of phases with spin-driven order in quantum materials.

\subsection{Spin-resolved TR-ARPES} \label{spin_TRARPES}
The advent of spin-resolved TR-ARPES \cite{cinchetti2006spin,melnikov2008nonequilibrium,gotlieb2013rapid,nie2019spin,fanciulli2020spin} has led to pioneering investigations of the transient evolution of the spin order of quantum materials upon light excitation. Despite significant advances in the efficiency of modern spin polarimeters, spin-resolved TR-ARPES remains a highly complex technique and requires long integration times. Consequently, most spin-resolved TR-ARPES studies have focused on extracting the transient spin dynamics for just one or at most a few momentum points. Section\,\ref{FM_spin} shows how spin-resolved TR-ARPES has helped to investigate the physical mechanisms underlying the ultrafast demagnetization process in ferromagnetic metals, while Section\,\ref{TI_spin} focuses on recent work on ultrafast spin dynamics in the surface resonant state and topological surface state of 3D topological insulators.

\subsubsection{Ferromagnetic materials} \label{FM_spin}
Since it was first observed experimentally in nickel \cite{beaurepaire1996ultrafast}, the light-induced ultrafast demagnetization of ferromagnets has intrigued the scientific community. Various proposals have been put forward for its underlying mechanism, from electron-magnon scattering \cite{carpene2008dynamics} to spin-lattice Elliott-Yafet events \cite{koopmans2010explaining} and superdiffusive currents \cite{battiato2010superdiffusive}. Most of the experimental efforts investigating ultrafast demagnetization processes in magnetic systems have historically relied on all-optical probes, such as magneto-optical Faraday and Kerr effects, second harmonic generation at the interface, as well as x-ray magnetic circular dichroism \cite{kirilyuk2010ultrafast}. Although all-optical probes can easily retrieve demagnetization timescales and amplitudes, they lack the momentum resolution needed to track the evolution of the spin-polarization of specific electronic states. 

By combining time-resolved high harmonic ARPES and magneto-optical probes, \onlinecite{tengdin2018critical,you2018revealing} showed that light-induced spin excitation in Ni actually occurs on a $\sim$20\,fs timescale, far faster than previously thought. By tracking the transient electronic temperature via fitting the ARPES spectra to a Fermi-Dirac distribution [Fig.\,\ref{FM_Ni}(a)-(b)], they revealed a critical behavior associated with a critical laser fluence ($F_c$) that is needed to drive the electronic temperature above the Curie temperature. As displayed in the top panel of Fig.\,\ref{FM_Ni}(c), as the laser fluence approaches $F_c$ the energy gained by the electron bath is reduced, due to the transfer of energy to the spin bath on ultrafast <20\,fs timescales. A related critical behavior is highlighted by the bottom panel of Fig.\,\ref{FM_Ni}(c), showing that the persisting collapse of the exchange splitting to longer timescales (>2\,ps) [as extracted by the ARPES maps, Fig.\,\ref{FM_Ni}(a)] occurs only for fluences above $F_c$. Therefore, while the spin bath can absorb sufficient laser pump pulse via high-energy spin excitations to store the magnetic energy -- and subsequently proceed through a magnetic phase transition --, the demagnetization and collapse of the exchange splitting occur on much longer $\sim$200\,fs timescales [see Fig.\,\ref{FM_Ni}(d)], after the magnon population has equilibrated.
These findings connect the light-induced out-of-equilibrium magnetic state to the equilibrium ferromagnetic-to-paramagnetic phase transition that occurs at the Curie temperature in Ni, showing that the electronic temperature alone dictates the magnetic response in this single-element material.

In another study, \onlinecite{eich2017band} showed that the ultrafast quenching of the magnetization in another 3d ferromagnet, Co/Cu(001), cannot be explained by a reduction of the exchange splitting. Instead, the authors ascribed the observed ultrafast demagnetization process to efficient emission of magnons as the leading underlying mechanism. Such emission of magnons would be associated with fluctuating spin-split electronic states, which result in a folding of states of different spins one on top of the other.
Despite more TR-ARPES studies on 3$d$ \cite{rhie2003femtosecond,scholl1997ultrafast,gort2018early,schmidt2010ultrafast,weber2011ultrafast,fognini2014ultrafast} and 4$f$ \cite{lisowski2005femtosecond,frietsch2020role,carley2012femtosecond} ferromagnets, which have successfully probed the ultrafast dynamics of spin-polarization or exchange splitting for a specific momentum upon light perturbation, further research is needed to completely unravel the microscopic mechanism and to reach a comprehensive description of the ultrafast demagnetization effect.

\subsubsection{Topologically-protected systems} \label{TI_spin}
\begin{figure}
\centering
\includegraphics[scale=1]{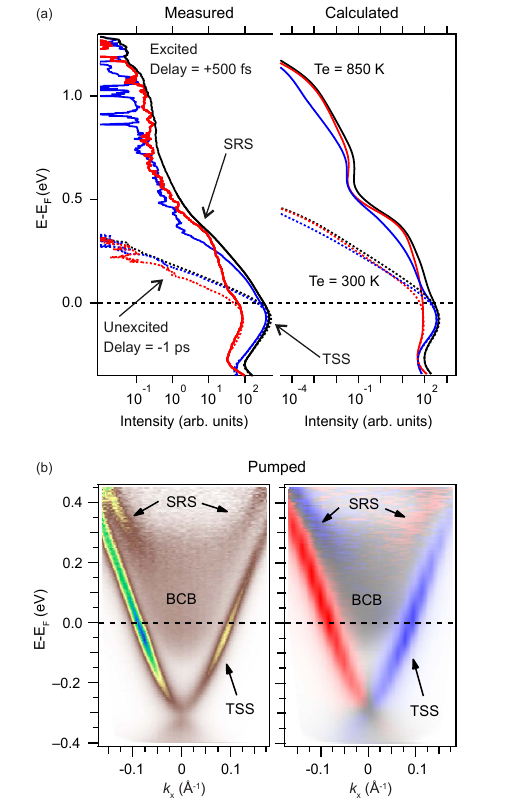}
\caption[Spin2]{Transient mapping of the spin-resolved surface resonance state (SRS) in Bi$_2$Se$_3$. (a) Measured and calculated spin-resolved EDCs along the $\Gamma$-K direction close to the TSS. Dotted (continuous) lines indicate opposite spin-resolved photoemission intensities before (after) the optical excitation, while TSS and SRS are indicated by arrows. Adapted from\,\onlinecite{cacho2015momentum}. (b) Spin-integrated and spin-resolved ARPES maps (left and right panels, respectively) along the $\Gamma$-K direction at 0.7\,ps pump-probe delay upon near-IR excitation. Adapted from\,\onlinecite{jozwiak2016spin}.}
\label{FM_SRS}
\end{figure}

Topological systems such as topological insulators or Weyl semimetals host topologically protected states with characteristic spin textures \cite{yan2017topological,hasan2010colloquium,cao2013mapping,zhu2013layer}. The unique capabilities of spin-resolved TR-ARPES allow to assess unoccupied spin-polarized states, as well as to track ultrafast carrier relaxation processes with spin resolution.
\onlinecite{cacho2015momentum}, followed by \onlinecite{jozwiak2016spin}, employed spin-resolved TR-ARPES to map a topologically trivial spin-polarized surface resonance state (SRS) within the unoccupied bulk conduction band of Bi$_2$Se$_3$, the prototypical 3D topological insulator. Figure\,\ref{FM_SRS}(a) compares measured and calculated spin-resolved EDCs, hinting at the presence of the SRS $\sim$\,0.4\,eV above the Fermi level \cite{cacho2015momentum}. A direct visualization of the SRS spin-polarization is given in the spin-resolved TR-ARPES maps of Fig.\,\ref{FM_SRS}(b), which clearly confirm the presence of the SRS, characterize its dispersion, and also reveal that its spin-polarization is opposite that of the TSS \cite{jozwiak2016spin}. Subsequent spin-resolved TR-ARPES experimental studies have also verified the presence of additional SRSs in the unoccupied bulk band gaps of Bi$_2$Te$_3$ \cite{sanchez2017subpicosecond}.
Spin-resolved TR-ARPES has also been recently employed to reveal the spin-polarization of electrons bound into a spatially indirect topological exciton in Bi$_2$Te$_3$, as discussed in Sec.\,\ref{exciton_map} and shown in Fig.\,\ref{Map_TopologialEx}(b).

Spin-resolved TR-ARPES can also characterize the transient evolution of the spin-polarization of the light-injected electron population. \onlinecite{jozwiak2016spin} demonstrated that the spin-polarization of the SRS does not evolve with time, ensuring that the spin-resolved spectrum of Fig.\,\ref{FM_SRS}(b) directly reflects the intrinsic spin-polarization of the SRS rather than a pump-induced selective occupation of those states. Indeed, if the transiently occupied eigenstates were spin unpolarized, such a population would tend to depolarize on a sub-picosecond timescale \cite{hsieh2011selective}, whereas the SRS in Bi$_2$Se$_3$ retains the same polarization during the whole relaxation process. 
However, note that the use of tailored optical excitation may transiently alter the polarization of carriers injected into a final state. As an emblematic example, \onlinecite{sanchez2016ultrafast} reported a $\sim$\,1\,ps decay time for the transient out-of-plane spin polarization of electrons excited into the topological surface state of p-doped Sb$_2$Te$_3$ upon circular near-IR excitation. This work not only demonstrated light-control of the out-of-plane spin-polarization of carriers photo-injected into the TSS, but the vanishing of the out-of-plane spin polarization on a picosecond timescale also hints at an inefficient depolarization mechanism via electron-phonon scattering.

\begin{figure*}
\centering
\includegraphics[scale=1]{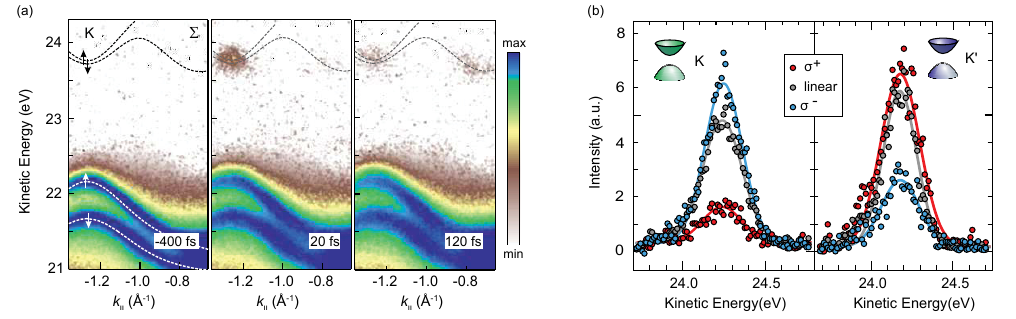}
\caption[SpinDicTMD]{Selective population of the K/K' valley of 2H-WSe$_2$ upon circularly polarized optical excitation. (a) TR-ARPES maps at various pump-probe delays acquired with linearly polarized pump pulses (logarithmic color scale). The dashed lines display the VB and CB dispersion obtained by DFT calculation for a bilayer compound, with the CB shifted 250\,meV upwards to match the experimental observations. (b) Energy distribution curves of the excited state signal at K and K', 15\,fs after excitation with linearly and circularly polarized optical pump. Note the strong valley-selectivity attainable by exciting with circular polarized light. Adapted from \onlinecite{bertoni2016generation}.}
\label{Spin_Dic_TMD}
\end{figure*}

Finally, a recent spin-resolved XUV-based TR-ARPES study has addressed the spin-polarization of the unoccupied states of the putative Weyl type-II semimetal WTe$_2$, in which Weyl points are predicted above the Fermi level \cite{fanciulli2020spin}. Although unable to clearly capture the presence of such Weyl points from the spin-polarized spectra, the authors reported a spin-selective bottleneck effect at 1\,ps in the region above their expected energy. This observation points toward a reduction of the available scattering phase space due to a defined spin texture of the unoccupied states, thus supporting the topological nature of WTe$_2$. 

\begin{figure*}
\centering
\includegraphics[scale=1]{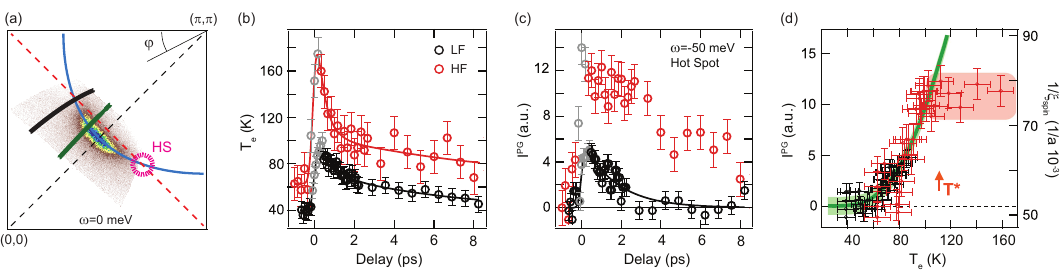}
\caption[Spin3]{Tracking the filling of the pseudogap driven by shortening of the spin-correlation length in the optimally-doped Nd$_{\text{2-x}}$Ce$_{\text{x}}$CuO$_{\text{4}}$ electron-doped cuprate. (a) Experimental Fermi surface measured with a 6.2\,eV at 10\, K. The solid blue line is a tight-binding constant energy contour at $\omega$\, =\, 0\,meV, the red dashed line marks the antiferromagnetic zone boundary, and the violet dotted circle indicates the hot spot (HS) position. Green and the black solid cuts mark the two momentum directions explored in the study: near-node and HS. 
(b) Transient electronic temperature extracted by fitting the Fermi edge broadening along the near-nodal direction for two fluence regimes: low-fluence (LF) and high-fluence (HF). Grey points indicate time delays for which a pure thermal fitting is not accurate. (c) Temporal evolution for both LF and HF of the photoemission intensity at the HS for $\omega$ \,=\, −50$\pm$10\,meV, representing the evolution of the pseudogap. (d) Photoemission intensity at the HS, $\omega$ \,=\, −50$\pm$10\,meV, as a function of the electronic temperature (black and red circles for LF and HF, respectively). The green line and shadow represent the inverse of the spin-correlation length $\xi_{\text{spin}}$ from neutron scattering studies. Saturation of the photoemission intensity at the HS (red shadow) marks the temperature T$^*$ at which a complete filling of the PG is observed. Adapted from \onlinecite{boschini2020emergence}.}
\label{SDW_PG}
\end{figure*}

\subsection{Dichroic TR-ARPES} \label{dichroic_TRARPES}
Although spin-resolved ARPES directly measures the spin degree of freedom by employing spin polarimeters, it confronts an intrinsic challenge of low efficiency due to the small spin-dependent scattering cross-section. One way to partially work around this challenge relies on the fact that the photoemission process is essentially an optical transition between an initial and final state, so the ARPES signal depends on the polarization of the incoming light via the matrix element term. Analyzing the difference in the photoemission signal acquired with different light polarizations (both linear and circular), a.k.a. dichroic signal, can provide insight into the symmetry of the initial state wavefunction. The polarization of the incoming light becomes a tunable parameter to selectively couple to electronic states of different total angular momentum (or \textit{pseudospin}) quantum number. Although in the case of spin-orbit coupled systems the dichroic signal may resemble the underlying spin texture as an indirect consequence of optical selection rules related to the orbital angular momentum of electronic states, the connection between dichroism and spin order of the initial states is not straightforward and its interpretation may be further complicated by the symmetry of the final photoemission states. 

Since TR-ARPES employs two separate light pulses, it is necessary to distinguish the cases of dichroic signals produced by differently polarized pump pulses vs. probe pulses. The latter is the transient counterpart of static dichroic ARPES \cite{schuler2022polarization,sobota2021angle,schuler2020circular,wang2013circular,beaulieu2020revealing,moser2022toy} and has been mainly applied to the study of topological insulators \cite{hedayat2021ultrafast, zhang2021probing} and Rashba materials \cite{mauchain2013circular,zhang2022dichroism}.
As mentioned above, linear or circular probe dichroic photoemission signals can offer information on whether the initial or final photoemission states have a well-defined spin or orbital angular momentum. Therefore, when applied to materials with strong spin-orbit coupling or spin-momentum locking, circular probe TR-ARPES dichroism may offer a qualitative mapping of the underlying spin-texture of unoccupied states and their related spin dynamics. However, a detailed interpretation of the TR-ARPES signal acquired with a dichroic probe is often a challenging undertaking (as is its equilibrium counterpart). Indeed, such a task requires the ability to unambiguously disentangle whether the evolution of the circular probe dichroic signal arises from transient changes of the spin or orbital contributions to the wavefunction of both the initial and final photoemission states. This involves also the evaluation of potential changes in the photoemission matrix elements \cite{boschini2020role}, which are not associated with any modifications of the spin polarization itself. For this reason, the following discussion focuses solely on TR-ARPES studies performed with dichroic pump excitation in which changes in the dichroism asymmetry can be directly related to the selective photoexcitation of polarized electrons. 

TMDs exemplify the potential of the pump-dichroic TR-ARPES technique. As mentioned in Sec.\,\ref{CB_semi} while discussing their unoccupied band structure, the two-atoms-basis and the breaking of inversion symmetry in TMDs lead to the in-equivalence of two adjacent K points, commonly identified as the K and K' valleys. Of particular importance, electrons in the CB and VB have different azimuthal quantum numbers (m): m$_{\text{VB}}$=0 and m$_{\text{CB}}$=$\pm$1 at both K and K', respectively \cite{cao2012valley}. Therefore, right (left) circularly polarized light can only promote K (K') transitions, prompting the chiral optical selection rule of TMDs between two bands sharing the same spin. Pump-dichroic TR-ARPES has verified this selection rule on bulk 2H-WSe$_2$ \cite{bertoni2016generation}, monolayer WS$_2$ \cite{ulstrup2017spin,beyer201980}, and bilayer MoS$_2$ \cite{volckaert2019momentum}.
Figure\,\ref{Spin_Dic_TMD}(b) displays the different photoemission signal acquired at K and K' upon pump pulses of different polarizations in bulk 2H-WSe$_2$, demonstrating that circularly-polarized optical excitations drive a layer-dependent valley-polarized population \cite{bertoni2016generation}. The pump-induced K/K' valley populations decay on a 100\,fs timescale into the $\Sigma$ valley (\emph{i.e.} the global CB minimum), as shown in the TR-ARPES maps of Fig.\,\ref{Spin_Dic_TMD}(a).
Moreover, by investigating singly-oriented monolayer WS$_2$ on Ag(111), \onlinecite{beyer201980} reported a light-induced valley-polarization of $\sim$85$\%$ for the top of the VB and only $\sim$55$\%$ for the bottom of the CB. This discrepancy of the valley-polarization between the initial and final optical states was understood in terms of a difference in the efficiency of K--K' intervalley scattering for the CB and VB (for the latter, the large spin-orbit split mitigates intervalley scattering events).
These examples show how circular pump light may be utilized to prepare the excited state with a well-defined spin/valley polarization, thus enabling TR-ARPES to track scattering processes involving specific polarized carriers.

In addition to TMDs, 3D topological insulators have been among the first systems to which pump dichroic TR-ARPES has been applied, due to the characteristic spin-texture of their topological surface state. \onlinecite{kuroda2017ultrafast} investigated linear and circular photogalvanic effects due to polarization-variable mid-IR excitation on the topological surface state of Sb$_2$Te$_3$. Relying on the selective population of electronic states with opposite pseudospin, \onlinecite{niesner2012unoccupied} first reported the presence of a spin-polarized second surface state in the unoccupied band structure of Bi$_2$Se$_3$, and this observation was then confirmed via 6-eV two-photon photoemission \cite{sobota2013direct,sobota2014ultrafast}. Subsequent TR-ARPES studies with circular optical excitation have shown a population asymmetry in this second surface state as a function of the different pump polarization \cite{bugini2017ultrafast,soifer2019band}. The observed dichroic signal has been proposed to be related to an ultrafast spin-polarized surface current, although its intrinsic depolarization time is sub-50\,fs.

\subsection{Spin Density Wave gaps} \label{SDW_spin}
Even without the use of spin polarimeters or dichroic signals to extract spin textures and dynamics, TR-ARPES can provide insights into spin-ordered phases. In analogy to the approach used to track the charge-order spectral gap (see Section\,\ref{CO_phase}), light-induced modifications of electronic gaps mediated by spin-density-waves (SDW) may offer direct insights into the transient evolution of underlying spin orders \cite{boschini2020emergence,suzuki2017ultrafast,rettig2012ultrafast,nicholson2016ultrafast}.
\onlinecite{nicholson2016ultrafast} have elegantly demonstrated this concept by tracking the transient evolution of the antiferromagnetic order parameter in Cr(110) and relating it to the transient evolution of the electronic temperature. This work suggested that the photoexcited electrons form a quasi-equilibrium state with spin order, thus allowing the use of thermodynamic concepts even on ultrafast timescales.

Along the same line, \onlinecite{boschini2020emergence} tracked the transient filling of the SDW gap at the antiferromagnetic hot spot of the optimally-doped Nd$_{\text{2-x}}$Ce$_{\text{x}}$CuO$_{\text{4}}$ electron-doped cuprate [see Fig.\,\ref{SDW_PG}(a)]. Although optimally-doped NCCO does not exhibit any long-range magnetic order, it is characterized by a short-range antiferromagnetic order with a spin-correlation length that decreases with temperature. For this reason, the SDW-mediated electronic gap at the antiferromagnetic hot spot is only partial, and it is often referred to as the pseudogap (PG) \cite{armitage2001anomalous}.
\onlinecite{boschini2020emergence} employed TR-ARPES to map the transient modifications of the PG spectral feature into an effective temperature evolution, revealing a filling rather than a closing of the PG with increasing temperature, as shown in Fig.\,\ref{SDW_PG}(b)-(d).
This result recalls the case of the superconducting gap in hole-doped cuprates: whereas in those systems the quenching of phase coherence drives the filling of the gap (see Sec.\,\ref{SC_phase}), in electron-doped NCCO the filling of the gap is instead associated with the temperature evolution of the spin correlation length [green line and shadow in Fig.\,\ref{SDW_PG}(d)]. The PG is completely filled at the crossover temperature T$^{*}$, for which the spectral broadening driven by the reduction of spin-correlation length overcomes the PG amplitude.
This work provided direct evidence for the primary role of short-range antiferromagnetic correlations in determining the partial suppression of spectral weight at the hot spots in electron-doped cuprates \cite{boschini2020emergence}, and showcased the transient tracking of SDW gap amplitude or filling as a reliable approach to retrieve information about the long- and short-range spin orders.

\subsection{Summary and Outlook}
In summary, extensive technical and analytical advances of the TR-ARPES technique over the past two decades have finally unlocked investigations of the spin degree of freedom directly in momentum space. This section reviewed how TR-ARPES has been adopted in different ways to gain such information, from the spin-resolved fashion to the analysis of transient dichroic signals and the tracking of spin-induced electronic gaps. While each of these approaches faces challenges and limitations (such as low detector efficiency and arduous data analysis), it is clear that TR-ARPES has the potential to expand our understanding of the role of electronic spins in defining the properties of quantum materials far beyond what can be learned from equilibrium ARPES. Looking ahead, the advent of momentum microscopes with two-dimensional spin filtering \cite{tusche2015spin,kutnyakhov2016spin} promises new insights into spin dynamics of quantum materials and spintronic devices.

\section{Concluding Remarks} \label{Conclusions_section}
TR-ARPES has emerged as one of the most powerful tools for investigating dynamical properties of quantum materials, as exemplified by the scientific achievements presented in this review. Although TR-ARPES has primarily relied on optical excitation in the near-IR/visible spectral range at its early stages, it nonetheless has offered fascinating insights into electron dynamics and light-induced manipulation of the electronic band structure with momentum resolution in disparate quantum materials, ranging from high-temperature superconductors \cite{perfetti2007ultrafast,smallwood2012tracking,boschini2018collapse} to charge-ordered systems \cite{schmitt2008transient,rohwer2011collapse}, as well as topological systems \cite{sobota2012ultrafast,hajlaoui2012ultrafast,wang2012measurement,crepaldi2012ultrafast} and monolayer transition metal dichalcogenides \cite{cabo2015observation}. In recent years, advances in the generation of mid-IR/THz pulses and the use of momentum microscopes have enabled the observation of Floquet-Bloch states \cite{wang2013observation,zhou2023pseudospin,ito2023build}, THz-driven currents \cite{reimann2018subcycle}, and bright/dark excitons in $\mu$m-sized exfoliated transition metal dichalcogenides \cite{madeo2020directly,karni2022structure,schmitt2022formation,dong2021direct}, thus inspiring development of next-generation TR-ARPES systems and advances in theoretical analysis of TR-ARPES data.

\begin{figure}
\centering
\includegraphics[scale=1]{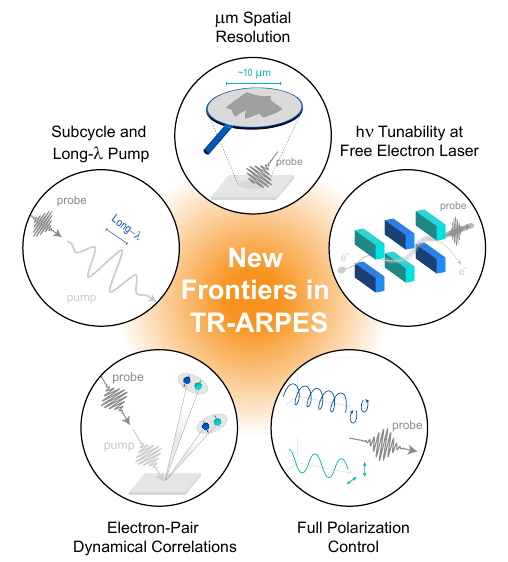}
\caption[FinalFig]{Illustration of the forthcoming new frontiers of the TR-ARPES technique in the coming years.}
\label{FinalFig}
\end{figure}

\vspace{2mm}

Since this panoramic review has covered what the TR-ARPES technique has accomplished to date, we would now like to offer a forward-looking view of what we believe will be the new frontiers of TR-ARPES in the coming years, focusing on five different ongoing or anticipated advancements that we expect will mature over the next decade (summarized in Fig.\,\ref{FinalFig}).

\subsection*{I) Subcycle and long-wavelength-pump TR-ARPES} 
In recent years, advances in the generation of pump pulses in the mid-infrared-to-THz range have enabled tracking electron dynamics beyond the conventional near-infrared excitation scheme \cite{gierz2013snapshots,kuroda2016generation,kuroda2017ultrafast,chavez2019charge}, such as the observation of Floquet-Bloch states \cite{wang2013observation,mahmood2016selective,reutzel2020coherent,aeschlimann2021survival,ito2023build}, as well as the first THz-pump TR-ARPES work on topological insulators \cite{reimann2018subcycle}. Furthermore, the two seminal papers of \onlinecite{reimann2018subcycle} and \onlinecite{ito2023build} have demonstrated the capability of tracking light-matter interaction processes with subcycle resolution, thus opening the way for subcycle TR-ARPES to explore highly out-of-equilibrium processes. We expect the TR-ARPES community to continue to advance tunable and intense long-wavelength pump sources (\textit{e.g.}, $\lambda$\,$>$\,4\,$\mu$m -- h$\nu$\,$<$\,300\,meV) to coherently excite collective modes and drive the emergence of new phases of matter with no equilibrium counterpart. Ideally, we contemplate TR-ARPES experiments where the pump excitation does not couple to the electronic bath at all but instead launches collective excitations, which in turn modify and renormalize the single-particle dispersion.

\subsection*{II) TR-ARPES with $\mu$m spatial resolution on exfoliated and stacked/twisted materials}
The recent development of TR-ARPES systems with $\mu$m spatial resolution (via momentum microscopy) has demonstrated the capability of measuring electron and exciton dynamics in exfoliated samples and Moir\'{e} heterostructures \cite{madeo2020directly,karni2022structure,schmitt2022formation,dong2021direct}. However, so far, momentum microscopes cannot easily achieve sub-20\,meV energy resolutions and are not well suited for measuring irregular/non-ideal cleaved surfaces. For this reason, we envision the development of TR-ARPES systems with $\mu$m UV/XUV spot sizes, coupled to hemispherical or time-of-flight detectors operating at high repetition rates to mitigate space-charge effects \cite{Dufresne2023RSI}. These efforts should enable the investigation of ultrafast electron dynamics in twisted Moir\'{e} structures \cite{andrei2021marvels,kennes2021moire,topp2019topological,topp2021light,rodriguez2021low}, and in general $\mu$m-size samples, with high enough energy resolution to resolve light-induced changes of the low-energy electronic structure.

\subsection*{III) TR-ARPES at free electron laser (FEL) facilities and with tunable $\Delta$E$\Delta \tau$ product}
FELs are rapidly upgrading their operation to the kHz-to-MHz range, thus enabling TR-ARPES studies that otherwise would have required unreasonably long acquisition times at sub-1\,kHz repetition rates \cite{rossbach201910,oloff2016time,kutnyakhov2020time}. FEL facilities promise access to fully tunable probe pulses in the XUV-to-soft-X-ray photon energy range, which would allow performing core-level photoemission studies \cite{hellmann2010ultrafast,hellmann2012timeFEL,dendzik2020observation,curcio2021ultrafast,pietzsch2008towards}, as well as tracking light-induced changes in electronic structure throughout the entire 3D Brillouin zone (\emph{e.g.}, by probing different k${_\text{z}}$ via a photon energy dependent scan), hence differentiating surface from bulk contributions to the photoemission intensity. When considering probe tunability, we expect that FEL facilities and new in-house TR-ARPES systems will also enable control of the $\Delta$E$\Delta \tau$ product. Indeed, this review covered several experimental studies that employed diametrically opposed working parameters, such as tracking the superconducting gap of cuprates \cite{parham2017ultrafast,boschini2018collapse} thanks to high energy resolution, or tracking the out-of-equilibrium electron dynamics on sub-30\,fs timescales \cite{rohde2018ultrafast,tengdin2018critical}. Undoubtedly, the TR-ARPES community would like to continuously tune $\Delta$E$\Delta \tau$ (for instance, by adjusting the probe pulse bandwidth and its compression) in an effort to reliably correlate ultrafast changes in the electronic distribution and photoemission intensity to changes in the underlying spectral function and orbital characters for all quantum materials. Alternative experimental strategies to overcome the Fourier uncertainty to achieve high energy and temporal resolutions simultaneously could rely on the use of pairs of entangled photons \cite{gu2023photoelectron} or double probe interferometric schemes \cite{randi2017bypassing}.

\subsection*{IV) TR-ARPES with full polarization control to explore orbital and spin textures} 
While tunable polarization of  low-energy pump pulses is already widely available, full control of the probe pulse polarization (linear and circular), especially in the XUV range, presents greater challenges. Advances in this area will likely enable careful studies on the symmetry of the orbital and spin degree of freedom of transient states in magnetic and spin-orbital coupled systems \cite{beaulieu2020revealing,schuler2022probing}, as well as full disentanglment of matrix element effects from the transient photoemission intensity \cite{boschini2020role}.
XUV light sources that can access the entire 3D Brillouin zone with full polarization control  are already under construction \cite{comby2022ultrafast}, and we expect that this new experimental capability will not only facilitate and expand the current understanding of what we precisely measure in a TR-ARPES experiment, but will also allow the detection of transient states that have so far eluded us.

\subsection*{V) Observation of dynamic electron correlations with momentum resolution}
TR-ARPES, in common with ARPES, probes the one electron-removal spectral function and infers underlying many-body correlations via analysis of the lineshape and dispersion of spectral features. However, it would be desirable to directly access the dispersion relation and correlation strength of two (or more) entangled particles (\emph{e.g.}, a Cooper pair) with momentum resolution. To this end, extensions of the ARPES technique such as two-electron ARPES (2e-ARPES) \cite{berakdar1998emission,mahmood2022distinguishing} or noise--correlation ARPES \cite{stahl2019noise} have been proposed. We anticipate that 2e-ARPES and noise--correlation ARPES (and other possible approaches that might go beyond the single-particle picture) will be naturally extended into the time domain, thus providing a comprehensive picture of how light-excitation may disrupt, enhance, or lock charge correlations among specific electronic states. Moreover, we foresee increasing interest in combining TR-ARPES with other complementary time-resolved techniques, such as ultrafast electron and X-ray diffraction \cite{filippetto2022ultrafast,elsaesser2014perspective,gerber2017femtosecond}, time-resolved inelastic X-ray scattering \cite{mitrano2020probing}, and time-resolved electron energy loss spectroscopy to link light-induced changes of the correlation strength between specific electronic states to those of the dynamic charge density response function.

\vspace{5mm}
In conclusion, while the work reviewed here has already established TR-ARPES as a mature technique and demonstrated its impact on various branches of physics and chemistry, further experimental and theoretical developments -- augmented by a portfolio of complementary time-resolved techniques -- strongly suggest that even more exciting times lie ahead.
\vspace{5mm}

\begin{acknowledgments}
We would first like to express our gratitude to the current and former members of the UBC ARPES group for their cooperation, daily discussions, and many useful comments: M. Bluschke, R. P. Day, P. Dosanjh, S. Dufresne, I. Elfimov, H.-H. Kung, G. Levy, I. Markovic, M. Michiardi, A. K. Mills, M. Na, E. Razzoli, M. Schneider, S. Smit, D. Wong, S. Zhadanovich, B. Zwartsenberg. Over the years, we have benefited tremendously from scientific discussions with D. J. Jones, G. A. Sawatzky, C. Giannetti, and F. Parmigiani. In addition, we are thankful to colleagues who provided comments and suggestions to this review, helping us to improve it greatly: E. Baldini, M. Bauer, S. Beaulieu, U. Bovensiepen, E. Carpene, A. Cavalleri, A. Crepaldi, K. M. Dani, C. Dallera, J. Freericks, N. Gedik, U. Hoefer, P. Hofmann, A. K. Kemper, P.  Kirchmann, M. Marsi, S. Mathias, C. Monney, M. M. Murnane, C. Nicholson, K. Rossnagel, M. Sch\"{u}ler, M. A. Sentef, J. Sobota, J. Staehler, K. Tanimura, S. Ulstrup, W. Zhang, and S. Zhou.
This project was undertaken thanks in part to funding from the Max Planck-UBC-UTokyo Center for Quantum Materials and the Canada First Research Excellence Fund, Quantum Materials and Future Technologies Program. This effort was also funded by the Gordon and Betty Moore Foundation’s EPiQS Initiative, Grant GMBF4779 to A.D.; the Killam, Alfred P. Sloan, and Natural Sciences and Engineering Research Council of Canada’s (NSERC’s) Steacie Memorial Fellowships (A.D.); the Alexander von Humboldt Foundation (A.D.); the Canada Research Chairs Program (A.D.); NSERC; the Department of National Defence (DND); Canada Foundation for Innovation (CFI); the British Columbia Knowledge Development Fund (BCKDF); and the CIFAR Quantum Materials Program (A.D.). F.B. acknowledges support from the Fonds de recherche du Qu\'{e}bec – Nature et Technologies (FRQNT) and the Minist\`{e}re de l’\'{E}conomie, de l’Innovation et de l’\'{E}nergie - Qu\'{e}bec.
\end{acknowledgments}

\bibliographystyle{apsrmp4-1}
\providecommand{\noopsort}[1]{}\providecommand{\singleletter}[1]{#1}
\end{document}